\documentclass[structabstract]{aa}  
\usepackage{amsmath}
\usepackage{graphicx}
\usepackage[varg]{txfonts}
\usepackage{longtable}
\usepackage{longtable,lscape}
\usepackage{lscape}
\usepackage{epsfig}
\usepackage{txfonts}
\begin{document}

\title{Protostars, multiplicity, and disk evolution in the Corona Australis region: A Herschel Gould Belt Study\thanks{{\it Herschel} is an ESA space observatory with science instruments provided by European-led Principal Investigator consortia and with important participation from NASA.}}

\author{Aurora Sicilia-Aguilar\inst{1}, Thomas Henning\inst{2}, Hendrik Linz\inst{2}, Philippe Andr\'{e}\inst{3}, 
Amy Stutz\inst{2}, Carlos Eiroa\inst{1}, Glenn J. White\inst{4,5}}

\institute{\inst{1}Departamento de F\'{\i}sica Te\'{o}rica, Facultad de Ciencias, Universidad Aut\'{o}noma de Madrid, 28049 Cantoblanco, Madrid, Spain\\
	\email{aurora.sicilia@uam.es}\\
	\inst{2}Max-Planck-Institut f\"{u}r Astronomie, K\"{o}nigstuhl 17, 69117 Heidelberg, Germany\\
	\inst{3}Laboratoire AIM, CEA/DSM--CNRS--Universit\'e Paris Diderot, IRFU/Service d'Astrophysique, CEA Saclay, 91191 Gif-sur-Yvette, France\\
	\inst{4}RAL Space, STFC Rutherford Appleton Laboratory, Chilton, Didcot, Oxfordshire, OX11 0QX, UK\\
	\inst{5}Department of Physical Sciences, The Open University, Walton Hall, Milton Keynes, MK7 6AA, UK\\
	}

  \date{Submitted August 6, 2012,  accepted 27 November 2012}

  \abstract
   {The CrA region and the Coronet cluster form a nearby (138 pc), young (1-2 Myr) star-forming region hosting a
   moderate population of Class I, II and III objects.}
   {We study the structure of the cluster and the properties
    of the protostars and protoplanetary disks in the region.}
   {We present Herschel PACS photometry at 100 and 160~$\mu$m, 
   obtained as part of the Herschel Gould Belt Survey.
   The Herschel maps reveal the cluster members within the cloud
   with both high sensitivity and high dynamic range.}
   {Many of the cluster members are detected, including some 
   embedded, very low-mass objects, several protostars (some of them extended), and
   substantial emission from the surrounding molecular cloud.
   Herschel also reveals some striking structures, 
   such as bright filaments around the IRS~5 protostar complex
   and a bubble-shaped rim associated with the Class I object 
   IRS~2. The disks around the Class II objects display a wide
   range of mid- and far-IR excesses consistent with different disk structures.
   We have modelled the disks using the RADMC radiative transfer code to
   quantify their properties. Some of them are consistent with flared, massive, 
   relatively primordial disks (S~CrA, T~CrA). Others display significant 
   evidence for inside--out evolution, consistent with the presence of inner 
   holes/gaps (G-85, G-87). Finally, we find disks with a dramatic small dust
   depletion (G-1, HBC~677) that, in some cases, could be related to 
   truncation or to the presence of large gaps in a flared disk (CrA-159). 
   The derived masses for the disks around the low-mass stars are
   found to be below the typical values 
   in Taurus, in agreement with previous Spitzer observations.}
   {The Coronet cluster appears to be an interesting compact region containing
   both young protostars and very evolved disks. 
   The Herschel data provide sufficient spatial resolution
   to detect small-scale details, such as
   filamentary structures or spiral arms associated with multiple star formation. 
   The disks around the cluster members range from massive, flared 
   primordial disks, to disks with substantial 
   small dust depletion or with evidence of inside-out evolution. 
   This results in an interesting mixture of objects for a young and 
   presumably coevally formed cluster. Given the high degree of multiplicity
   and interactions observed among the 
   protostars in the region, the diversity of disks may be a consequence of
   the early star formation history, which should also be taken into account when
   studying the disk properties in similar sparsely populated clusters.}

\authorrunning{Sicilia-Aguilar et al.}

\titlerunning{Herschel/PACS~CrA observations}

\maketitle

\section{Introduction \label{intro}}

\begin{figure*}
   \centering
   \resizebox{\hsize}{!}{\includegraphics{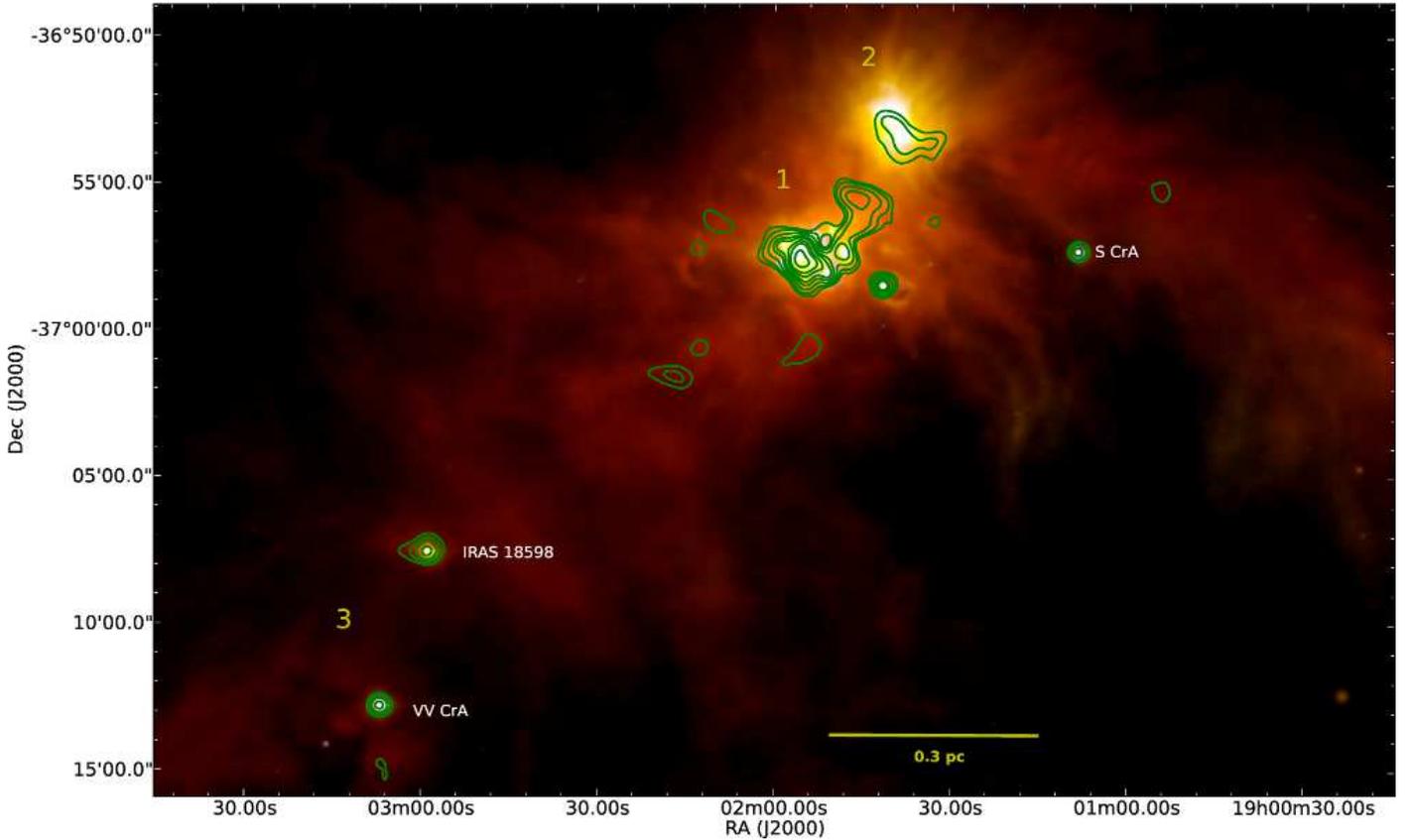}}
\caption{Combined 3-color image (with MIPS/Spitzer 24~$\mu$m, 
Herschel PACS 100 and 160~$\mu$m as blue, green,
and red, respectively) of the Coronet/CrA region, with the 870~$\mu$m LABOCA contours 
(at 0.1, 0.2, 0.4, 0.7, 1.0,1.5, 2.0,3.0,4.0,5.0, and 6.0 Jy/beam) in green (Sicilia-Aguilar et al. 2011a). 
To facilitate visual identification of the objects, we have labeled the prominent ones in white,
and added yellow numbers to mark the three subclusters (Section \ref{sources}), with 1 denoting the
central cluster, 2 the northern subcluster, and 3 the southern subcluster. \label{map-fig}}
\end{figure*}

The Corona Australis (CrA) star-forming region is a nearby,
large ($\sim$13pc; Graham 1992) molecular cloud, located out of the galactic plane, and
populated by a few young,
intermediate-mass stars and a sparse cluster, known as the Coronet 
cluster. Marraco \& Rydgren (1981) identified a compact group of pre-main sequence stars, and 
Ortiz et al. (2010) confirmed a distance of 138$\pm$16 pc based on light echoes,
placing it among the closest, brightest and
most compact star-forming regions. 
The Coronet cluster is an obscured, compact ($<$1.1~pc across) region
associated with the Ae star R~CrA (Taylor \& Storey 1984) 
and a dense molecular core (Loren 1979; Harju et al. 1993).
The presence of several far-IR sources revealed some embedded young stars (de Muizon et al. 1980; Wilking
et al. 1985; Henning et al. 1994). X-ray observations with Einstein, XMM-Newton, ROSAT and Chandra
(Walter 1986, 1997; Neuh\"{a}user et al. 1997, 2000; Garmire \& Garmire 2003; 
Hamaguchi et al. 2005: Forbrich \& Preibisch 2007), which are less
affected by dust extinction, extended the initial identification of intermediate-mass stars to a large number of low-mass
members, characterizing a quite rich cluster with several Class I protostars, some of them showing strong
X-ray flares. Combined X-ray and radio observations (Forbrich, Preibisch \& Menten 2006) confirmed these
results. Deep X-ray observations revealed a large number ($>$ 70, Garmire \& Garmire 2003) of sources that
could be associated with young stars, many of which are deeply embedded. Independently, IR observations
(including Spitzer) reveal the presence of more than 100 sources (Class I, Class II and Class III objects),
some of them highly extincted, suggesting ages well under 2-3 Myr (Nisini et al. 2005; Peterson et al. 2011;
Sicilia-Aguilar et al. 2011). Such very young ages are in agreement with the detection of ongoing star formation and
Class 0 candidates (Wang et al. 2004; Nutter et al. 2005; Chen \& Arce 2010).
Finally, optical spectroscopy of some of the
X-ray and Spitzer sources revealed a large variety of accreting and non-accreting T Tauri Stars (TTS) and
brown dwarf (BD) candidates with spectral
types K7-M8 and different disk morphologies (Sicilia-Aguilar et al. 2008, 2011a). The disk fraction
was suggested to be about 65\% (Sicilia-Aguilar et al. 2008), although the discovery of new 
diskless members lowered it to $\sim$50\% for the M-type objects (L\'{o}pez-Mart\'{\i} et al. 2010).
Isochrone fitting to the extinction-corrected JHK photometry revealed an age below 2 Myr 
for the low-mass TTS in the region (Sicilia-Aguilar et al. 2011a).

Given that the Coronet cluster is one of the most interesting 
regions for studies of low-mass star formation, several
millimetre and submillimetre studies have concentrated on identifying the 
earlier phases of the process (Henning
et al. 1994; Chini et al. 2003; Groppi et al. 2004, 2007; Nutter et al. 2005). The 1.3~mm study of
Henning et al. (1994) had a mass sensitivity limit of around 5$\times$10$^{-4}$~M$_\odot$ in cold dust, revealing the presence of disks
and structures around the Herbig Ae/Be stars and massive TTS. The study of Chini et al. (2003) reached higher sensitivity at 1.3mm,
identifying 25 individual emission sources in a 15'$\times$30' area containing the cluster center 
and its southern part around VV CrA. Comparing with other observations, these objects could be traced back to
disks around K and earlier-type stars, embedded protostellar sources,  
as well as more diffuse ambient cloud emission. Groppi et al. (2004)
mapped the densest part of the cloud in its 870~$\mu$m continuum emission, 
but the survey avoided the less dense regions 
that contain most of the Class II/III objects.
Nutter et al. (2005) and Groppi et al. (2007) used SCUBA to obtain a 
higher resolution map of the region, which covered the
central part around the emission peak in IRS~7. They resolved several peaks in this region,
including a potential Class 0 protostar.
APEX/LABOCA mapping of the region at 870~$\mu$m confirmed the presence of 
dense structures associated with
embedded objects, but also revealed starless regions with lower column densities,
suggestive of inefficient star formation, although
the low spatial resolution did not allow to study
the details in the cluster center (Sicilia-Aguilar et al. 2011a). 

Here we present observations of the CrA region, obtained with the ESA Herschel Space Observatory
(Pilbratt et al. 2010) using the PACS bolometer (Poglitsch et al. 2010) at 100 and 160~$\mu$m.
The observations are part of the Herschel Gould Belt Survey
(cf. Andr\'{e} et al. 2010 and http://gouldbelt-herschel.cea.fr).
The Herschel/PACS data reveal the structure of the CrA star-forming region and the properties of
its protostellar and disk population with unprecedent mapping speed, sensitivity, dynamic range, and
spatial resolution.
In Section \ref{data} we describe the Herschel observations and data reduction, as well as
other complementary data.
In Section \ref{sources} we study the properties of the individual objects and 
Section \ref{analysis} presents the implications for intermediate-mass
and low-mass star formation. We also derive disk parameters by fitting the SEDs of
the objects with the radiative transfer code RADMC. 
Finally, we summarize our results in Section \ref{conclu}.

\section{Observations and data reduction \label{data}}

\subsection{Herschel/PACS observations \label{pacs}}

%\begin{landscape}
\begin{table*}
\caption{PACS 100~$\mu$m photometry.} 
\label{green-table}
\begin{tabular}{l c c c l}
\hline\hline
Source ID & RA(J2000) & DEC (J2000) &  Flux  & Comments \\
       & (deg) & (deg) &  (Jy) &   \\
\hline
S~CrA & 285.285750 & -36.955429 & 17.7$\pm$2.6 & SMM7 (Nutter et al. 2005) \\
CrA-466 & 285.32887 & -36.97450 & 0.32$\pm$0.05 & G-113 \\
\#6 & 285.385710 & -36.894553 & 0.34$\pm$0.08 &  Chini et al. (2003), extended \\
G-85 & 285.390750 & -36.962428 & 0.42$\pm$0.06 & Sicilia-Aguilar et al. (2008) \\
\#8 & 285.409360 & -36.890428 & 12$\pm$2 &  Chini et al. (2003), extended \\
G-122 & 285.421760 & -36.953843 & 0.61$\pm$0.09 & Sicilia-Aguilar et al. (2008) \\
G-65 & 285.417650 & -36.862488 & 19$\pm$3 & Sicilia-Aguilar et al. (2008), mostly cloud emission \\
IRS~2 & 285.422790 & -36.975106 & 82$\pm$12 &  \\
IRS~2+ring & 285.422790 & -36.975106 & 104$\pm$15 &  Including ring-like structure \\
HBC~677 & 285.423070 & -36.997754 & 0.089$\pm$0.014 &  \\
IRS 5a+b & 285.450130 & -36.955705 & 19.4$\pm$2.8 & Unresolved binary \\
FP-25 & 285.45193 & -36.95413 & 15$\pm$3 & Probably extended, IRS~5~N \\
IRS 5ab + FP-25 & 285.450130 & -36.955705 & 71$\pm$10 & Multiple system without spiral structure \\
IRS 5+spiral & 285.450130 & -36.955705 & 92$\pm$13 & Multiple system plus spiral structure  \\
IRS 6a+b & 285.459790 & -36.943242 & 9.2$\pm$1.3 & Unresolved binary  \\
V~710 & 285.46066 & -36.96934 & 97$\pm$14 & SMM3 (Nutter et al. 2005), IRS 1 \\
IRS 7w & 285.47973 & -36.95577 & 476$\pm$69 & Uncertain due to proximity of IRS 7e, FP-34  \\
IRS 7e & 285.48414 & -36.95843 & 162$\pm$24 & Uncertain due to proximity of IRS 7w , FP-34\\
SMM~1A & 285.48138  & -36.961325 & 148$\pm$22 & Nutter et al. (2005), source \#13 (Chini et al. 2003) \\
SMM~2 & 285.493130 & -36.952151 & 12$\pm$2 & Nutter et al. (2005)\\
T~CrA & 285.494140 & -36.963643 & 12$\pm$2 &  \\
B185839.6-3658 & 285.50792 & -36.899722 & 0.12$\pm$0.03 & BD candidate \\
CrA-159 & 285.637520 & -36.972469 & 0.111$\pm$0.016 &  \\
IRAS 18598 & 285.743990 & -37.126549 & 52$\pm$5 & IRAS 18598 \\
VV CrA & 285.778000 & -37.213581 & 80$\pm$12 &  Extended? \\
CrA-45 & 285.816690 & -37.235688 & 0.92$\pm$0.13 & YSO (Peterson et al. 2011)\\
R~CrA & 285.47343 & -36.952274 & 71$\pm$10 & Strongly contaminated by IRS 7 \\
SMM~1A~s & 285.47859 & -36.97195 & 13$\pm$2 & Groppi et al. (2007)\\
\#19 & 285.5521 & -36.9544 &  0.72$\pm$0.11 & Chini et al. (2003), part of large cloud structure  \\
\hline
\end{tabular}
\tablefoot{Photometry of the known cluster members detected by Herschel/PACS at 100~$\mu$m. In case of extended sources,
the photometry was obtained around the peak (or peaks), but the presence of extended material
results in a large uncertainty. In the "Comments" column, we include the references of the source name
and/or other naming conventions, and whether the source appears extended or potentially contaminated. }
\end{table*}
%\end{landscape}

\begin{table*}
\caption{PACS 160~$\mu$m photometry.} 
\label{red-table}
\begin{tabular}{l c c c l}
\hline\hline
Source ID & RA(J2000) & DEC (J2000) &  Flux  & Comments \\
       & (deg) & (deg) &  (Jy) &    \\
\hline
S~CrA & 285.285790 & -36.955485 & 13.6$\pm$2.7 &  \\
G-122 & 285.421180 & -36.953821 & 2.2$\pm$0.4 & Sicilia-Aguilar et al.(2008) \\
IRS 2 & 285.422310 & -36.974903 & 61$\pm$12 &\\
IRS 2+ring &  285.422310 & -36.974903 & 103$\pm$20 & Including ring-like structure  \\
IRS 5a+b + FP-25 & 285.451320 & -36.954874 & 23$\pm$5 & Multiple system without spiral structure \\
IRS 5a+b/FP-25 + spiral & 285.451320 & -36.954874 & 131$\pm$26 & Multiple system plus spiral structure \\
IRS 6a+b & 285.460040 & -36.943701 & 11$\pm$2 & \\
V~710 & 285.461110 & -36.968953 & 70$\pm$14 & IRS 1, extended?  \\
SMM~1A~s & 285.47914 & -36.97314 & 30$\pm$7 & Groppi et al.(2007), extended?\\
IRS 7w+e & 285.4812 & -36.9564 & 880$\pm$170 & Binary plus surrounding structure \\
SMM~2 & 285.49313 & -36.95201 & 25$\pm$5 &  \\
T~CrA & 285.49493 & -36.96386 & 13$\pm$3 & Uncertain due to proximity of IRS~7w/e \\
APEX 13+15 & 285.552040 & -37.008617 & 0.46$\pm$0.10 & Sicilia-Aguilar (2011a), extended\\
APEX 6 & 285.577930 & -37.023607 & 0.40$\pm$0.08 & Sicilia-Aguilar (2011a), extended \\
APEX 16 & 285.599610 & -36.942940 & 0.28$\pm$0.06 & Sicilia-Aguilar (2011a), extended \\
APEX 19 & 285.635820 & -37.033284 & 0.12$\pm$0.03 & Sicilia-Aguilar (2011a), extended \\
IRAS 18598 & 285.744070 & -37.126434 & 42$\pm$3 & Extended? \\
APEX 7+10+24 & 285.770620 & -37.242373 & 0.24$\pm$0.05 & Sicilia-Aguilar (2011a), extended\\
VV CrA & 285.778020 & -37.213602 & 66$\pm$13 &  \\
CrA-45 & 285.816670 & -37.235619 & 0.81$\pm$0.16 & Peterson et al. (2011)\\
\#19 North & 285.5520 & -36.9549 &  13$\pm$3 &  Part of source \#19 from Chini et al.(2003), extended \\
\#19 South & 285.5646 & -36.9631 & 2.4$\pm$0.5 &   Part of source \#19 from Chini et al.(2003), extended \\
\hline
\end{tabular}
\tablefoot{Photometry of the known cluster members detected by Herschel/PACS at 160~$\mu$m. In case of extended sources,
the photometry is obtained around the peak (or peaks), but the presence of extended material
results in a large uncertainty. In the "Comments" column, we include the references of the source name
and/or other naming conventions, and whether the source appears extended or potentially contaminated. }
\end{table*}
%\end{landscape}

Broad-band continuum data were taken with the Photodetector Array Camera
\& Spectrometer (PACS, Pilbratt et al. 2010) on board the Herschel
spacecraft. The data were obtained as part of the Herschel Gould
Belt Survey  (P.I. P. Andr\'{e}) in a similar way to the
rest of Gould Belt targets. Square scan maps with an extent of
almost 89 arcmin on a side were obtained simultaneously in the 100 and 160 ~$\mu$m
filters on April 18, 2011. The nominal scan speed was 20$''$/s. For the first
stage of the data reduction we used customized Jython scripts within the {\sc
HIPE}\footnote{HIPE is a joint development by the Herschel Science Ground Segment 
Consortium, consisting of ESA, the NASA Herschel Science Center, and the HIFI, 
PACS and SPIRE consortia.} environment, version 7.3.0 (Ott et al. 2010). 
In addition to the standard steps to calibrate the raw data
from internal units to Jy/pixel and to remove
glitches and other bad data in the scan time lines, we applied additional
corrections for electronic cross talk as well as for non--linearity in the
PACS detector response for the brightest compact sources. These so-called
Level1 data were further processed with
{\sc Scanamorphos} (Roussel 2012). This program has its own
heuristic algorithms to remove artifacts caused by detector flickering noise
as well as spurious bolometer temperature drifts, being better suited to the
case of extended emission, compared to the standard
approach of high--pass filtering that is implemented in {\sc HIPE}. We 
used version 14 of {\sc Scanamorphos}. The field of view of the
entire maps is large and contains large areas with low emission (especially
in the 100~$\mu$m map). Since we are mainly interested in the CrA point sources,
we did not apply the {\sc galactic} option that is used in case of high
background levels and extended emission dominating the maps. The final maps were
projected onto an 2 $''$/pixel grid (100~$\mu$m) and 3$''$pixel 
(160~$\mu$m) within {\sc Scanamorphos} and written out as fits files for further
analysis. 

The photometry was performed on the final map (see Tables \ref{green-table} and \ref{red-table}).
Given the characteristics of our objects (usually either relatively faint
targets or extended sources), we chose to make aperture photometry.
For non-extended (or marginally extended), point-like sources, we used 
IRAF\footnote{IRAF is distributed by the National Optical Astronomy Observatories,
which are operated by the Association of Universities for Research
in Astronomy, Inc., under cooperative agreement with the National
Science Foundation.}  tasks $daofind$ and $apphot$
to extract the positions of the sources and to perform aperture photometry. 
We used the apertures and sky parameters used for the DUNES program (Mora, private communication)
to maximize S/N,
namely apertures of 5" and 8" for the 100~$\mu$m and 160~$\mu$m bands, respectively, with
their corresponding aperture corrections of 1.949 (100~$\mu$m) and 1.919 (160~$\mu$m).
The sky regions were selected according to the local background of each
individual sources, given the enormous variability of the cloud and background emission
in few-arcsecs scales. In general, large separations between the source and
the sky annulus would ensure that the fraction of source flux within the sky annulus
is minimal, but in the cases with nearby sources or variable cloud emission, this
option is not viable. Therefore, we used sky annuli in the range 12-30" (100~$\mu$m) and 
18-120" (160~$\mu$m), depending on each specific source, with widths of the sky annulus
ranging 2-20" (100~$\mu$m) and 3-30" (160~$\mu$m), respectively.
The flux errors were derived considering the local background rms and the
values of the correlated noise derived for the DUNES program, which depend on the
pixel size of the maps (in our case, 2" and 3" for the 100~$\mu$m and 160~$\mu$m maps, respectively).
In order to account for any other potential sources of error, including flux calibration 
and eventual aperture effects,
we also included an empirical 10\% error in quadrature. This is the dominant source of error for bright
sources, while the faint sources are dominated by the
sky noise.

\begin{figure*}
   \centering
   \resizebox{\hsize}{!}{\includegraphics{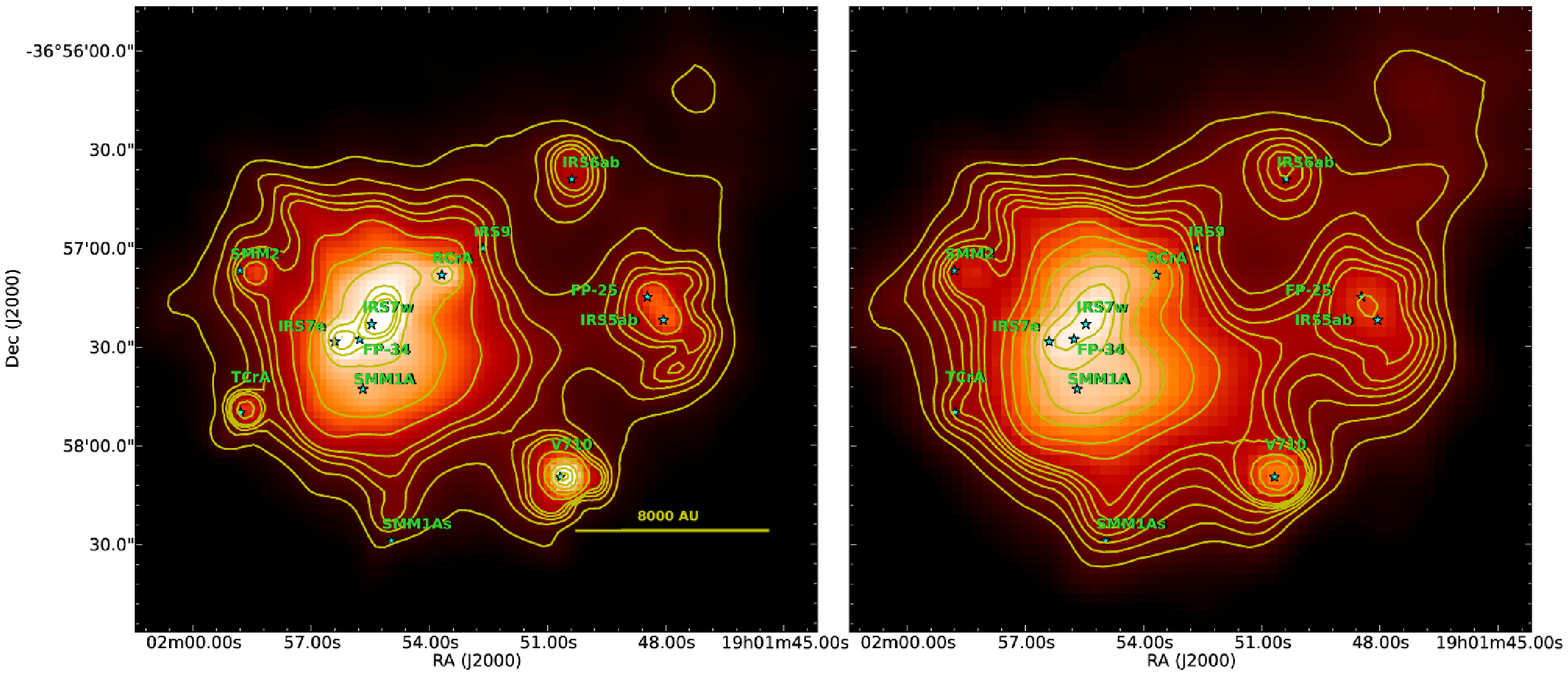}}
\caption{The region around IRS~7w/e at 100 and 160 ~$\mu$m 
(left and right, respectively). Cyan stars mark the position of the known
cluster members, labeled in green. The X-ray source FP-34
appears between the
two far-IR peaks. IRS~7w is the brightest source in the field at both wavelengths,
and it appears clearly extended, although IRS~7e becomes increasingly bright at
longer wavelengths. Both IRS~7w and IRS~7e are surrounded by a strongly emitting
extended structure. The binary IRS~5 and the X-ray source FP-25 appear surrounded by a distinct structure
that could be a common filament or associated spiral arms. IRS~6 appears as an extended
source. IRS~2 and V~710 are rather compact, albeit surrounded by extended structures. The contours mark the 
Herschel/PACS levels at 0.1,0.15,0.20,0.25,0.3,0.5,1,2,3,4,5 Jy/beam (for the 100$\mu$m image) and 
0.4,0.5,0.6,0.7,0.8,0.9,1.0,1.5,2.0,3,4,5,6 Jy/beam (for the 160$\mu$m image).}
\label{IRS7-fig}
\end{figure*}

For the extended and relatively bright sources, we used the CLASS/GREG application 
from the GILDAS software
package\footnote{See http://www.iram.fr/IRAMFR/GILDAS/} in order to 
accurately select the emission structure, to measure it flux, and the corresponding sky.
In this case, no aperture correction was used, and the errors were obtained in the same way 
than for the point-like sources\footnote{In case of isolated extended sources, the
size of the selected area encloses all the emission. Nevertheless, in case of nearby/blended sources,
part of the flux may be lost and some flux from nearby sources may be present, so the error
estimates include the flux variations observed when modifying the selected contours}.
The central part of the cluster (see Figures \ref{map-fig} and \ref{IRS7-fig}) contains several very close, extended
sources that are not easy to separate (e.g. IRS~7w/e, IRS~5a/b). 
The flux for these sources is thus uncertain, as we will discuss when individual sources are presented. 
There are several sources that appear point-like with
some additional extended structure (e.g. IRS~2, the IRS~5 complex). For the analysis of the source 
spectral energy distribution (SED),
we measured the compact part, and the extended structure will be discussed separately.
Some of the multiple sources do not appear resolved at all wavelengths. For instance, in the
IRS~5 complex, containing IRS~5a, IRS~5b, and FP-25 (also known as IRS~5~N), IRS~5a/b appear
blended, but FP-25 is resolved at 100~$\mu$m. The complex appears as an elongated
structure containing the three objects at 160~$\mu$m.
IRS~7w/e are also resolved at 100~$\mu$m but not at 160~$\mu$m, and both objects are very close
to (but not blended with) the candidate Class 0 object SMM~1~A. The brightness of all these sources ensures they are
all detected, but the presence of substantial nebular extended emission
around them and the nearby objects is the main source of error in their photometry.

In general, the Herschel data offer an excellent view of the cluster center, where Spitzer surveys
suffered from saturation and artifacts due to the presence of very bright sources,
and submillimetre/millimetre studies lack enough spatial resolution to
separate the emission from individual sources. It also reveals substantial cloud
structure where evidence for extended material was not conclusive at other
wavelengths (Figure \ref{map-fig}).

\subsection{Other data \label{otherdata}}

In order to complete the study of the different sources, we used the
available optical, IR, and millimetre data for the cluster members.
The summary of all the data for the sources studied in this paper is shown in Table \ref{sed-table}.
Optical photometry is available in Bibo et al. (1992) for the bright sources,
and L\'{o}pez-Mart\'{\i} et al. (2004) for the fainter members. 
Most of the IR Spitzer photometry was published in Sicilia-Aguilar et al. (2008)
and Currie \& Sicilia-Aguilar (2011), including several IRS spectra, and 
Peterson et al. (2011) presents detailed Spitzer IRAC and MIPS
photometry for several of the sources already mentioned in the previous literature, plus
some new member candidates. Submillimetre/millimetre data is listed in
Groppi et al. (2004, 2007), Nutter et al. (2005), Chini et al. (2003),
and Peterson et al.(2011). 

We also revised the archive Spitzer/MIPS 70~$\mu$m data (AOR 3664640)
for all the sources using smaller apertures and improved aperture corrections,
since some of the  Sicilia-Aguilar et al. (2008) fluxes at 70~$\mu$m were larger than
expected considering the MIPS 24~$\mu$m point and the Herschel data
(strongly suggesting contamination by nebular emission). On the other hand, the
70~$\mu$m fluxes of
many of the objects listed in Peterson et al. (2011) appeared lower than
expected from the 24 and 100~$\mu$m fluxes. An important problem for the Spitzer 70~$\mu$m
photometry is that the field presents zones of substantial nebular emission, plus ghosts
produced by the bright objects in the field, together with the fact that there is a
non-negligible part of the flux at distances of 100" (aperture correction of 1.10 for
an annulus of 100" radius, according to the MIPS manual). We thus followed different
procedures depending on the background of each object. For faint objects in the
proximity of other sources or ghosts, we used a small aperture of 14.2" (sky annulus 16-26.5") with a large
aperture correction (2.79). For isolated sources, we used a 40.4"
aperture with a 1.27 aperture correction
and sky annulus 152-168". Both aperture corrections were obtained for isolated, known young
stellar objects (YSO) in the field,
to avoid contamination by cloud material, and to minimize the temperature dependence of the
aperture correction. In the case of objects in crowded fields, the sky annulus had to
be set to avoid the presence of nearby emission, with the problem that there is non-negligible
emission from the object at these close distances. Nevertheless, for faint objects usually the background
emission and standard deviation dominate over the object flux at 16-26.5". This results in a
typical uncertainty around 10\%, which exceeds the uncertainties derived from calibration
and background subtraction. For the final fluxes, we used the zeropoint in the MIPS manual (0.778$\pm$0.012 Jy). 

\onllongtab{3}{
\begin{longtable}{lcccl}
\caption{\label{sed-table} SED information for CrA members.\\
For each object, name and position (and extinction, is relevant) are listed,
followed by the available observations and their references. The references are
this work (TW); Wilking et al. (1997; W97);  Bibo et al. (1992; B92);
L\'{o}pez-Mart\'{\i} et al. (2005; LM05); Cutri et al. (2003; 2MASS); Forbrich \& Preibisch (2007; FP07);
Acke \& Van den Ancker (2004); Groppi et al.(2007; G07); Sicilia-Aguilar et al. (2008; SA08);
Sicilia-Aguilar et al.(2011a; SA11); Currie \& Sicilia-Aguilar (2011; CSA11); Peterson et al. (2011; P11);
Kataza et al. (2010; AKARI); Chini et al. (2003; Ch03); Nutter et al. (2005; N05); 
IRAS database (IRAS); Bast et al. (2011; B11). 
The extinction of the sources comes either from the literature, or from our SED fitting
(in this case, it is labeled with $^f$). Potentially saturated photometry is marked by $^s$. 
Fluxes with large uncertainties due to the presence of other
objects or cloud structures that make it difficult to delimit individual objects (IRS~7w/e, IRS~5a/b) 
are marked with $^u$. } \\
\hline\hline
ID, RA DEC (J2000),  A$_V$ (mag)  & $\lambda$ (~$\mu$m) &  Flux (Jy) &  Uncertainty (Jy) & References/Comments \\
\hline
\endfirsthead
\caption{Continued.}\\
\hline\hline
ID, RA DEC (J2000),  A$_V$ (mag)  & $\lambda$ (~$\mu$m) &  Flux (Jy) &  Uncertainty (Jy) & References/Comments \\
\hline
\endhead
\hline
CrA-432,19:00:59.74 -36:47:10.9, A$_V$=2.7$^f$ & 0.66 & 7e-5 & 1e-5 & LM05 \\
 & 0.80 & 5e-4 & 1e-4 & LM05 \\
& 1.22 & 3.3e-3 & 1e-4 & 2MASS \\
& 1.63 & 4.7e-3 & 2e-4 & 2MASS \\
& 2.19 & 4.7e-3 & 2e-4 & 2MASS \\
& 3.6 & 4.5e-3 & 2e-4 &  SA08  \\
& 4.5 & 3.8e-3 & 3e-4 &  SA08  \\
& 5.8 & 3.5e-3 & 2e-4 &   SA08 \\
& 8.0 & 3.3e-3 & 3e-4 &  SA08  \\
& 23.9 & 3.1e-3 & 4e-4 & SA08   \\
& 100 & $<$0.019 & --- & TW \\
\\
S~CrA,19:01:08.62 -36:57:20.3, A$_V$=4.0$^{f}$ & 1.22 & 0.828 & 0.019 &  2MASS \\
 & 1.63 & 1.54 & 0.04 &  2MASS \\
& 2.19 & 2.29 & 0.05 &  2MASS \\
& 3.6 & 2.66 & 0.01 & CSA11 \\
& 4.5 & 3.16 & 0.01 & CSA11 \\
& 5.8 & 3.81 & 0.02 & CSA11 \\
& 8.0 & 4.17 & 0.02 & CSA11 \\
& 9 & 4.41 & 0.08 & AKARI \\
& 18 & 7.19 & 0.10 & AKARI \\
& 65 & 13.8 & 4.3 & IRAS \\
& 70 & 12.4 & 1.2 & TW \\
& 100 & 17.7 & 2.6 & TW  \\
& 160 & 13.6 & 2.7 & TW \\
& 450 & 3 & 0.3 & N05 \\
& 850 & 0.7 & 0.2 & N05 \\
& 870 & 0.77 & 0.02 & SA11 \\
& 1200 & 0.29 & 0.03 & Ch03 \\
\\
CrA-4110, 19:01:16.29 -36:56:28.2, A$_V$=0.41$^{SA08}$  & 0.66 & 4e-4 & 1e-4 & LM05 \\
 & 0.80 & 2.2e-3 & 1e-4 &  LM05 \\
 & 1.22 & 0.0103 & 2e-4 &  2MASS \\
& 1.63 & 0.0121 & 2e-4 &  2MASS \\
& 2.19 & 0.0111 & 2e-4 &  2MASS \\
& 3.6 & 9.1e-3 & 2e-4 &  SA08 \\
& 4.5 & 8.2e-3 & 2e-4 &  SA08 \\
& 5.8 & 5.7e-3 & 4e-4 &  SA08 \\
& 8.0 & 5.3e-3 & 2e-4 &  SA08 \\
& 23.9 & 7-9e-3 & 7e-4 &  SA08 \\
& 100 & $<$0.025 & --- & TW \\
\\
CrA-466, 19:01:18.93 -36:58:28.2, A$_V$=8.1$^{SA08}$ & 0.66 & 1e-46 & 1e-4 & LM05 \\
 & 0.80 & 1e-3 & 1e-4 &  LM05 \\
& 1.22 & 0.0115 & 3e-4 & 2MASS \\
& 1.63 & 0.0324 & 9e-4 & 2MASS \\
& 2.19 & 0.0419 & 1e-3 & 2MASS  \\
& 3.6 & 0.0402 & 4e-4 &  SA08  \\
& 4.5 & 0.0374 & 4e-4 &  SA08  \\
& 5.8 & 0.0356 & 1.1e-3 &  SA08  \\
& 8.0 & 0.0331 & 5e-4 &  SA08  \\
& 23.9 & 0.0522 & 1e-4 &  SA08  \\
& 70.0 & 0.146 & 0.027 & TW \\
& 100 & 0.32 & 0.05 & TW \\
\\
G-87, 19:01:32.32 -36:58:03.0, A$_V$=16$^{SA11}$ & 1.22 & 1.80e-3 & 7e-5 & 2MASS \\
 & 1.63 & 9.1e-3 & 3e-4 & 2MASS \\
 & 2.19 & 0.0170 & 4e-4 & 2MASS \\
& 3.6 & 0.020 & 1e-3 & SA08 \\
& 4.5 & 0.020 & 1e-3 & SA08 \\
& 5.8 & 0.0164 & 8e-4 & SA08 \\
& 8.0 & 0.0137 & 7e-4 & SA08 \\
& 23.9 & 0.022 & 2e-3 & SA08 \\
& 100 & $<$0.082 & --- & TW \\
\\
G-85, 19:01:33.85 -36:57:44.8, A$_V$=19$^{SA11}$ & 1.22 & 2e-361 & 9e-5 & 2MASS \\
 & 1.63 & 0.0175 & 4e-4 & 2MASS \\
 & 2.19 & 0.0411 & 8e-4 & 2MASS \\
& 3.6 & 0.0687 & 3e-4 & SA08 \\
& 4.5 & 0.0760 & 3e-4 & SA08 \\
& 5.8 & 0.0661 & 3e-4 & SA08 \\
& 8.0 & 0.0691 & 3e-4 & SA08 \\
& 23.9 & 0.162 & 2e-3 & SA08 \\
& 70 & 0.41 & 0.04 &  TW \\
& 100 & 0.42 & 0.06 & TW \\
& 160 & $<$0.69 & --- & TW \\
\\
G-122, 19:01:40.9 -36:57:15 & 3.6 & 2e-30 & 2e-4 & SA08 \\
 & 4.5 & 2.1e-3 & 2e-4 & SA08  \\
& 5.8 & 6.8e-3 & 8e-4 & SA08 \\
& 8.0 & 0.0147 & 4e-4 & SA08 \\
& 23.9 & 0.123 & 6-3 & SA08 \\
& 70 & 0.550 & 0.023 & TW \\
& 100 & 0.61 & 0.09 & TW \\
& 160 & 2.2 & 0.4 &  TW \\
\\
IRS~2, 19:01:41.56 -36:58:31.2 & 1.22 & 5e-30 & 1e-4 & 2MASS \\
 & 1.63 & 0.129 & 4e-3 & 2MASS  \\
& 2.19 & 0.949 & 0.023 & 2MASS  \\
& 8.0 & 7.712:$^s$ & 0.018 & CSA11 \\
& 70 & 32.8 & 3.3 & TW  \\
& 100 & 82 & 12 & TW  \\
& 160 & 61 & 12 &  TW \\
& 450 & 12 & 4 & N05 \\
& 850 & 2.0 & 0.2 & N05 \\
& 870 & 1.44 & 0.01 & SA11 \\
& 1200 & 1.32 & 0.20 &  Ch03\\
\\
HBC~677, 19:01:41.62 -36:59:52.7, A$_V$=4.5$^{f}$ & 0.44 & 3e-4 & 1e-4 & SIMBAD \\
 & 0.55 & 1.1e-3 & 1e-4 & SIMBAD \\
& 1.22 & 0.102 & 3e-3 & 2MASS \\
& 1.63 & 0.247 & 6e-3 & 2MASS \\
& 2.19 & 0.37 & 0.01 & 2MASS \\
& 3.6 & 0.302 & 2e-3 & CSA11 \\
& 4.5 & 0.292 & 2e-3 & CSA11 \\
& 5.8 & 0.279 & 2e-3 & CSA11 \\
& 8.0 & 0.291 & 1e-3 & CSA11 \\
& 100 & 0.089 & 0.014 & TW \\
\\
IRS~5a+b$^u$, 19:01:48.06 -36:57:22.0 & 1.63 & 3e-35 & 4e-4 & 2MASS \\
 & 2.19 & 0.060 & 3e-3 & 2MASS \\
& 3.6 & 0.2925 & 4e-4 & CSA11 \\
& 4.5 & 0.5464 & 6e-4 & CSA11 \\
& 8.0 & 1.706 & 5e-3 & CSA11 \\
& 70 & 17.3 & 1.7 & TW \\
& 100 & 19.4 & 2.8 & TW \\
& 160 & 23 & 5 & TW, IRS~~5ab+FP-25$^u$ \\
\\
FP-25, 19:01:48.46 -36:57:14.9 & 2.19 & 6e-3 & 1e-3 & 2MASS \\
 & 3.6 & 8.1e-3 & 1e-4 & CSA11 \\
& 4.5 & 0.0209 & 1e-4 & CSA11 \\
& 5.8 & 0.0360 & 2e-4 & CSA11 \\
& 8.0 & 0.0529 & 2e-4 & CSA11 \\
& 100 & 15 & 3 & TW \\
& 160 & 23 & 5 & TW, IRS~5ab+FP-25$^u$\\
& 450 & 12  & 4 & N05 \\
& 850 & 1.8 & 0.2 & N05\\
& 1300 & 0.095 & 0.007 & P11 \\
\\
IRS~6, 19:01:50.48 -36:56:38.4 & 1.22 & 4e-4 & 1e-4 & 2MASS  \\
 & 1.63 & $<$0.014 & --- &  2MASS \\
& 2.19 & $<$0.048 & --- &  2MASS \\
& 3.6 & 0.1139 & 3e-4 & CSA11 \\
& 4.5 & 0.1513 & 5e-4 & CSA11  \\
& 8.0 & 0.1962 & 6e-4 & CSA11  \\
& 23.9 & 0.348 & 0.035 & P11 \\
& 70 & 5.30 & 0.5 & TW \\
& 100 & 9.2 & 1.3 & TW \\
& 160 & 11 & 2 & TW \\
\\
V~710, 19:01:50.68 -36:58:09.6 & 1.22 & 6.5e-4 & 6e-5 & 2MASS \\
 & 1.63 & 0.036 & 1e-3 & 2MASS \\
& 2.19 & 0.570 & 0.026 & 2MASS \\
& 4.5 & 6.178 & 0.018 & CSA11 \\
& 70 & 108 & 11 &  TW \\
& 100 & 97 & 14 & TW  \\
& 160 & 70 & 14 &  TW \\
& 450 & 9  & 3 & N05 \\
& 850 & 1.5 & 0.2 & N05\\
\\
CrA-465, 19:01.53.74 -37:00:33.9, A$_V$=0.08$^{SA08}$ & 0.66 & 7e-5 & 1e-5 & LM05 \\
 & 0.80 & 5e-4 & 1e-4 &  LM05 \\
& 1.22 & 3.7e-3 & 1e-4 &  2MASS  \\
& 1.63 & 4.5e.3 & 2e-4 &  2MASS  \\
& 2.19 & 4.0e-3 & 2e-4 &  2MASS  \\
& 3.6 & 3.4e-3 & 3e-4 & SA08 \\
& 4.5 & 2.9e-3 & 2e-4 &  SA08 \\
& 8.0 & 2.5e-3 & 3e-4 &  SA08 \\
& 23.9 & 0.010 & 1e-3 &  SA08 \\
& 100 & $<$0.027 &  --- & TW \\
\\
SMM~1A~s, 19:01:54.9 -36:58:19 & 100 & 13 & 2 & TW \\
& 160 & 30 & 7 & TW \\ 
& 450 & 6.9 & 1.7 & G07 \\
& 850 & 1.2 & 0.2 & G07 \\
\\
IRS~7w, 19:01:55.32 -36:57:21.9 & 1.22 & 0.015 & 5e-3 & 2MASS \\
 & 1.63 & 5e-3 & 2e-3 & 2MASS  \\
& 2.19 & 0.031 & 0.01 & 2MASS \\
& 3.6 & 0.0988 & 3e-4 &  CSA11 \\
& 4.5 & 0.2662 & 7e-4 & CSA11  \\
& 70 & 234 & 36 & TW \\
& 100 & 476 & 69 & TW$^u$  \\
& 160 & 880 & 170 & TW$^u$ \\
& 450 & 45  & 15 & N05 \\
& 850 & 5.6 & 0.6 & N05\\
\\
SMM~1~A, 19:01:55.5 -36:57:41 & 100 & 148 & 22 & TW \\
& 450 & 151 & 30 & G07 \\
& 850 & 14.5 & 2.2 & G07 \\
\\
SMM~2,19:01:58.54 -36:57:08.5 & 1.22 & 2.3e-4 & 2e-5 & 2MASS \\
 & 1.63 & 6.3e-4 & 8e-5 & 2MASS \\
& 2.19 & 2.01e-3 & 7e-5 & 2MASS \\
& 3.6 & 0.0225 & 1e-33 & P11 \\
& 4.5 & 0.0721 & 3e-38 &  P11 \\
& 5.8 & 0.136 & 8e-3 &  P11 \\
& 8.0 & 0.195 & 0.011 &  P11 \\
& 23.9 & 0.807 & 0.078 &  P11 \\
& 100 & 12 & 2 & TW \\
& 160 & 25 & 5 & TW \\
& 450 & 10  & 3 & N05 \\
& 850 & 1.5 & 0.2 & N05\\
& 450 & 5.8  & 1.5 & G07 \\
& 850 & 1.2 & 0.2 & G07 \\
& 1300 & 0.135 & 0.027 & P11 \\
\\
T~CrA,19:01:58.78 -36:57:49.9, A$_V$=2.45$^{AVA04}$ & 0.36 & 9e-3 & 5e-3 & B92 \\
 & 0.44 & 0.037 & 0.018 & B92 \\
 & 0.55 & 0.069 & 1e-3 & B92 \\
& 1.22 & 0.421 & 0.014 & 2MASS \\
& 1.63 & 0.846 & 0.033 & 2MASS \\
& 2.19 & 1.460 & 0.035 & 2MASS \\
& 4.5 & 2.124 & 9e-3 & CSA11 \\
& 5.8 & 2.365 & 0.011 & CSA11 \\
& 100 & 12 & 2 & TW \\
& 160 & 13 & 3 & TW \\
\\
B185839.6-3658, 19:02:01.94 -36:54:00.1, A$_V$=13$^{FP07}$ & 1.63 & 1.6e-4 & 2.5 & W97 \\
 & 2.19 & 3.3e-4 & 2e-5 & W97 \\
 & 3.6 & 2.9e-4 & 3e-5 & P11 \\
 & 4.5 & 2.6e-4 & 2e-5 & P11 \\
 & 5.8 &  2e-4 & 1e-4 & P11 \\ 
 & 8.0 & 1.0e-3 & 1e-4 & P11 \\
 & 23.9 & 3.3e-3 & 4e-4 & P11 \\
 & 100 & 0.12 & 0.03 & TW \\ 
\\
G-14, 19:02:12.01 -37:03:09.3, A$_V$=1.9$^{SA08}$ & 1.22 & 6.8e-3 & 2e-4 & 2MASS  \\
 & 1.63 & 9.5e-3 & 3e-4 & 2MASS  \\
& 2.19 & 8.8e-3 & 2e-4 & 2MASS  \\
& 3.6 & 6.7e.3 & 3e-4 & SA08 \\
& 4.5 & 5.7e-3 & 3e-4 & SA08 \\
& 5.8 & 3.7e-3 & 3e-4 & SA08 \\
& 8.0 & 3.6e-3 & 3e-4 & SA08 \\
& 23.9 & 4.6e-3 & 5e-4 & SA08 \\
& 100 & $<$0.020 & --- & TW \\
\\
CrA-4109, 19:02:16.67 -36:45:49.3, A$_V$=0$^f$  & 0.66 & 1e-33 & 1e-4 & LM05 \\
 & 0.80 & 6.6e-3 & 1e-4 & LM05 \\
 & 1.22 & 0.0247 & 6e-4 &  2MASS \\
& 1.63 & 0.0307 & 7e-4 &  2MASS \\
& 2.19 & 0.0249 & 6e-4 &  2MASS \\
& 3.6 & 0.0147 & 2e-4 &  SA08 \\
& 5.8 & 6.7e-3 & 6e-4 &  SA08 \\
& 23.9 & 0.0117 & 4e-4 &  SA08 \\
& 100 & $<$0.018 & --- & TW \\
\\
G-1, 19:02:27.08 -36:58:13.2, A$_V$=3.4$^{SA08}$ & 1.22 & 0.297 & 8e-3 & 2MASS \\
 & 1.63 & 0.492 & 0.020 & 2MASS \\
 & 2.19 & 0.440 & 8e-3 & 2MASS \\
& 3.6 & 0.311 & 2e-3 & SA08 \\
& 4.5 & 0.269 & 1e-3 & SA08 \\
& 5.8 & 0.245 & 2e-3 & SA08 \\
& 8.0 & 0.246 & 1e-3 & SA08 \\
& 23.9 & 0.151 & 1e-3 & SA08 \\
& 100 & $<$0.017 & --- & TW \\
\\
CrA-159, 19:02:33.08 -36:58:21.2,  A$_V$=5.0$^f$ & 1.22 & 0.0916 & 2.9e-3 & 2MASS \\
& 1.63 & 0.201 & 5e-3 &   2MASS \\
& 2.19 & 0.266 & 5e-3 &  2MASS \\
& 3.6 & 0.356 & 0.019 & P11 \\
& 4.5 & 0.335 & 0.020 & P11 \\
& 5.8 & 0.347 & 0.017 & P11 \\
& 8.0 & 0.339 & 0.017 & P11 \\
& 23.9 & 0.238 & 0.022 & P11  \\
& 70 & 0.29 & 0.03 &  TW \\
& 100 & 0.111 & 0.016 &  TW \\
& 160 & $<$0.24 & --- & TW \\
\\
CrA-4107,19:02:54.64 -36:46:19.2, A$_V$=0$^f$ & 0.66 & 7e-4 & 1e-4 & LM05 \\
 & 0.80 & 3.9e-3 & 1e-4 & LM05 \\
 & 1.22 & 0.0166 & 4e-4 &  2MASS \\
& 1.63 & 0.0195 & 5e-4 &  2MASS \\
& 2.19 & 0.0174 & 4e-4 &  2MASS \\
& 3.6 & 0.0143 & 1e-4 & SA08 \\
& 5.8 & 0.0112 & 4e-4 & SA08 \\
& 23.9 & 0.0114 & 4e-4 & SA08 \\
& 100 & $<$.018 & --- & TW \\
\\
IRAS~18598, 19:02:58.56 -37:07:35.6, A$_V$=5$^{f}$ & 1.63 & 5e-4 & 1e-4 & 2MASS \\
 & 2.19 & 1e-3 & 1e-4 & 2MASS \\
 & 3.6 & 9.8e-3 & 4e-5 & CSA11 \\
& 4.5 & 0.0224 & 4e-4 & CSA11 \\
& 5.8 & 0.022 & 1e-3 & CSA11  \\
& 8.0 & 0.0138 & 4e-4 & CSA11 \\
& 23.9 & 2.474 & 1e-3 & CSA11 \\
& 18 & 0.884 & 0.026 & AKARI \\
& 70 & 32.2 & 3.2 & AKARI \\
& 870 & 3.39 & 0.33 & SA11 \\
& 1200 & 0.67 & 0.34 & Ch03 \\
& 100 & 52 & 5 & TW \\
& 160 & 42 & 3 &  TW\\
& 450 & 9 & 3 & N05 \\
& 850 & 1.5 & 0.2 & N05 \\
& 1300 & 0.103 & 0.008 & P11 \\
\\
VV~CrA,19:03:06.80 -37:12:49.1, A$_V$=15$^{f,B11}$ & 1.22 & 0.179 & 5e-3 & 2MASS, SIMBAD  \\
 & 1.63 & 0.729 & 0.030 & 2MASS, SIMBAD   \\
& 2.19 & 2.028 & 0.040 & 2MASS, SIMBAD  \\
& 3.6 & 4.753 & 0.017 & CSA11 \\
& 4.5 & 6.160 & 0.022 & CSA11 \\
& 5.8 & 15.21 & 0.05 & CSA11 \\
& 9 & 24.1 & 0.2 & AKARI \\
& 18 & 39.7 & 4.3 & AKARI \\
& 12 & 31.9 & 3 & IRAS \\
& 25 & 69.1 & 7 &  IRAS \\
& 60 & 131.0 & 13 & IRAS \\
& 100 & 95 & 10 & IRAS \\
& 70 & 55 & 6 & TW \\
& 100 & 80 & 12 & TW \\
& 160 & 66 & 13 & TW  \\
& 450 & 12  & 4 & N05 \\
& 850 & 2.0 & 0.2 & N05\\
& 870 & 1.66 & 0.06 & SA11 \\
& 1200 & 0.58 & 0.06 & Ch03 \\
\\
CrA-45, 19:03:16.09 -37:14:08.2, A$_V$=10$^{f}$ & 1.22 & 0.0271 & 7e-4 & 2MASS \\
 & 1.63 & 0.0775 & 2e-3 & 2MASS  \\
 & 2.19 & 0.1160 & 3e-3 & 2MASS \\
& 3.6 & 0.223 & 0.012 & P11\\
& 4.5 & 0.224 & 0.012 & P11  \\
& 5.8 & 0.256 & 0.013 & P11 \\
& 8.0 & 0.278 & 0.014 & P11 \\
& 23.9 & 0.676 & 0.063 & P11 \\
& 70 & 1.18 & 0.12 & TW \\
& 100 & 0.92 & 0.13 & TW \\
& 160 & 0.81 & 0.16 & TW \\
\hline
\end{longtable}
}

\section{The Herschel view of the CrA members \label{sources}}

In this section, we describe the most important sources in the region 
related to known YSOs, as seen from the Herschel/ PACS observations.
 We focus the discussion on those objects that are known to be 
 young cluster members from optical, X-ray, Spitzer, 
 and submillimetre studies (see references in the
Introduction and Section \ref{otherdata}). In addition
to these objects, we have found several other sources in the mapped 
region that have not previously been classified as YSOs in the literature, 
and some extended galaxies. Some of the new PACS sources may be extragalactic objects, 
given that their positions appear unrelated to other cloud emission. 
Conversely, a number of other sources appear spatially coincident to the more general cloud emission,
so they could be candidate members (either TTS or protostars), but the lack
of further data on them does not allow us to unambiguously determine their
cluster membership. 
These sources will be discussed in a follow-up publication by the
Gould Belt Survey group.

\begin{figure}
   \centering
  \includegraphics[width=9cm]{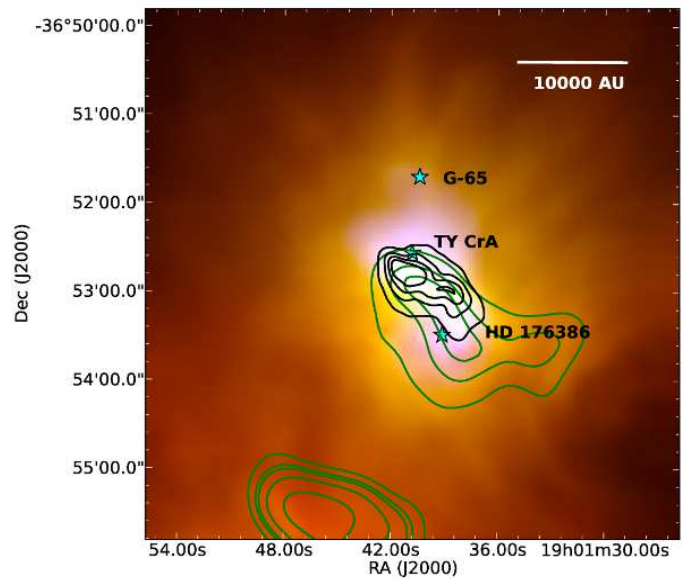}
\caption{Composite image (24, 100, 160~$\mu$m) of the HD~176386/TY~CrA region,
with the LABOCA 870~$\mu$m contours (at 0.1,0.2,0.25,0.4,0.7 Jy) marked in green, 
the 100~$\mu$m contours in black (at 0.4,0.6,0.8,1.0 Jy), and the positions of the sources
as white stars. The 24~$\mu$m data
traces the cavities cleared by the two massive stars. The 870~$\mu$m data reveals that
the peak of the submm emission (and thus the expected higher mass column density) 
is located around HD~176386, however the star itself is not affected by high extinction.
The Herschel data reveals a peak in the zone between the two
mid-IR illuminated cavities, which could indicate the formation of condensations. 
To the South, some details of another submillimetre emission region that lacks far-IR sources
can also be appreciated.}
\label{hd176-fig}
\end{figure}

The Herschel PACS data presents a new view of the CrA star-forming region (Figure \ref{map-fig}).
The emission at 100 and 160~$\mu$m follows well our previous APEX/LABOCA
observations at 870~$\mu$m (Sicilia-Aguilar et al. 2011a), 
but the much higher spatial resolution and sensitivity
of Herschel reveals several new interesting structures. 
As in the APEX/LABOCA data, the brightest part of the cloud appears associated
with the surroundings of the embedded sources IRS~7w and SMM~1~A (see Figure \ref{map-fig} and 
\ref{IRS7-fig}). This is also in agreement with the results of
Chini et al. (2003) and Groppi et al. (2007). 
The brightest part of the cloud contains two subclusters,
the first (and brightest) central cluster, 
containing IRS~7e/w,  SMM~1~A/As, R~CrA, T~CrA, IRS~5, IRS~6, and V~710 (Figure \ref{IRS7-fig}), and the 
second located to the North, around TY~CrA and HD 176386, with G-65 (Figure \ref{hd176-fig}). 
A dark lane (at Herschel wavelengths), coincident with the source SMM~6 in Nutter et al.(2005) and with 
the secondary 870~$\mu$m peak in our LABOCA map (Sicilia-Aguilar et al. 2011a) 
is visible between these two groups (see Figure \ref{map-fig}). In the first subcluster, the emission peaks
are clearly related to the embedded protostars (the IRS~7 and IRS~5 complexes, SMM~1~A/As, 
IRS~6, V~710; see Figure \ref{IRS7-fig}).
The peak in the second one is not related to TY~CrA nor HD~176386, but located
between them. This is substantially different from the
APEX/LABOCA observations, which showed a extended emission peaking at a position
coincident with HD~176386, where the cloud density should be maximal (Figure \ref{hd176-fig}).
 
In addition to the two subclusters, we have also detected strong emission from several 
less-embedded, independent
sources in the region (IRS~2, S~CrA, G-85, CrA-159, CrA-466, among
others), including a third subcluster to the South, associated with several 
bright young objects (VV CrA, CrA-45, and IRAS 18598) and some 870~$\mu$m 
extended emission (Sicilia-Aguilar et al. 2011a).
Besides the individual YSOs, we confirm the presence of extended cloud emission covering the
area where most of the known cluster members are located, including nebular
emission associated with each subcluster, and extending in the
directions suggested by the Graham (1992) observations and the extinction
maps by Kainulainen et al.(2009). Furthermore, we mention the occurrence of radial 
striations mainly visible in the Herschel emission blob associated with the 
HD176386/TY CrA region (Figures \ref{map-fig} and \ref{hd176-fig}). 
These might show an imprint of large scale collapse motions or the accretion of 
additional material on the central molecular clump, as seen for instance in the 
DR21 filament (Hennemann et al. 2012).

\subsection{The IRS~7 complex}

The brightest far-IR peak corresponds to the IRS~7 area (Figure \ref{IRS7-fig}).
IRS~7 is a binary embedded protostar, composed of IRS~7w (the brightest one
at IR wavelengths) and
IRS~7e. To the South of both sources, we find the submillimetre source SMM~1~A
(Nutter et al. 2005; Groppi et al. 2007). The PACS 100~$\mu$m data resolves all three peaks, but the 160~$\mu$m image
shows an elongated structure with two peaks corresponding to IRS~7e/w, and
a third emission structure related to SMM~1~A. 
Due to the difficulties to separate
both objects from each other and from the extended cloud emission,
the uncertainties in their fluxes are large (Table \ref{sed-table}).
The PACS positions of IRS~7w/e are in good agreement with the MIPS 24~$\mu$m and Chandra X-ray positions. 
In addition, the X-ray emission source FP-34 (source number 34 of Forbrich \& Preibisch 2007; see also 
Garmire \& Garmire 2003) is located close to IRS~7e, but cannot be identified with IRS~7e itself (see also
Peterson et al. 2011).  FP-34 appears as a single X-ray source 
located between the PACS emission peaks associated with IRS~7w, IRS~7e, and SMM~1~A.
Although the bright emission in the region does not allow to determine whether
a fourth, fainter embedded object is present at the position of FP-34, it
could also indicate that the X-ray emission is produced by other means
(e.g. shocks; Favata et al. 2002) in the interface between IRS~7w and IRS~7e,
or a jet related to the less-evolved SMM~1~A source.
At least one Herbig Haro object in the region is known to be source
of X-ray emission (G-80; Sicilia-Aguilar et al. 2008).

The submillimetre source SMM~1A, identified by Nutter et al. (2005) as the brightest,
most extended peak in the region, is detected by Herschel as a single
peak at 100~$\mu$m, and appears as a bright, extended emission
structure at 160~$\mu$m. This source and IRS~7 conform the brightest
peak of our LABOCA observations, which could indicate the presence of a further highly embedded
source, or a region of very high dust density in the proximity of the known protostars.
To the South of SMM~1A we detect a further PACS source, identified as SMM~1As by Groppi et al.(2007)
and without Spitzer counterparts. Both sources 
are Class 0 candidates and will be discussed
in detail in Section \ref{proto}.

\subsection{IRS~2}

\begin{figure*}
   \centering
   \resizebox{\hsize}{!}{\includegraphics{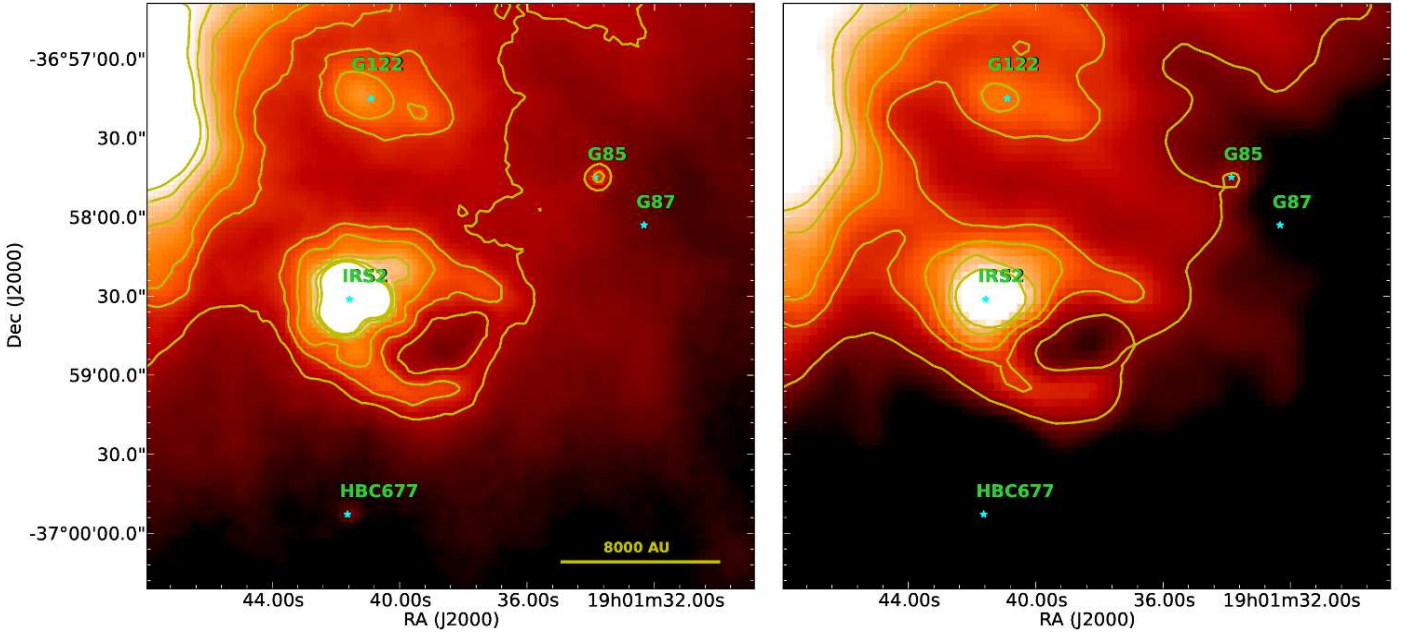}}
   \caption{The region around IRS 2 at 100 and 160 ~$\mu$m (left and right, respectively).
Stars mark the location of known cluster members, labeled in green.
G-85 and HBC~677 are well detected at 100 $\mu$m, but G-85 is only a marginal
detection at 160$\mu$m. Two  
small condensations to the north of IRS~2 are also apparent. The brightest one 
corresponds to G-122. The diameter of the IRS~2 ring is 
27", corresponding to about 4600 AU. The condensations are about 10" in radius,
or a diameter of $\sim$3500 AU at 170 pc. Another disked object in the region, G-87, is
not detected in either image. }
    \label{IRS2-fig}
\end{figure*}

The embedded IRS~2 source, classified as a Class I protostar with a spectral type K2
by Forbrich \& Preibisch (2007), is one of the most interesting objects in the
PACS data (Figure \ref{IRS2-fig}). In both the 100 and 160~$\mu$m images, it
appears as a bright point-like source  (with its 3-lobe structure 
closely resembling the PACS PSF) in the rim of a $\sim$5000 AU diameter ring-
or bubble-like structure.
The ring is clearly detected at the PACS wavelengths, but is not visible in the
Spitzer/MIPS maps. At 870~$\mu$m, our LABOCA map shows the object to be
slightly elongated in the direction of the ring, but the elongation is
marginal considering the beam size of 19.2" (Sicilia-Aguilar et al. 2011a).
The best explanation is the presence of an envelope or cloud 
structure that is excavated by the (large opening-angle) stellar winds of IRS~2,
that appears asymmetric with respect to this object due to a non-uniform distribution
of material (for instance, if IRS~2 were located at the edge of the cloud). There is no 
evidence of extended emission in the other direction. Despite the relative proximity of the bright V~710 source
and the cloud emission, any potential structure
to the East of IRS 2 with brightness similar to the observed ring should be clearly 
identifiable and resolved.

\subsection{The IRS~5 complex}

Next to IRS~2, the binary pair IRS~5a/b and the nearby X-ray source FP-25 (source
number 25 from Forbrich \& Preibisch 2007, also known as IRS~5N; Peterson et al. 2011) 
offer the next interesting structure resolved
in the Herschel/PACS maps (Figure \ref{IRS7-fig}). The two components, IRS~5a and IRS~5b,
with a projected separation of 78 AU for a distance of 130 pc (Nisini et al. 2005),
are not resolved.  
FP-25 and IRS 5a/b are surrounded by substantial extended emission, in particular,
a common filamentary structure that resembles a two-arm spiral with a resolved size corresponding
to $\sim$4000 AU, considering a distance of 138 pc. The extended structure is 
not detected in the Spitzer maps, but it is consistently
detected in the two PACS filters and follows the contours in our 870$\mu$m LABOCA maps. 
IRS~5a/b dominates the Spitzer and PACS 100~$\mu$m
maps, but the single peak observed at 160~$\mu$m is closer to FP-25, which
could indicate that the latter is not much fainter, but more embedded than IRS~5a/b.
Based on their Spitzer and SMA observations, Peterson et al.(2011) classified FP-25
as a Class I object, which is also consistent with our Herschel observations.
The appearance of the spiral structure around the system is very similar to the
ring detected around the young binary star SVS20 (which has a size of $\sim$6800$\times$4000 AU; Eiroa et al. 1997) and to the
predictions for binary formation (Bate 2000; Kley \& Burkert 2000), 
although the size of the spiral arms around the IRS~5 complex is about 7 times
larger than in Bate's simulation. The initial size and mass of the cloud,
the initial angular momentum, and the binary separation are important parameters in
determining the final structure. The projected separation between IRS~5a/b and FP-25 is 
approximately 800 AU (instead of 60-100 AU as in Bate et al. 2000), which would 
imply that the structure of the system is more affected by the properties of the surrounding cloud
and directly dependent on the initial angular momentum (Bate et al. 2000) and, in
general, on the initial conditions (Kley \& Burkert 2000). Therefore,
the observations of a larger spiral structure are in principle consistent with
the results of hydrodynamical simulations. Nevertheless,
the scale of the structure and the presence of substantial cloud
material around the sources does not exclude a complex,
non-spherical structure like those that have been observed 
associated to the envelopes of some Class 0/I objects
(Stutz et al. 2009; Tobin et al. 2011) or even 
heated material swept 
by the jets or winds of the embedded protostars, 
similar to what is seen near IRS~2.

\subsection{Other Class I candidates within the central cluster: IRS~6, V~710, and SMM~2}

IRS~6 has been identified as a binary with a projected separation slightly larger than
IRS~5a/b (97 AU for a distance 130 pc; Nisini et al. 2005),
but the pair is not resolved with Herschel/PACS. 
The source appears as a relatively faint (compared to the rest of protostellar
candidates in the region) and extended source (Figure \ref{IRS7-fig}). Contrary
to IRS~5 and IRS~2, the central emission is extended but there is no evidence for
further structure like extended rings or filaments, nor a compact core. The lack of an emission
peak and its SED suggest that the Herschel emission is consistent with
an extended envelope around the protostars, maybe a common envelope for both 
IRS~6 components.

V~710 (Figure \ref{IRS7-fig}) has been classified as a K5-M0 Class I protostar (Forbrich \& Preibisch 2007). 
At PACS wavelengths, it appears as a point-like,
bright object near the IRS~7/IRS~5 complex. It is surrounded by strong extended
emission associated with the central part of the cluster, although the object is bright enough
to be clearly distinguished. The SED of the object (Figure \ref{proto-fig}) 
is consistent with an embedded protostar, with a peak suggestive of temperatures between
500 and 100 K.

SMM~2, detected by Nutter et al. (2005) at SCUBA wavelengths (450 and 850~$\mu$m)
and by Peterson et al.(2011) at 1.3~mm
is also seen as an independent source in our Herschel maps (Figure \ref{IRS7-fig}). It is also 
marginally detected at IRAC wavelengths, well-detected at MIPS 24~$\mu$m,
and very bright at MIPS 70~$\mu$m. The Herschel detection appears point-like,
although surrounded by substantial extended emission. The SED is consistent
with an embedded protostar, although the high fluxes measured at PACS wavelengths
could indicate contamination by the surrounding cloud. 

\subsection{Very low-mass Class I objects: G-122 and nearby condensation}

\begin{figure*}
\centering
\begin{tabular}{ccc}
\epsfig{file=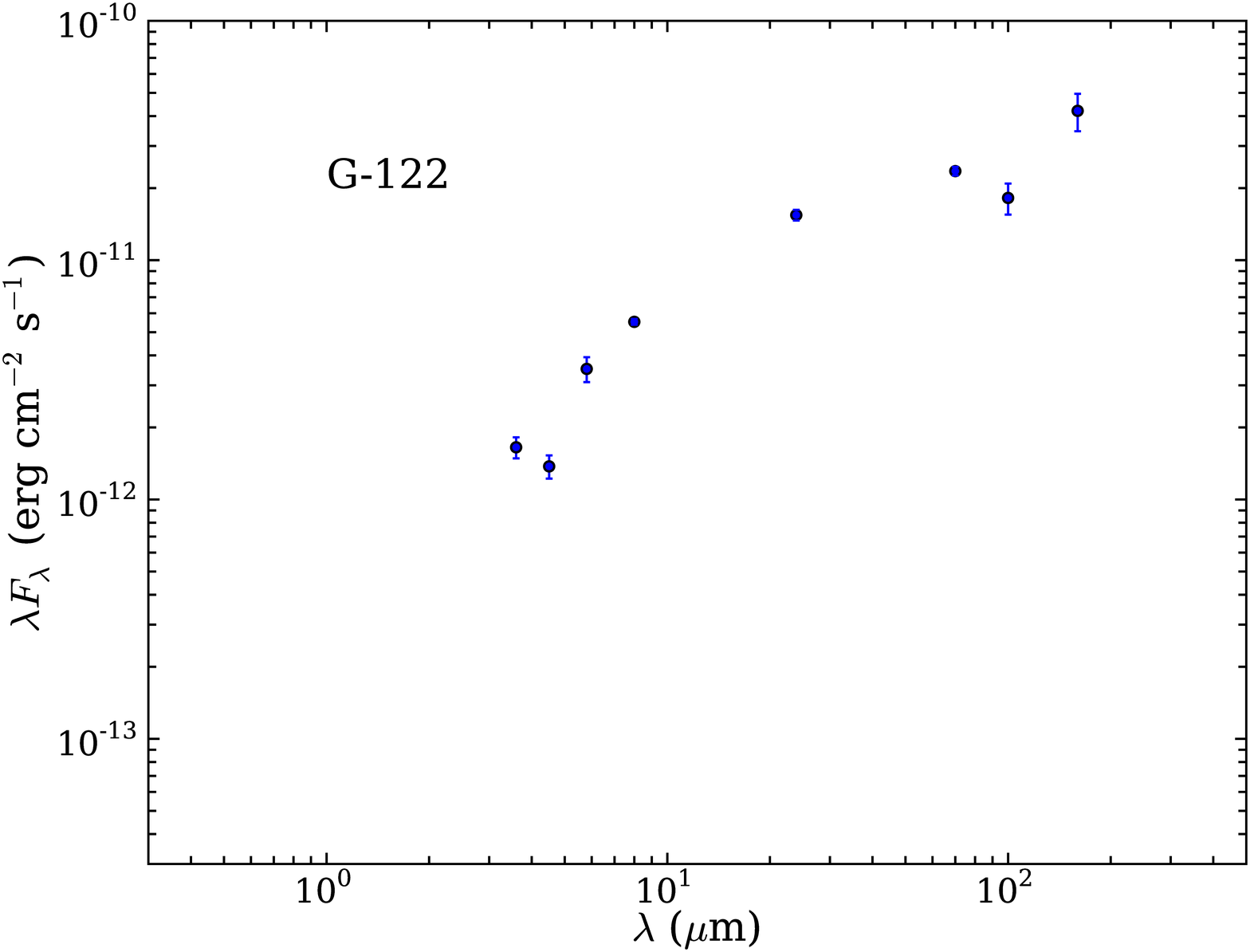,width=0.32\linewidth,clip=} &
\epsfig{file=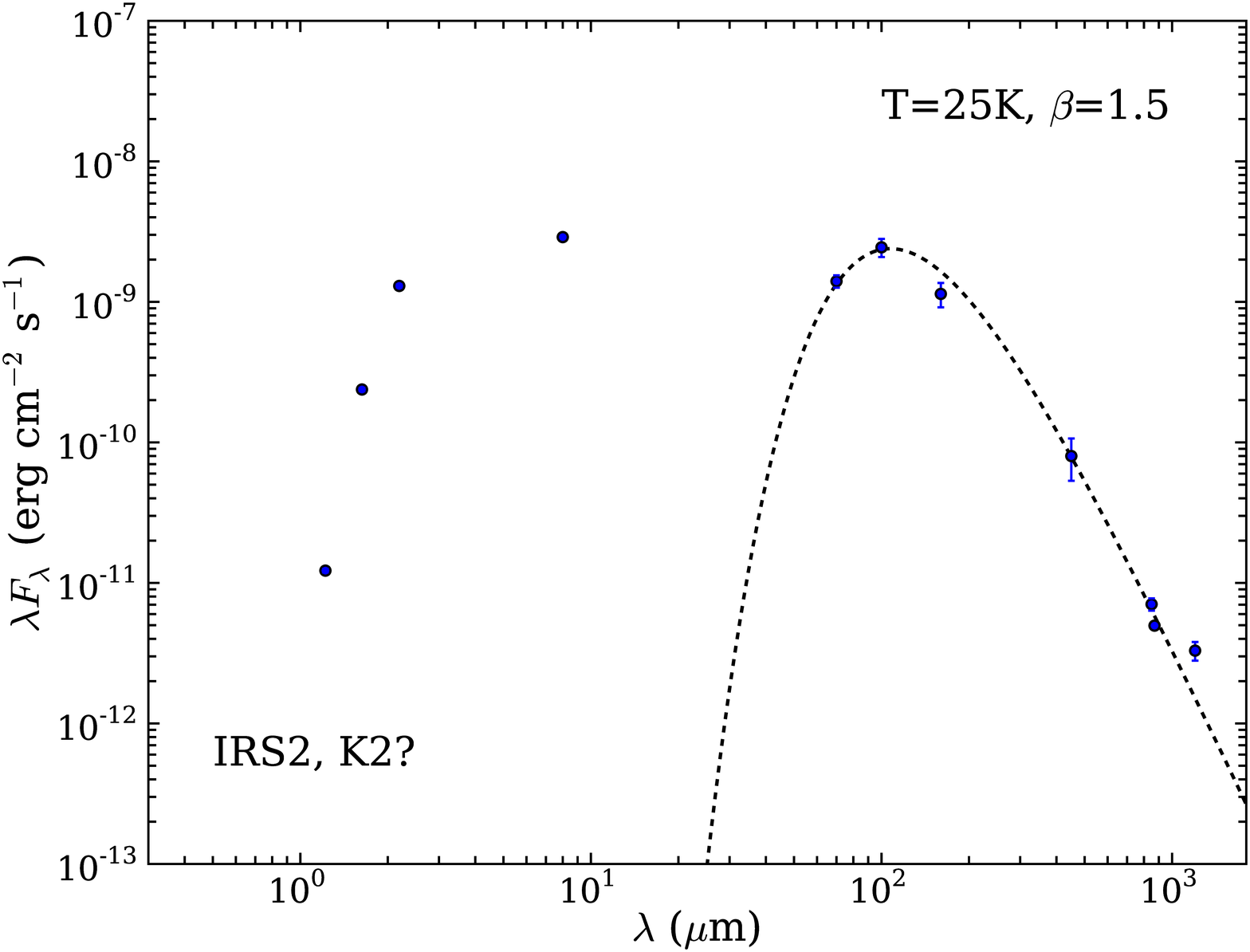,width=0.32\linewidth,clip=} &
\epsfig{file=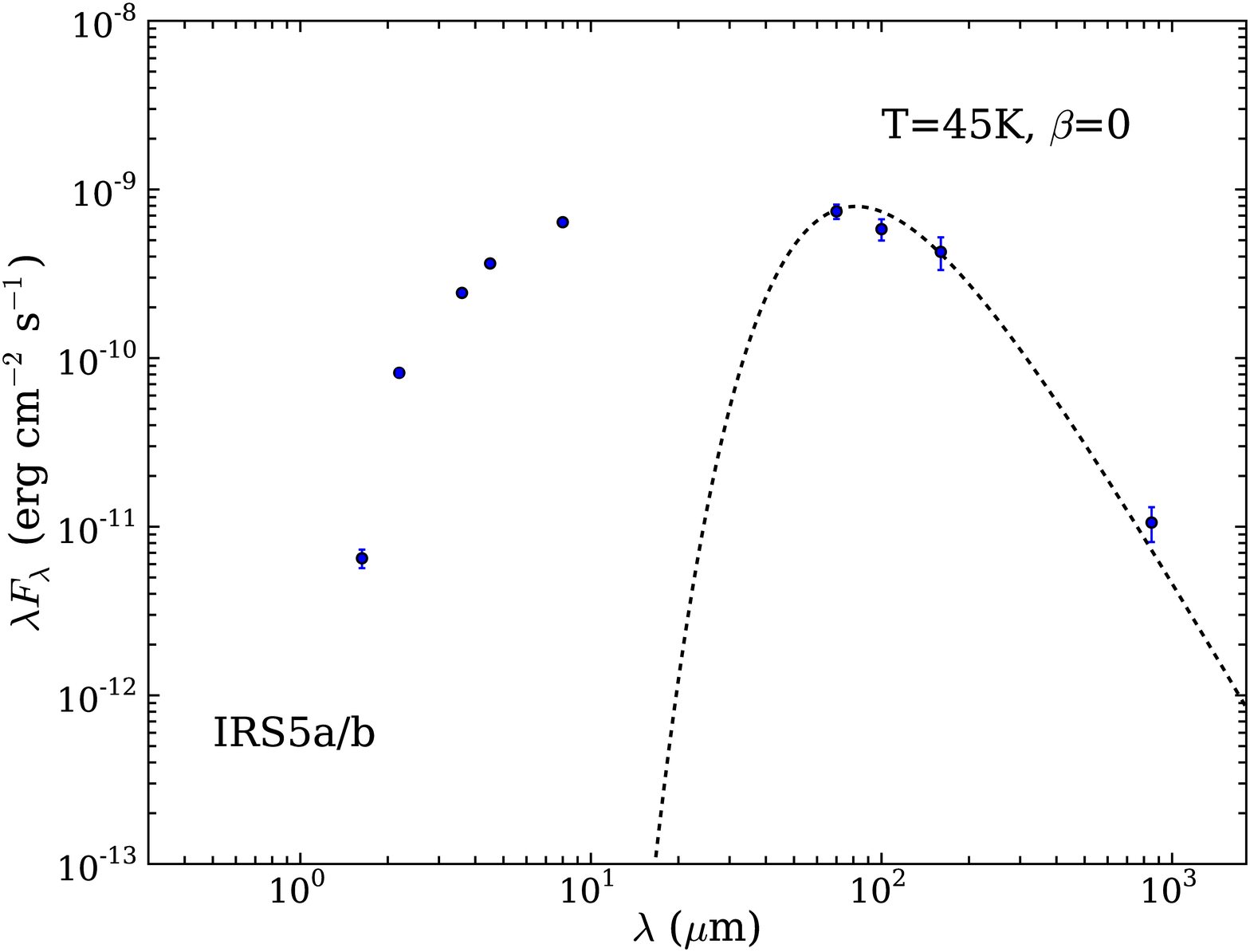,width=0.32\linewidth,clip=} \\
\epsfig{file=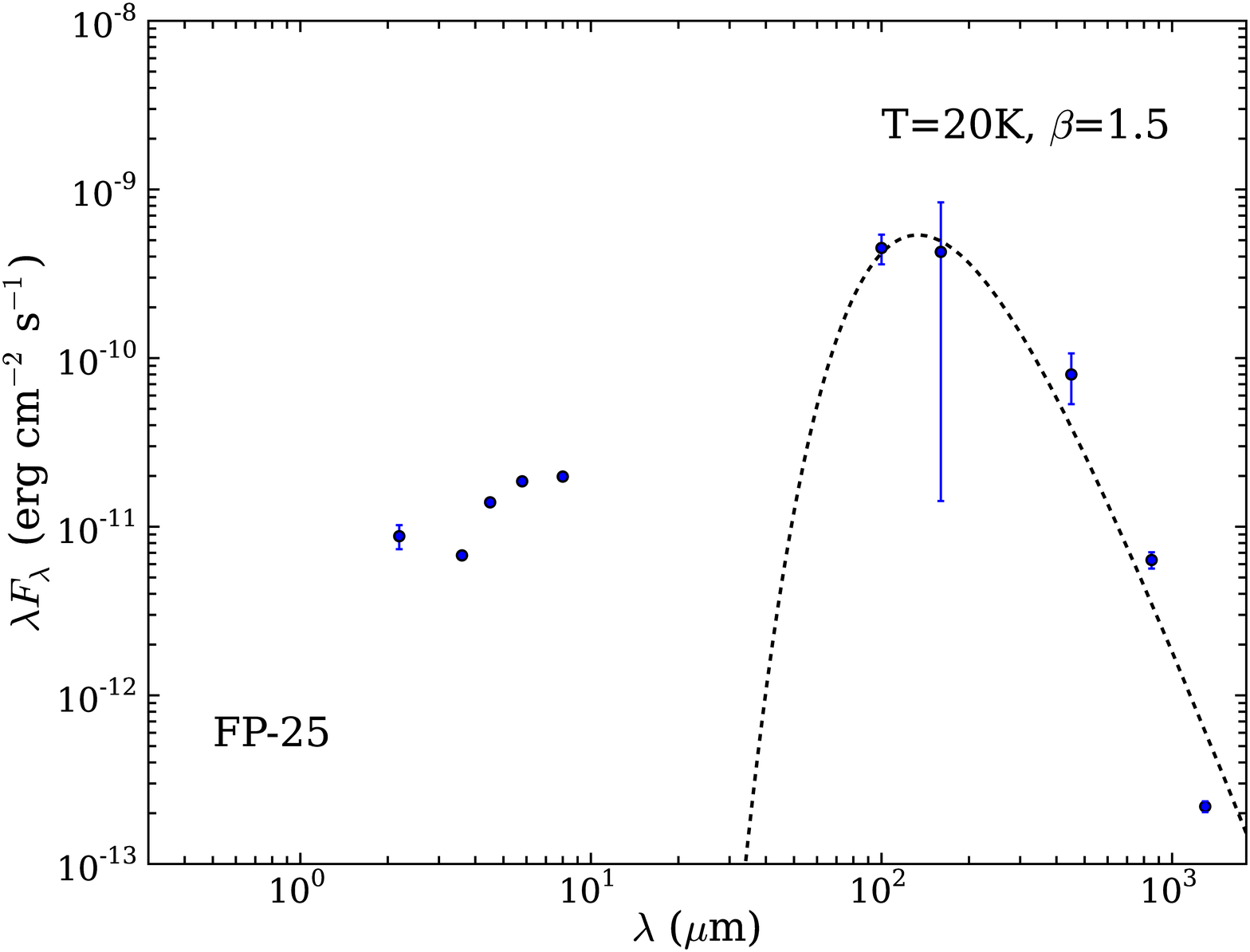,width=0.32\linewidth,clip=} &
\epsfig{file=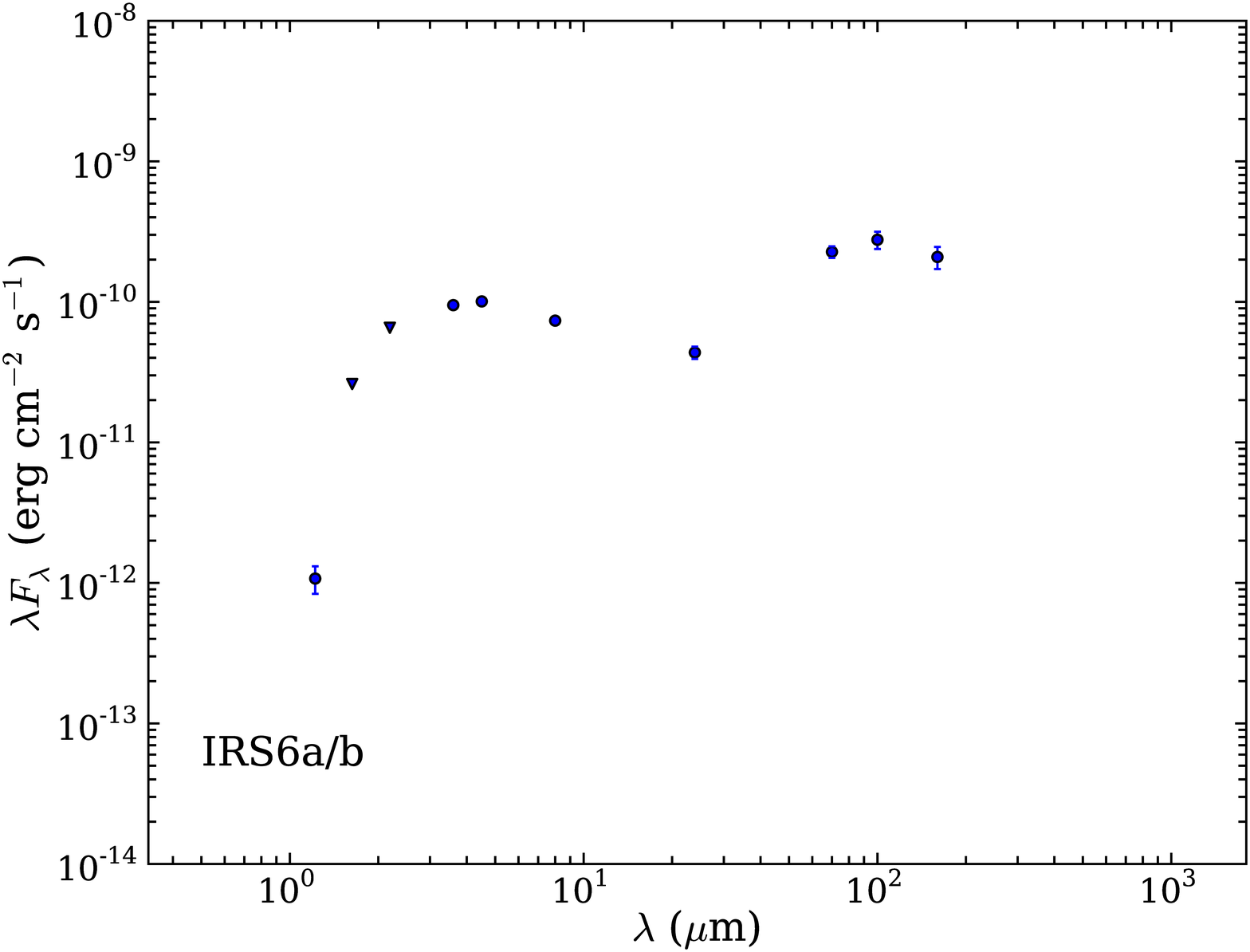,width=0.32\linewidth,clip=} &
\epsfig{file=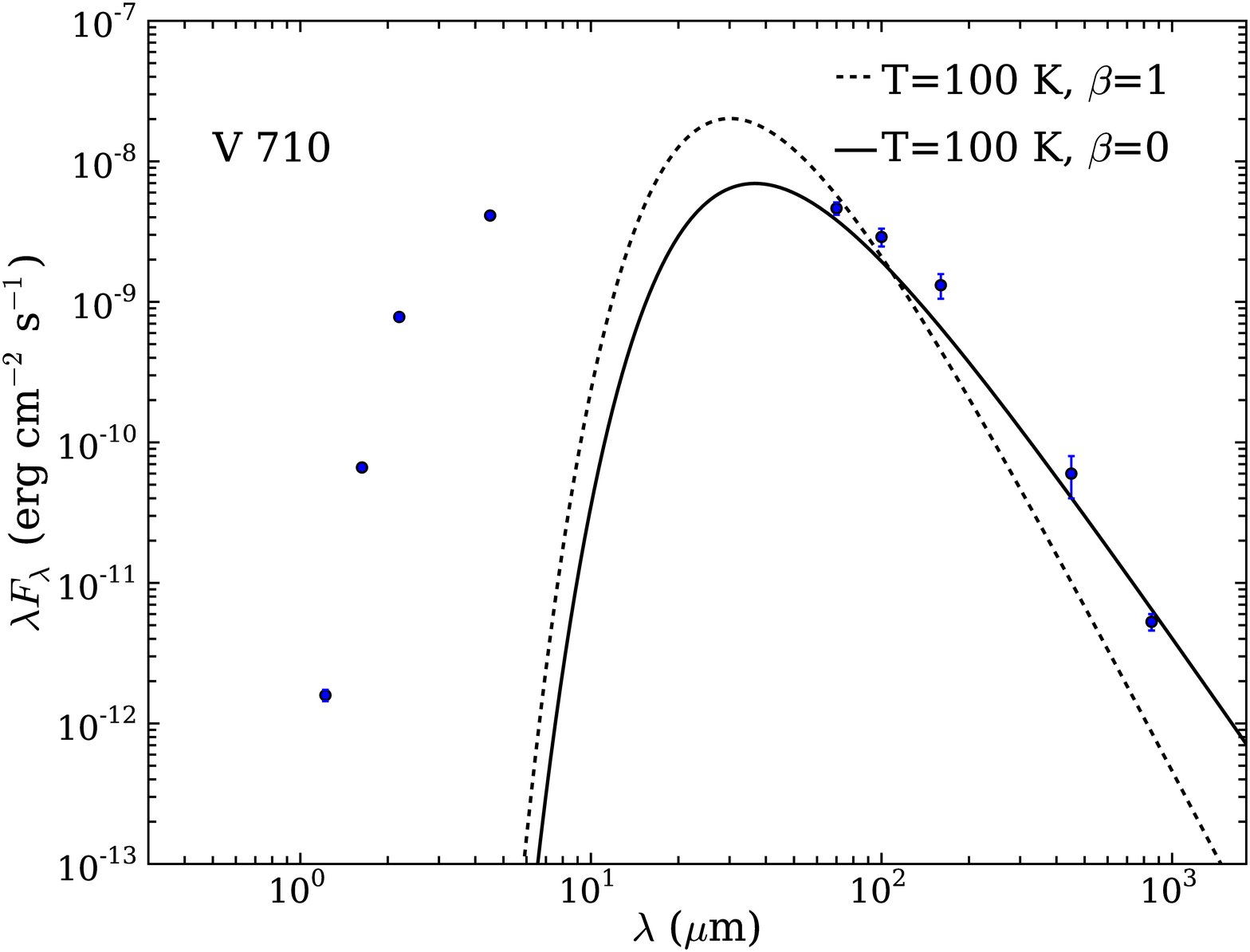,width=0.32\linewidth,clip=} \\
\epsfig{file=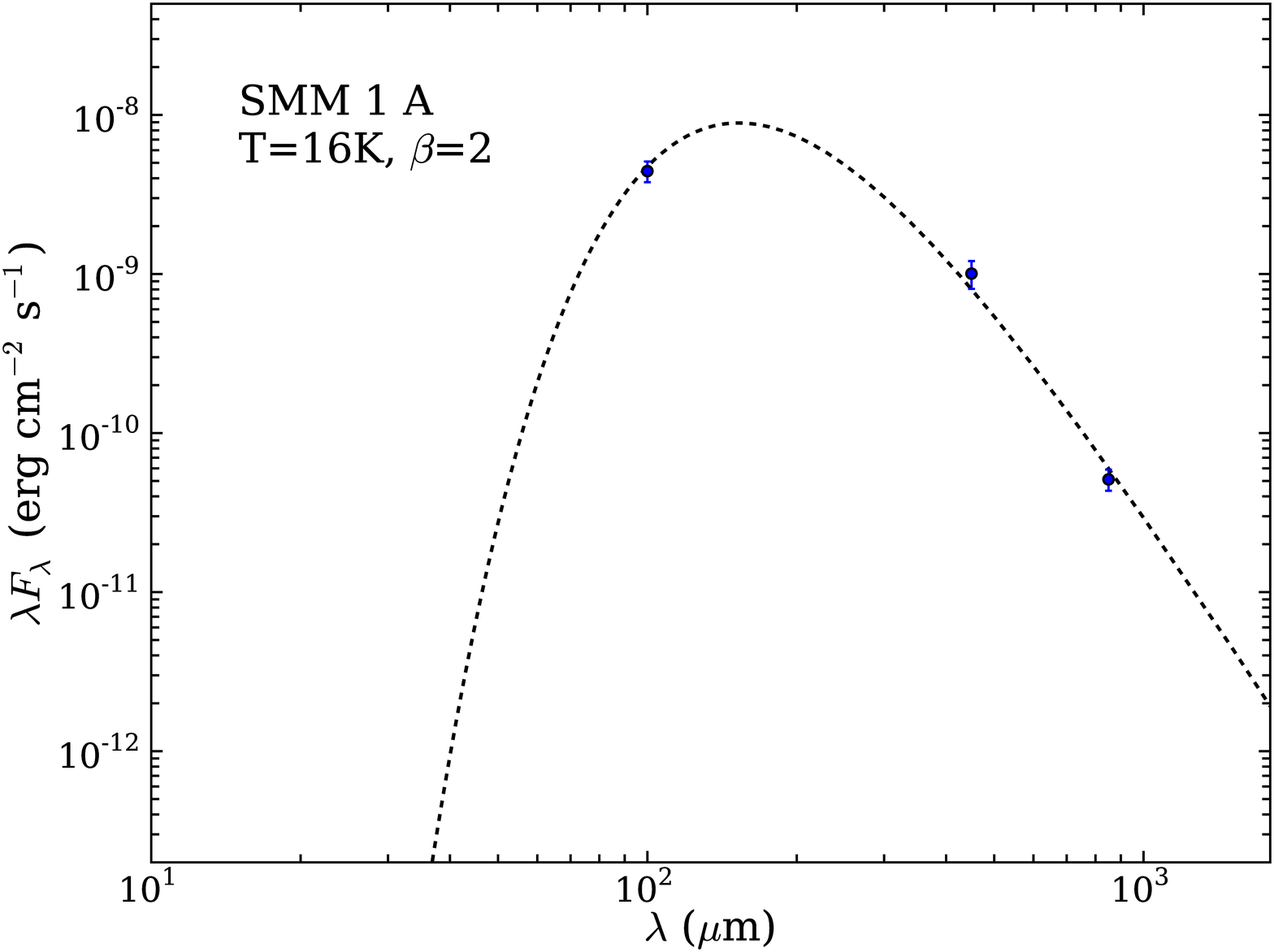,width=0.32\linewidth,clip=} &
\epsfig{file=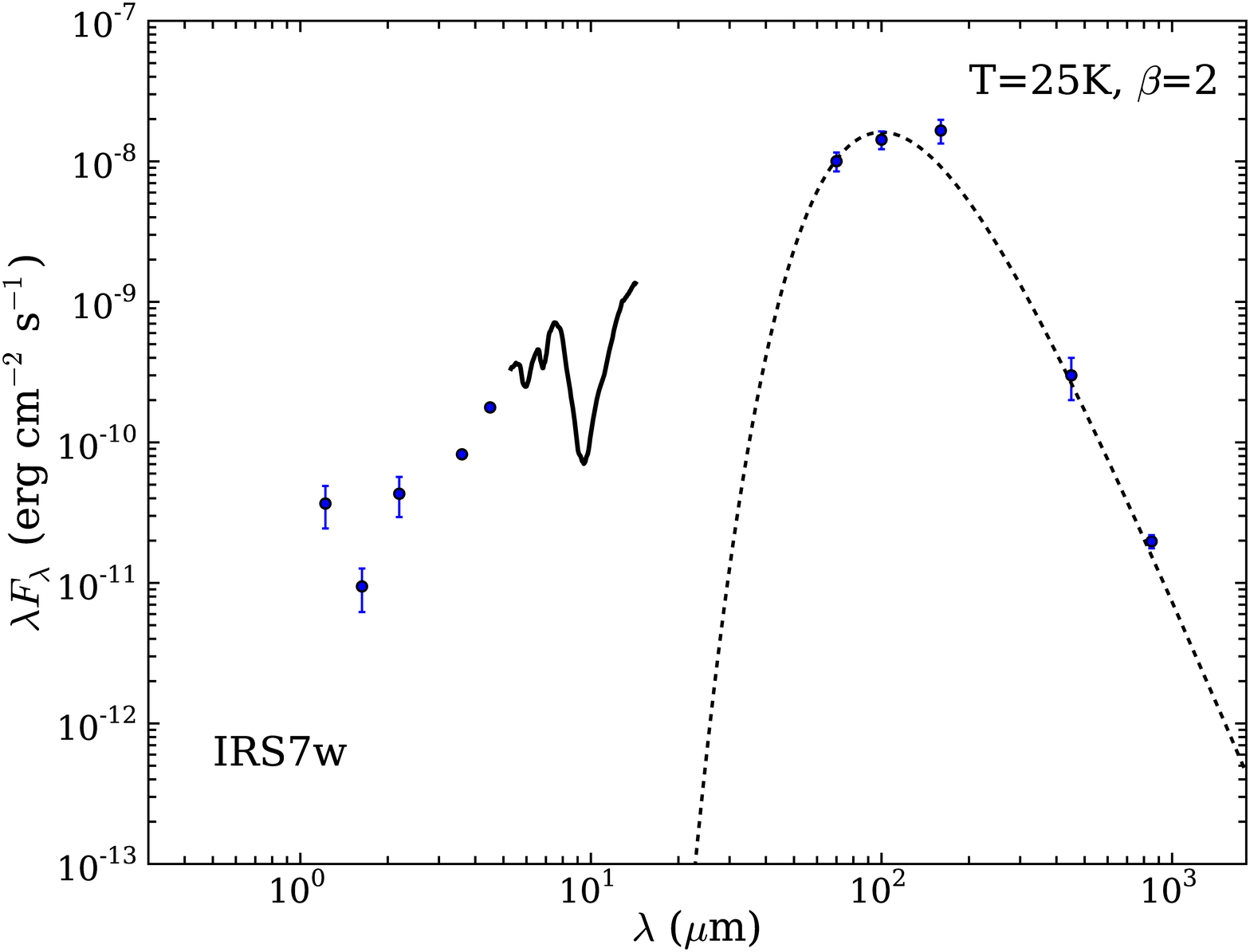,width=0.32\linewidth,clip=} &
\epsfig{file=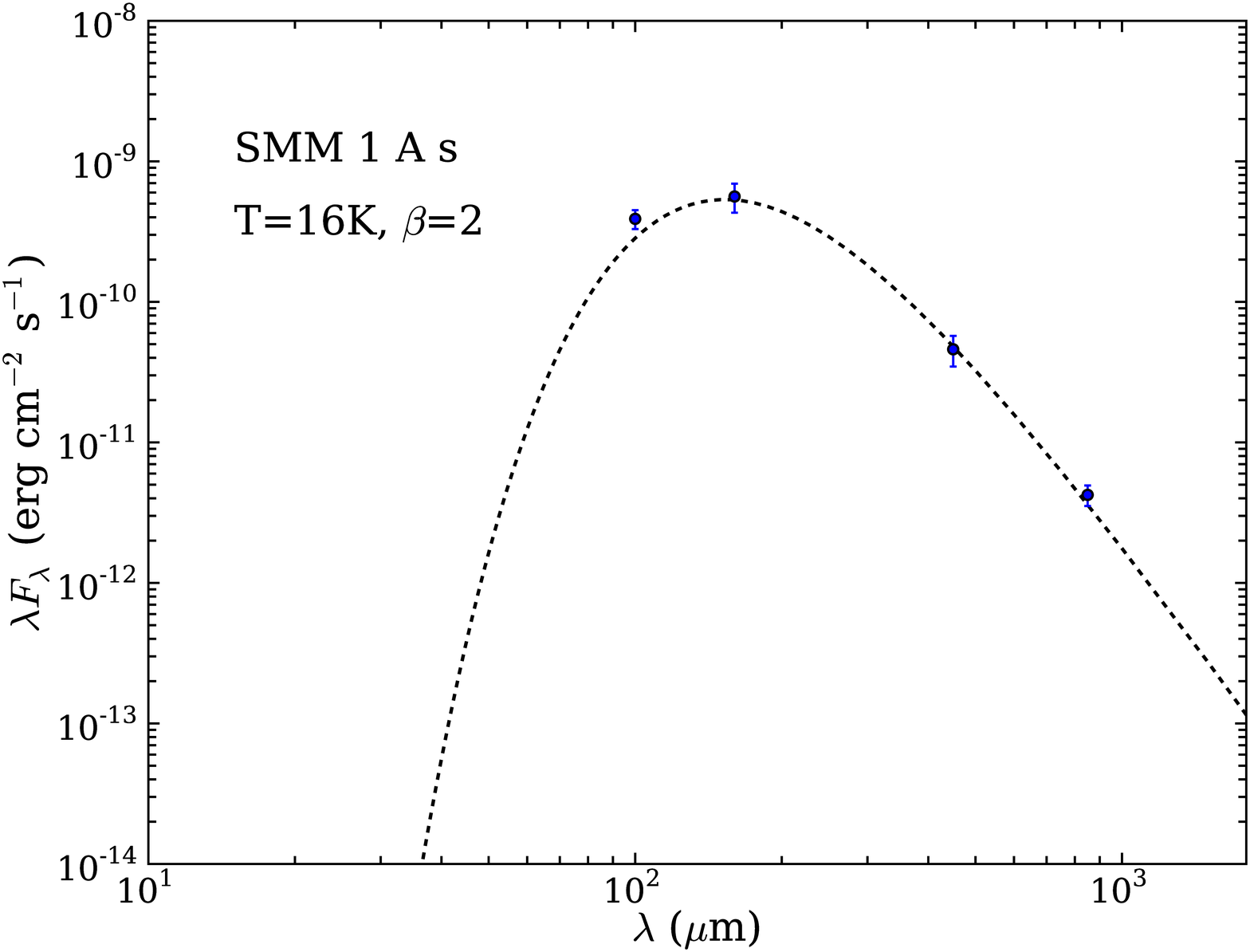,width=0.32\linewidth,clip=} \\
\epsfig{file=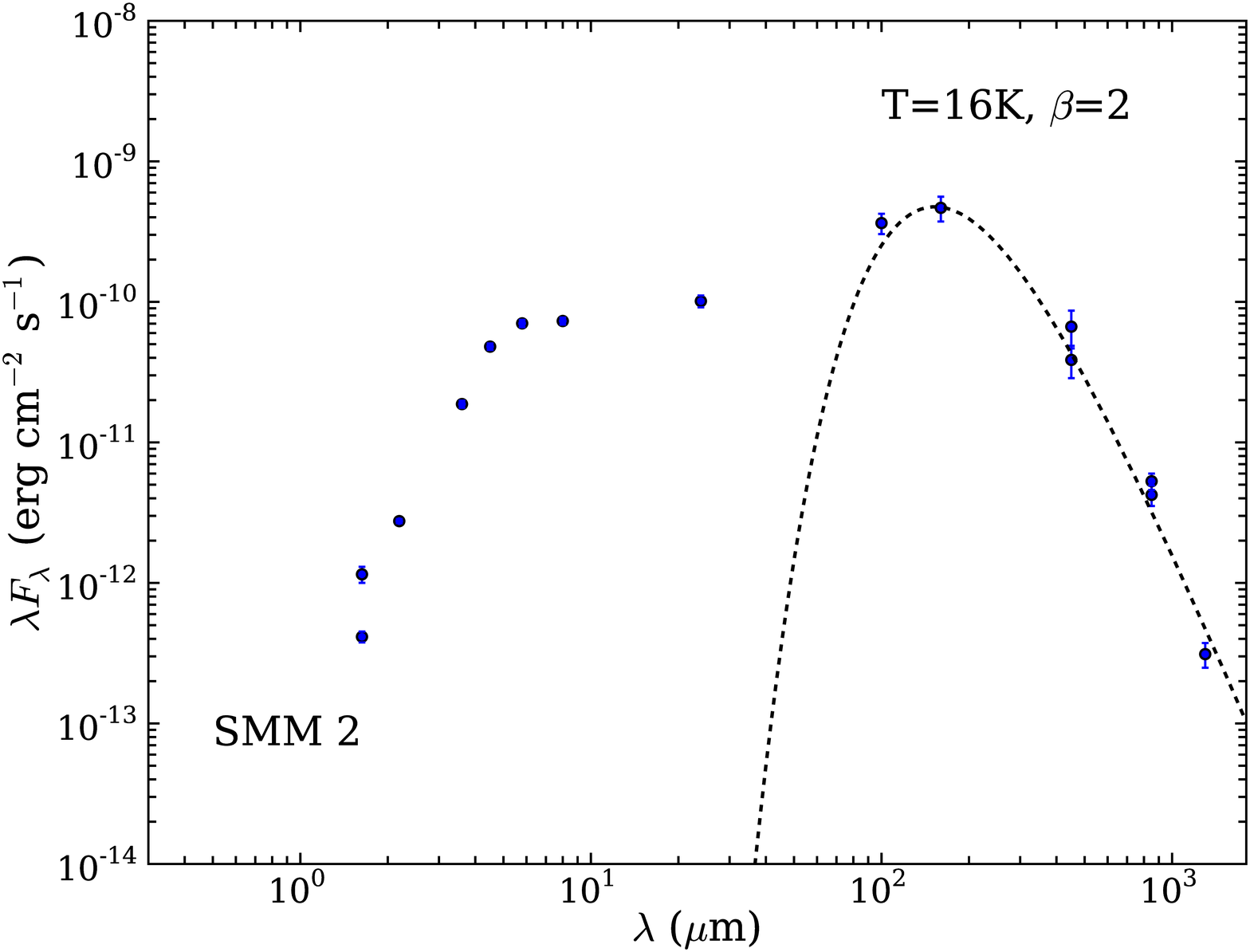,width=0.32\linewidth,clip=} &
\epsfig{file=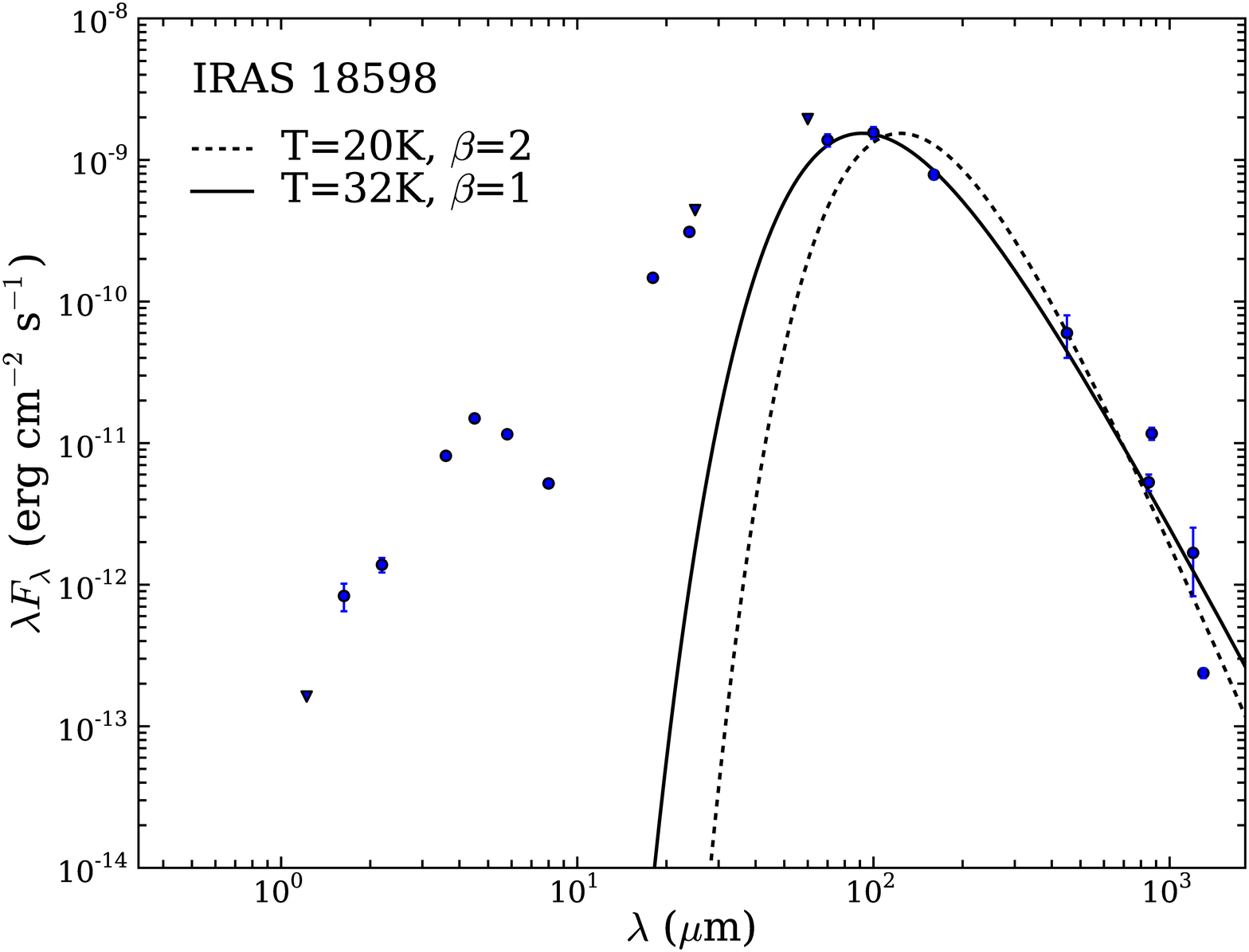,width=0.32\linewidth,clip=}&
 \\
\end{tabular}
\caption{SEDs of protostar candidates.  Photometry detections are marked as circles, with upper limits
marked as inverted triangles. The spectra correspond to Spitzer/IRS observations,
when available.
See Table \ref{sed-table} for details and references 
regarding the photometry data. In the relevant cases, a modified black-body is fitted to the data
(as a whole for SMM~1~A and SMM~1~As; in the submillimetre range only for the rest of
cases) and the corresponding values of temperature (T) and $\beta$ are
indicated in the figure, see Section \ref{proto} for further details. 
Note that IRAS~18598 is probably composed of two
sources.\label{proto-fig}}
\end{figure*}

The PACS data reveal two compact condensations to the North of IRS~2 (Figure \ref{IRS2-fig}).
One of them is associated with the X-ray source G-122, classified by Sicilia-Aguilar et al. (2008)
as a probable Class I protostar. The other one is marginally detected at 100~$\mu$m, but becomes
brighter at 160~$\mu$m. Both sources appear extended, with approximate sizes of the order
of 1600 AU at 138 pc. Due to the presence of extended cloud emission, the photometry of
the faintest source is uncertain, but G-122 is well detected with PACS. Combining Spitzer
data, G-122 appears as a cold condensation that could correspond to a very low-mass Class I
protostar (see Figure \ref{proto-fig}). The
nearby faint object is not detected at any Spitzer wavelengths, but the similarities 
with G-122 could indicate it
is another (fainter) very low-mass protostar. The projected separation between the centers of both
objects is $\sim$26" or about 3600 AU at 138 pc.

\subsection{The solar- and intermediate-mass disked stars: S~CrA, R~CrA, and T~CrA}

The binary source S~CrA appears as a bright object at PACS wavelengths 
(see Figure \ref{map-fig}). The source is
relatively compact, but differences of 10-20\% of the flux depending on the aperture
selection suggest that it could be surrounded by some extended material, either of
cloud or envelope origin. Its very complete SED (including optical, IR, and millimetre/submillimetre
data) reveals a very massive and flared disk. Disagreement in the near-
and mid-IR fluxes between the IRAC/MIPS and IRS data suggest that the short-wavelength
emission could be variable, as it has been seen in other actively accreting
young sources. 

\begin{figure*}
\centering
\begin{tabular}{ccc}
\epsfig{file=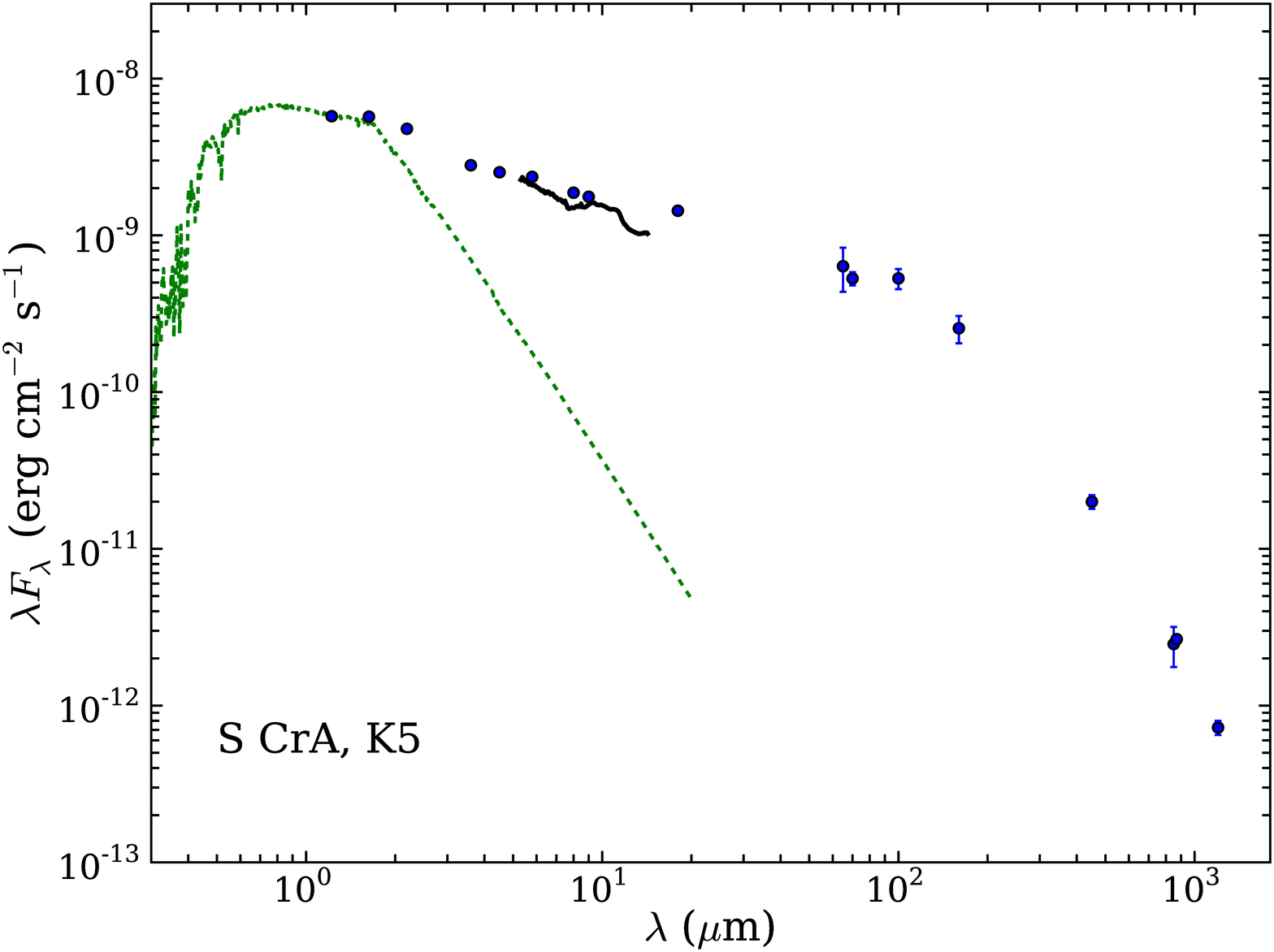,width=0.32\linewidth,clip=}&
\epsfig{file=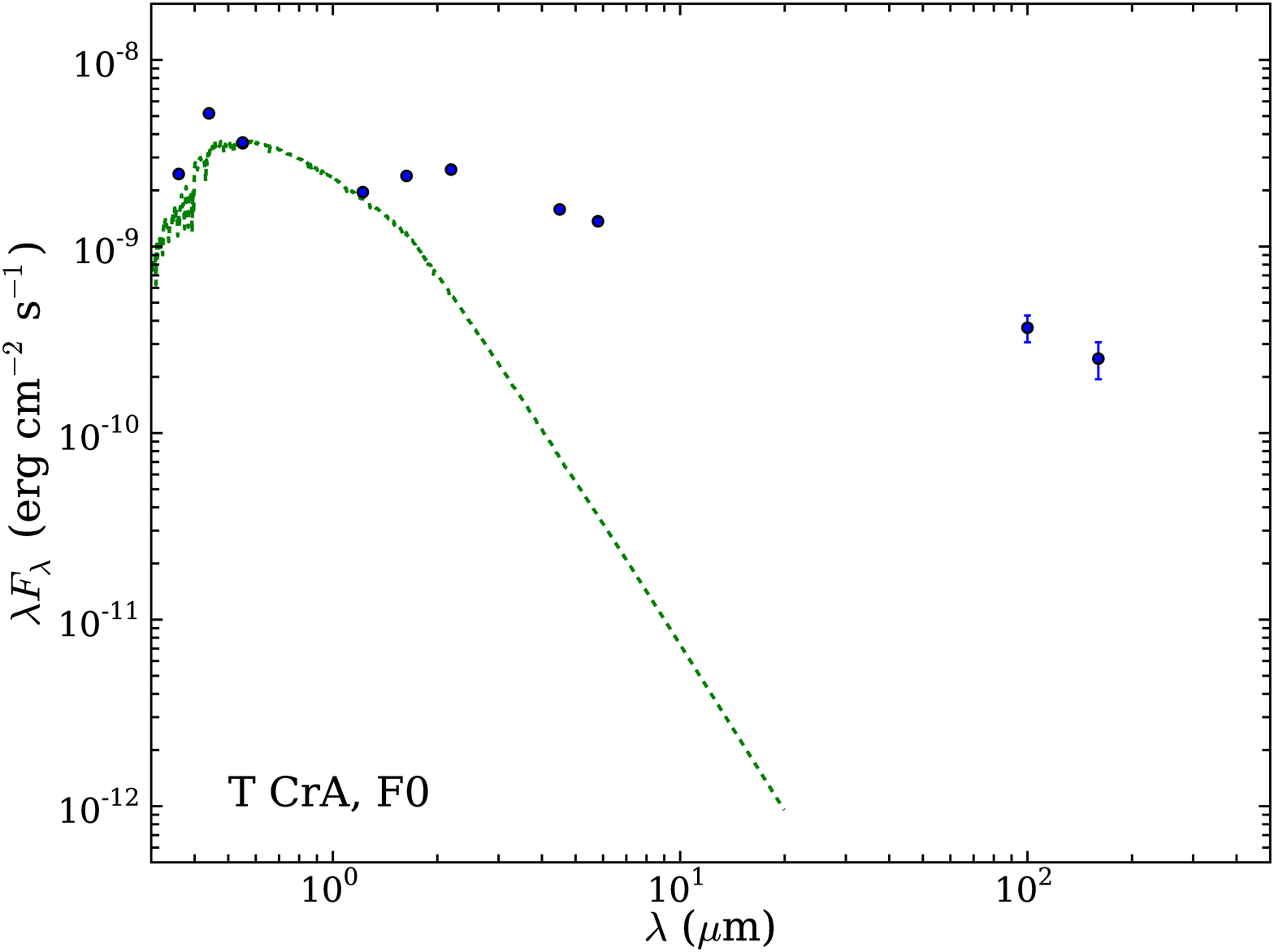,width=0.32\linewidth,clip=}&
\epsfig{file=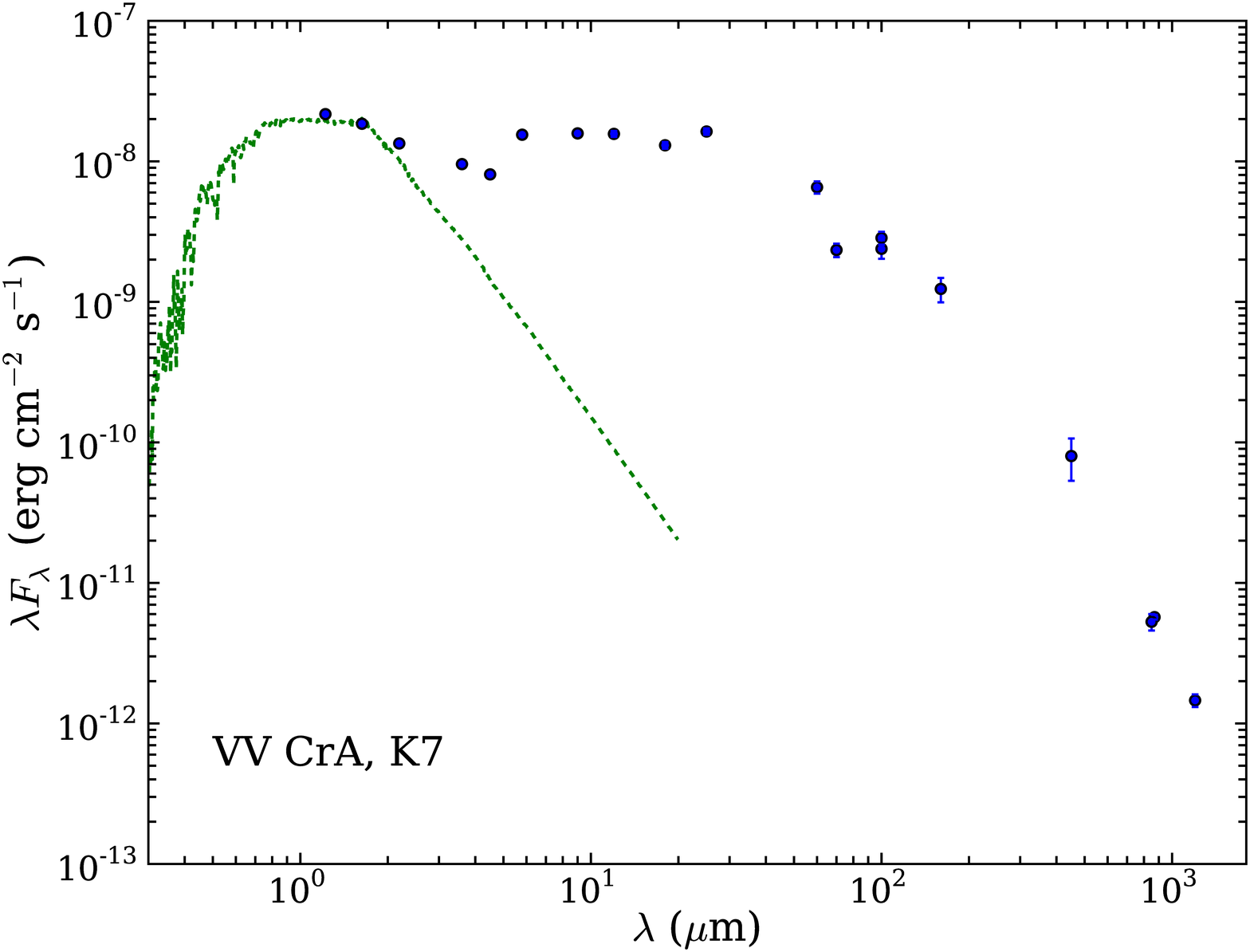,width=0.32\linewidth,clip=}\\
\end{tabular}
\caption{SEDs of detected intermediate-mass stars and solar-type TTS. Photometry detections are marked as circles, with upper limits
marked as inverted triangles. The spectra correspond to Spitzer/IRS observations,
when available. See Table \ref{sed-table} for details regarding 
the photometry data. For comparison, a photospheric MARCS model (Gutafsson et al. 2008)
with a similar spectral type is shown for each object (dotted line).  \label{interm-fig}}
\end{figure*}

R~CrA (spectral type A5; Bibo et al. 1992) and T~CrA (spectral type F0; Acke \& van den Ancker 2004) 
are the two intermediate-mass stars
with disks in the densest part of the
cluster (see Figure \ref{IRS7-fig}). They are both clearly detected in the PACS images, although the
strength of the nearby sources (especially, the IRS~7 complex and the extended cloud
structure) make it hard to quantify their fluxes. In particular, the
measured flux for R~CrA appears evidently contaminated by the nearby
objects and cannot be used to constrain its disk properties.
T~CrA, being more distant from the main emission peaks, is measurable
within reasonable limits. Its SED is consistent with
the presence of a flared, massive disk, and in agreement with the 
spectral type of F0 and extinction A$_V$=2.45 mag derived by
Acke \& van den Ancker (2004). The error in the flux at 160~$\mu$m
is nevertheless large, due to the presence of surrounding extended emission
(Figure \ref{interm-fig}).

\subsection{The low-mass CTTS: G-85, CrA-466 (G-113), CrA-159, HBC~677, and B18598.6-3658}

\begin{figure*}
\centering
\begin{tabular}{ccc}
\epsfig{file=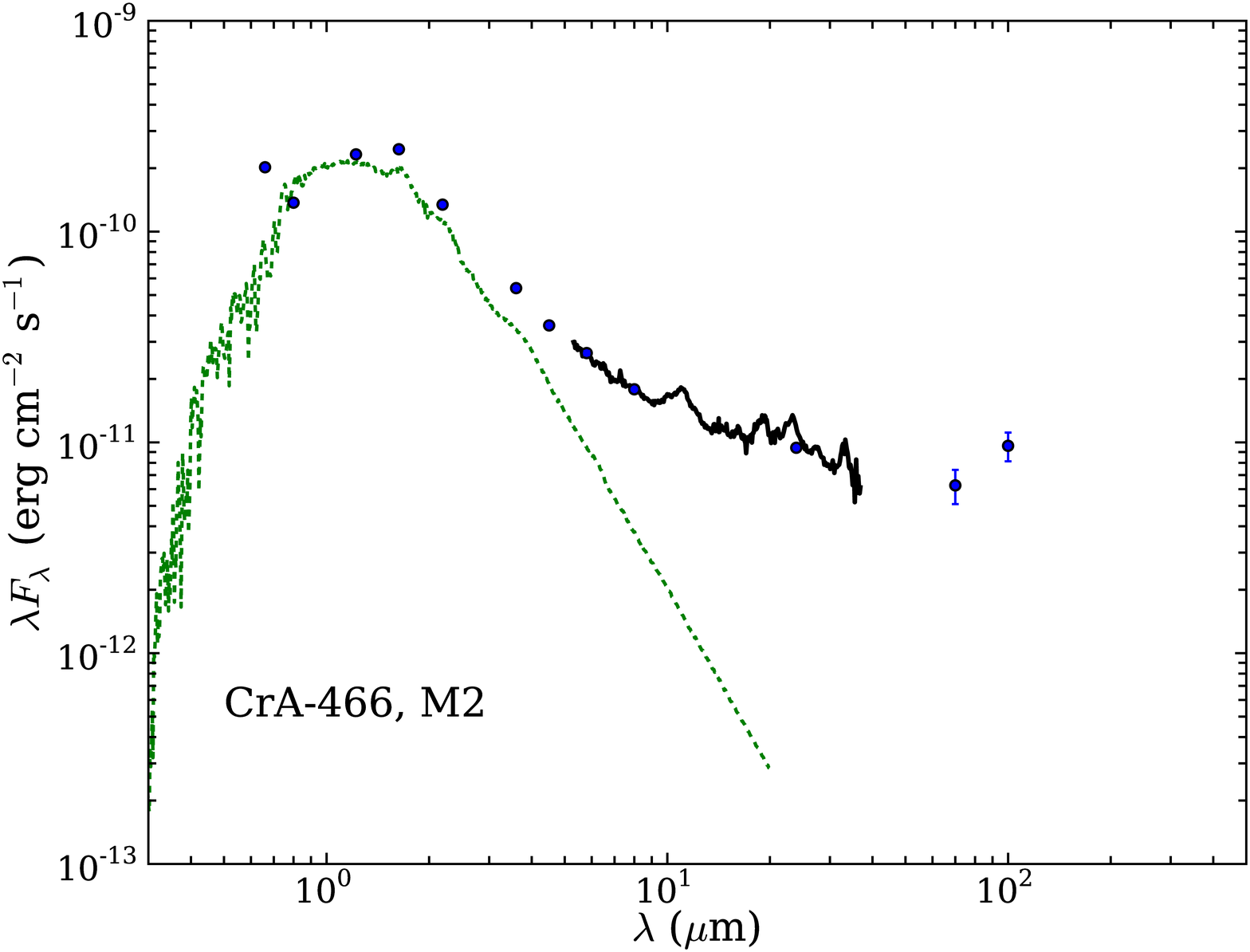,width=0.32\linewidth,clip=}&
\epsfig{file=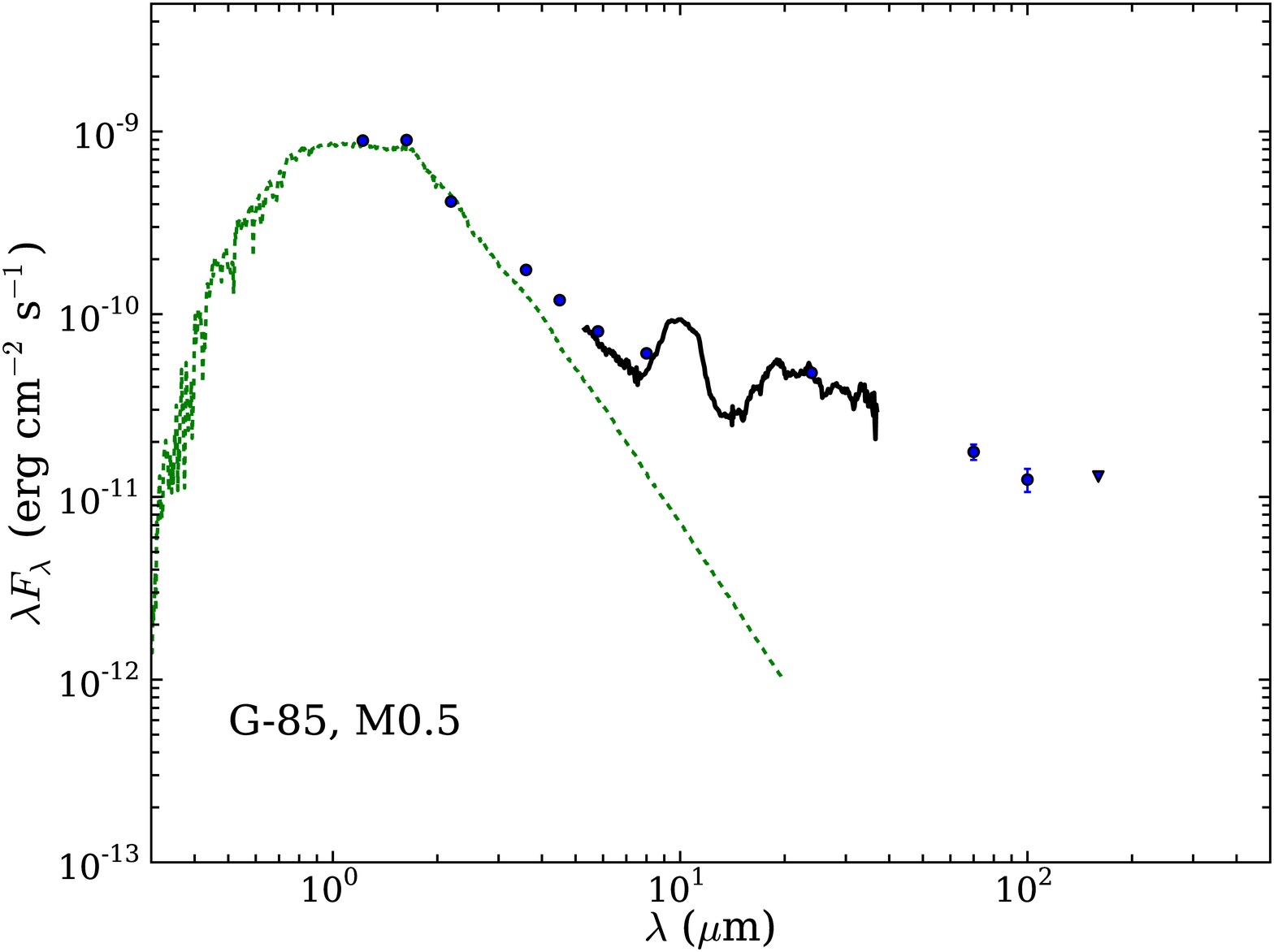,width=0.32\linewidth,clip=}&
\epsfig{file=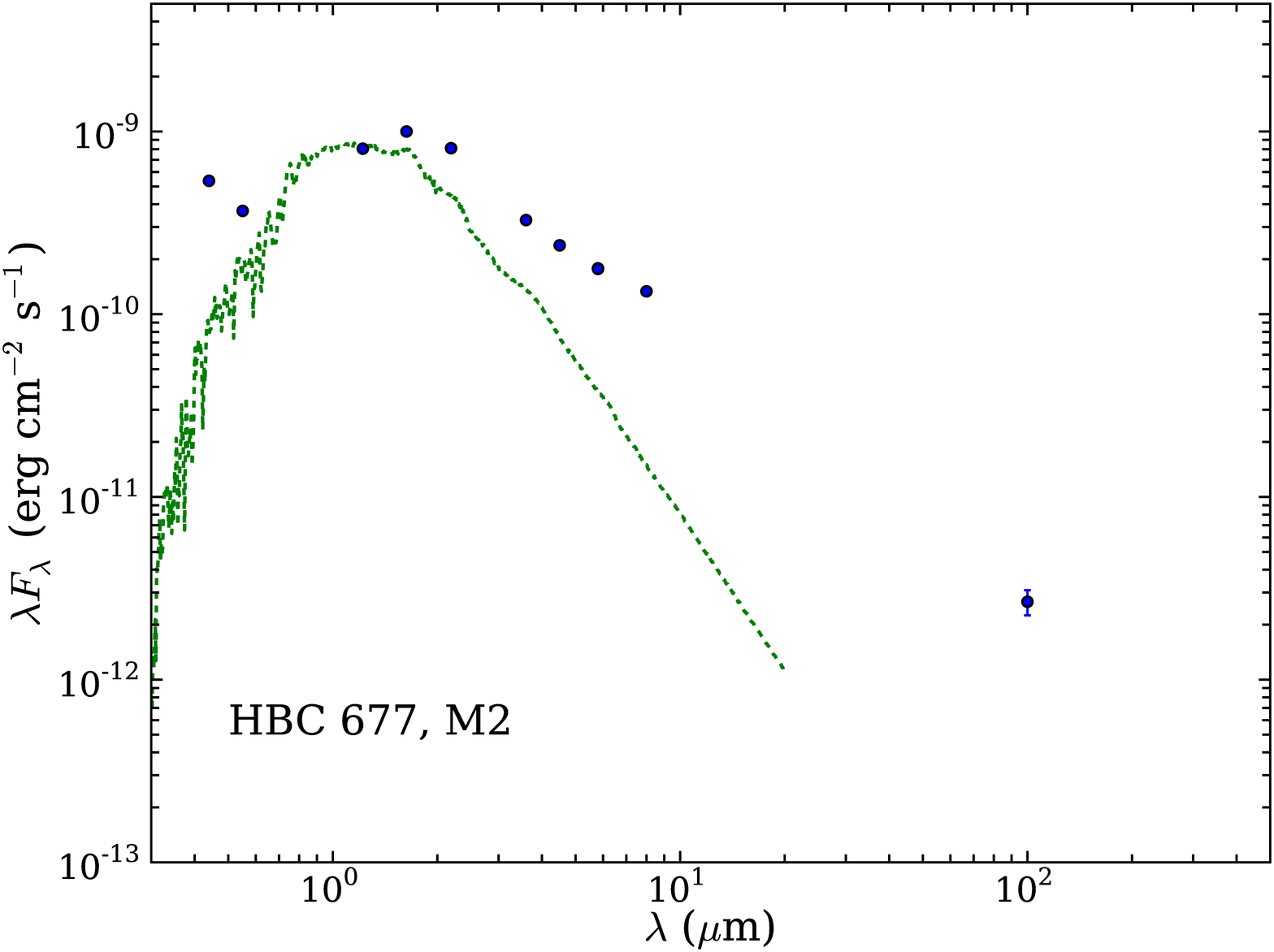,width=0.32\linewidth,clip=}\\
\epsfig{file=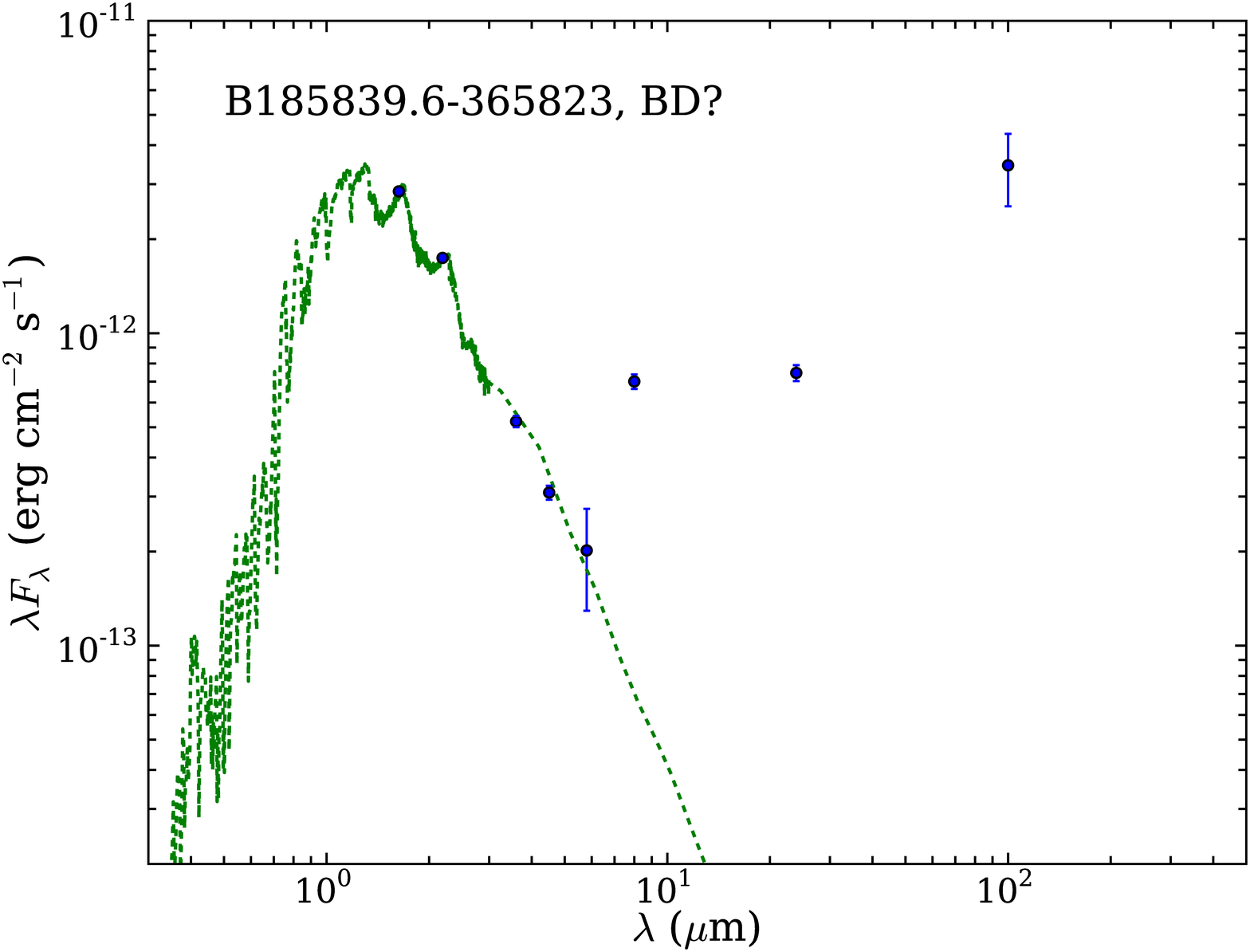,width=0.32\linewidth,clip=}&
\epsfig{file=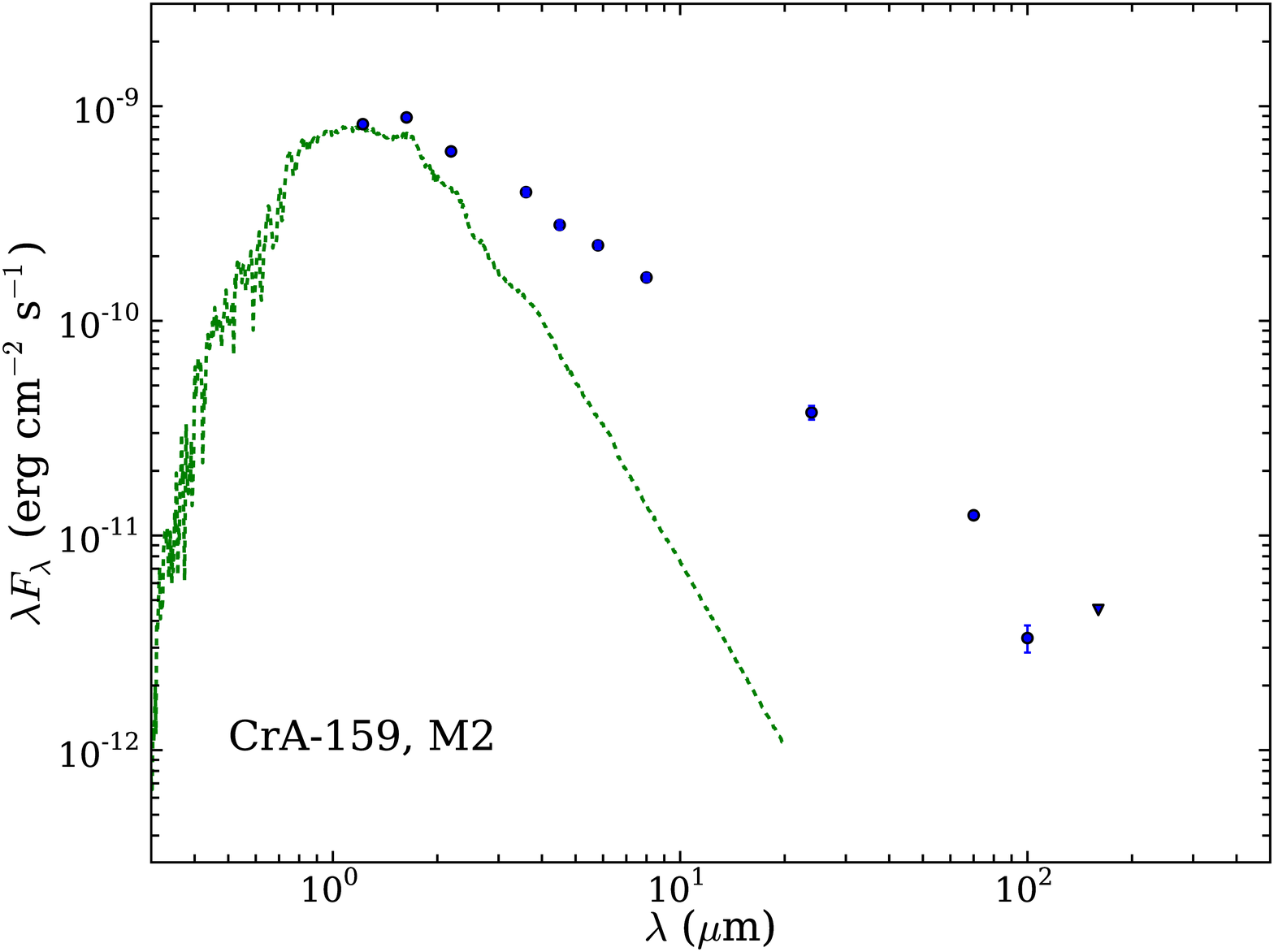,width=0.32\linewidth,clip=} &
\epsfig{file=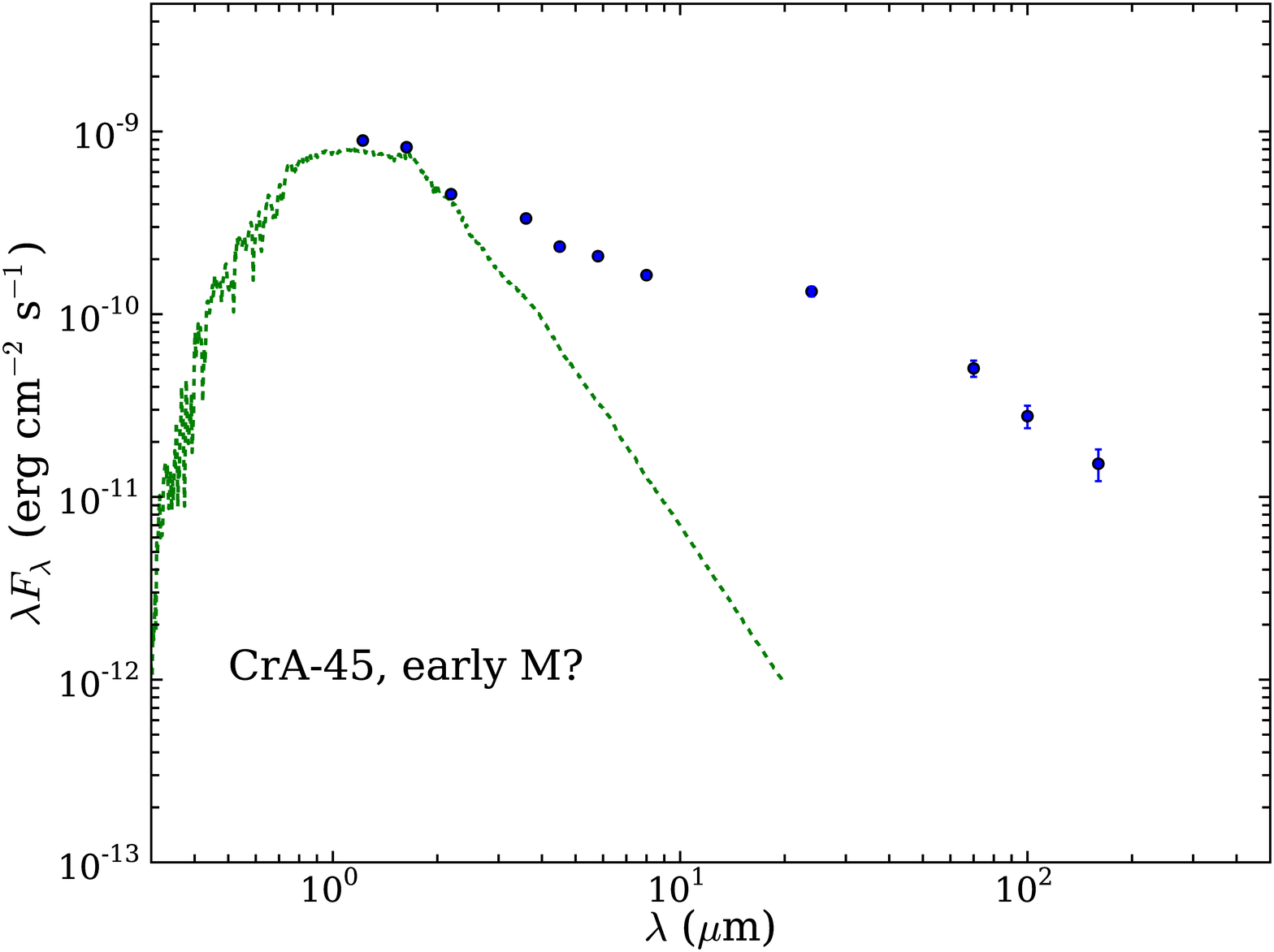,width=0.32\linewidth,clip=} \\
\end{tabular}
\caption{SEDs of detected low-mass TTS. Photometry detections are marked as circles, with upper limits
marked as inverted triangles. The spectra correspond to Spitzer/IRS observations,
when available. For comparison, a photospheric MARCS model (Gutafsson et al. 2008)
with a similar spectral type is shown for each object (dotted line). See Table \ref{sed-table} for details regarding 
the photometry data. \label{tts-fig}}
\end{figure*}

A few of the low-mass CTTS in the Coronet cluster are detected in the PACS images
(Figure \ref{tts-fig}).
The extincted M0.5 source G-85 (Sicilia-Aguilar et al. 2008, 2011a; Figure
\ref{IRS2-fig}) is well-detected
at 100~$\mu$m, and marginally detected at 160~$\mu$m. The M2 star CrA-466 (also known
as X-ray source G-113) is also detected in both PACS channels, although the 
160~$\mu$m detection is marginal. Two further M2-type sources, CrA-159 and
HBC~677, are well detected at 100~$\mu$m, but do not appear in the 160~$\mu$m images
(Figure \ref{IRS2-fig}).

There is emission detected towards the brown dwarf (BD) candidate B18598.6-3658 (Wilking et al. 1997;
Forbrich \& Preibisch 2007), although its SED is hard to interpret. Considering the
near-IR data from Wilking et al. (1997), the extinction estimate from Forbrich \& Preibisch (2007),
 and the Spitzer detections (Peterson et al. 2011),
the SED is consistent with a late M (M5-M7) star or BD with a remarkably bright transition disk.
Further data (e.g. confirmation of its spectral type) is required to determine
the true nature of the source. In particular, although the agreement between the PACS 
detection and the near-IR position of the source is excellent and it is fully consistent
with point-like emission like the rest of detected low-mass disks, contamination by nebular emission
or potential background objects should be ruled out.

The SEDs of the four detected known members (G-85, CrA-466, CrA-159, and 
HBC~677), very complete thanks to the availability of optical
and Spitzer data, reveal different types of protoplanetary disks. 
We will discuss their disk morphologies in more detail in Section \ref{models}.

\subsection{The northern subcluster: HD~176386, TY~CrA, and G-65}

The stars TY~CrA (B8) and HD~176386 (B9) are multiple systems that 
lie to the North of the main star-forming region (Bibo et al. 1992). 
Both of them have been detected at optical and IR wavelengths, although
the presence of substantial cloud material in the region has not 
unambiguously resolved the issue of whether they have circumstellar disks. MIPS/Spitzer data
reveal extended cloud emission, and a cavity-like structure presumably
created by TY~CrA, and since they do not have
an excess in the near-IR, they are most likely diskless (Sicilia-Aguilar et al. 2011a; 
Currie \& Sicilia-Aguilar 2011).  
The PACS images reveal substantial emission in the region, but unlike the MIPS
image (that showed emission centered around each one of the objects) and the
870~$\mu$m LABOCA map (that revealed extended emission centered around
HD 176386), the PACS data reveal an elongated structure located between
the two optical stars (Figure \ref{hd176-fig}). There is no detectable emission 
(at least, not over the extended background emission)
at the positions of the two stars. The extended emission is composed of a
brighter, larger blob closer to TY~CrA and to the edge of the MIPS cavity,
and a weaker extended tail that coils towards HD 176386. The full structure
is about 10000 AU in size (for a distance of 138 pc),
which could correspond to an embedded object or a heated clump 
located between the two intermediate-mass stars (see Figure \ref{hd176-fig}).
We cannot fully exclude the possibility that also an enhanced contribution from 
atomic lines (e.g., [CII] at 158~$\mu$m) to the total emission within the 
continuum filters can play a role here.
Some weaker emission is observed
around the structure, in particular, towards the low-mass object G-65, although
G-65 itself is not clearly detected due to the strong
background gradient. 
The fact that the PACS source(s) appear
undetected at MIPS 24~$\mu$m (MIPS 70~$\mu$m shows extended
emission in the area without differentiated peaks) 
suggest the presence of cold and dense material
compressed by the winds of the two nearby optical stars.
Such a structure, if dense enough, could be the ideal
environment for a new small episode of triggered star formation.

\subsection{The southern subcluster: VV CrA, IRAS 18598, and CrA-45}

To the south of the main star-forming region, we find three bright objects 
(IRAS 18598, VV CrA, and CrA-45; see Figure \ref{map-fig}). IRAS 18598 has some evidence of extended
emission at PACS wavelengths, which is consistent with the extended
structure observed at 870~$\mu$m (Sicilia-Aguilar et al. 2011a). Its SED is
hard to interpret, since it probably includes a near-IR source dominating the
Spitzer IRAC/MIPS emission, plus an extended structure that is responsible
for most of the far-IR and submillimetre emission. Further observations are
required to determine the nature of this object.

VV CrA is a known binary system with an IR companion (see
the detailed modeling by Kruger et al. 2011). The optical object has been
classified as K7 (SIMBAD; Bast et al. 2011). The PACS flux at 160~$\mu$m
shows extended emission, which could be associated with remnant cloud material
or to the envelope of the IR companion.  Small differences in the measured flux
at 100~$\mu$m depending on the 
aperture suggest that the IR emission could be extended. Simple SED fitting, assuming
that the optical companion is a K7 star, suggests that the extinction to the
optical source is rather A$_V$=15 mag, instead of the 26 mag suggested by Kruger
et al. (2012), but further modeling of the SED is hindered by the presence of
the dominant IR companion.

To the South/East of VV CrA we find the source CrA-45 (according to the IDs in
Peterson et al. 2011; also known as 
2MASS J19031609-3714080), which had been
previously identified as a YSO candidate (Gutermuth et al. 2009). The object
has been detected at Spitzer wavelengths, and its SED is consistent
with a low-extinction, early-M type CTTS.

\subsection{Other sources in the cluster}

The source SMM~6 from Nutter et al. (2005), which is one of the strongest peaks peak in our
LABOCA map (Sicilia-Aguilar et al. 2011a), shows very little emission at 160~$\mu$m
and is darker than the surroundings at 100~$\mu$m,
suggesting that it is a cold starless condensation. We will discuss it in more detail in Section \ref{proto}.

The source \#19 from Chini et al.(2003) appears as a complex emitting structure
in the Herschel images. It presents two relatively extended peaks (North and
South in Table \ref{red-table}, which could
simply be denser parts of the cloud), together with filamentary structures that
seem to extend from the central peaks out.

Finally, the extended emission seen with MIPS to the 
North-West of CrA-4111 is detected at PACS wavelengths as cloud emission with at least one, but
probably up to three point sources. At 100~$\mu$m, only one of the sources is clearly resolved
as an independent object (at 19:01:18.73, -37:02:60.7), 
while two more faint compact sources
are resolved at 160~$\mu$m (at 19:01:19, -37:03:22 and 19:01:19, -37:02:51). Only the first of 
these is detected (marginally) in IRAC 
at 4.5~$\mu$m (which could indicate the presence of an outflow) and MIPS/24~$\mu$m. 
The association with extended cloud material in all
these cases and the presence of nearby sources makes them good candidates to be new embedded cluster members
or very low-mass protostars,
although the lack of further data does not allow us to rule out the possibility
of extragalactic sources seen in projection against the cloud.
The complete list of newly detected Herschel sources and their analysis will be
presented in a follow-up publication by the Gould Belt Survey group.

\subsection{Non-detected YSOs}

\begin{figure*}
\centering
\begin{tabular}{ccc}
\epsfig{file=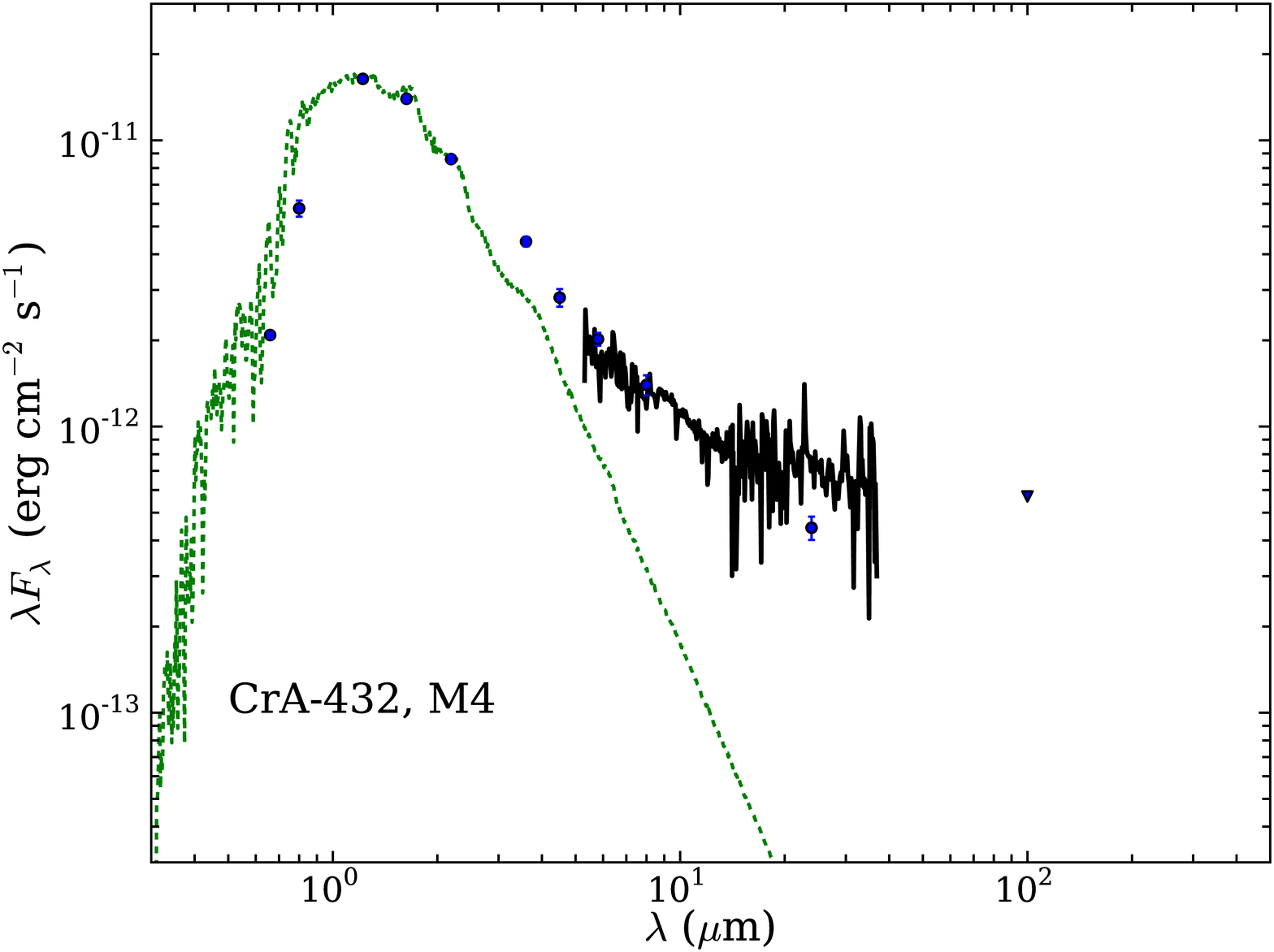,width=0.32\linewidth,clip=}&
\epsfig{file=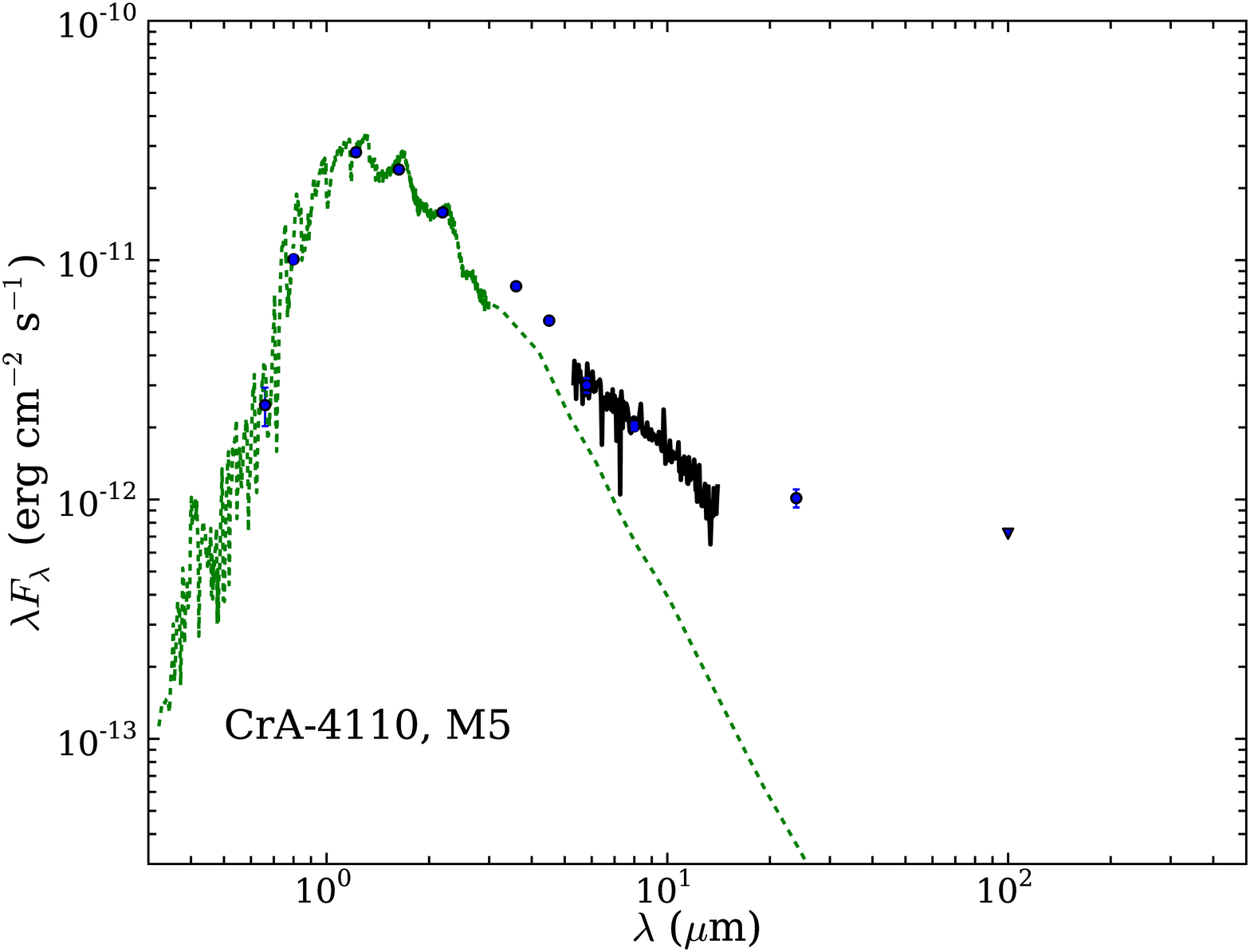,width=0.32\linewidth,clip=}&
\epsfig{file=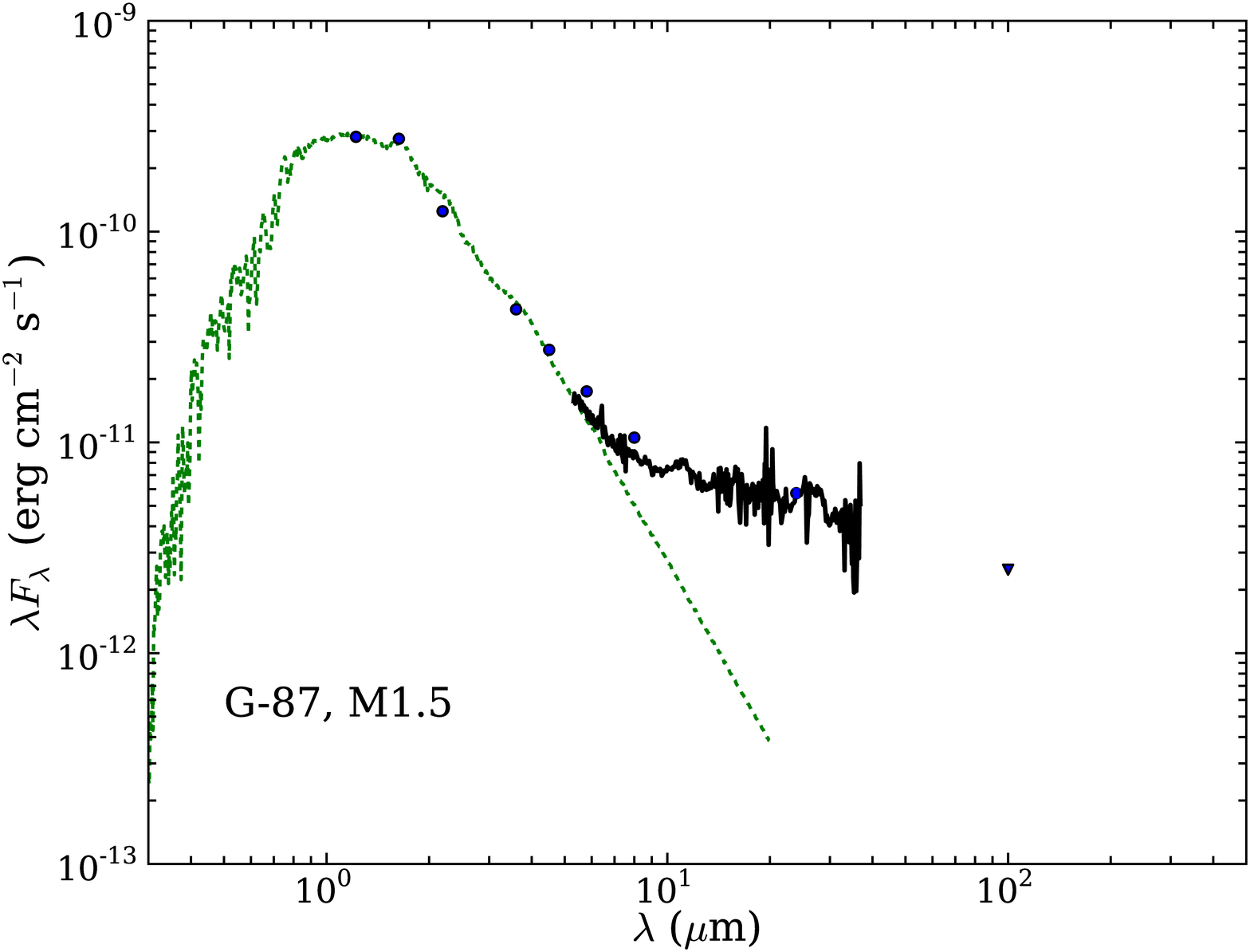,width=0.32\linewidth,clip=} \\
\epsfig{file=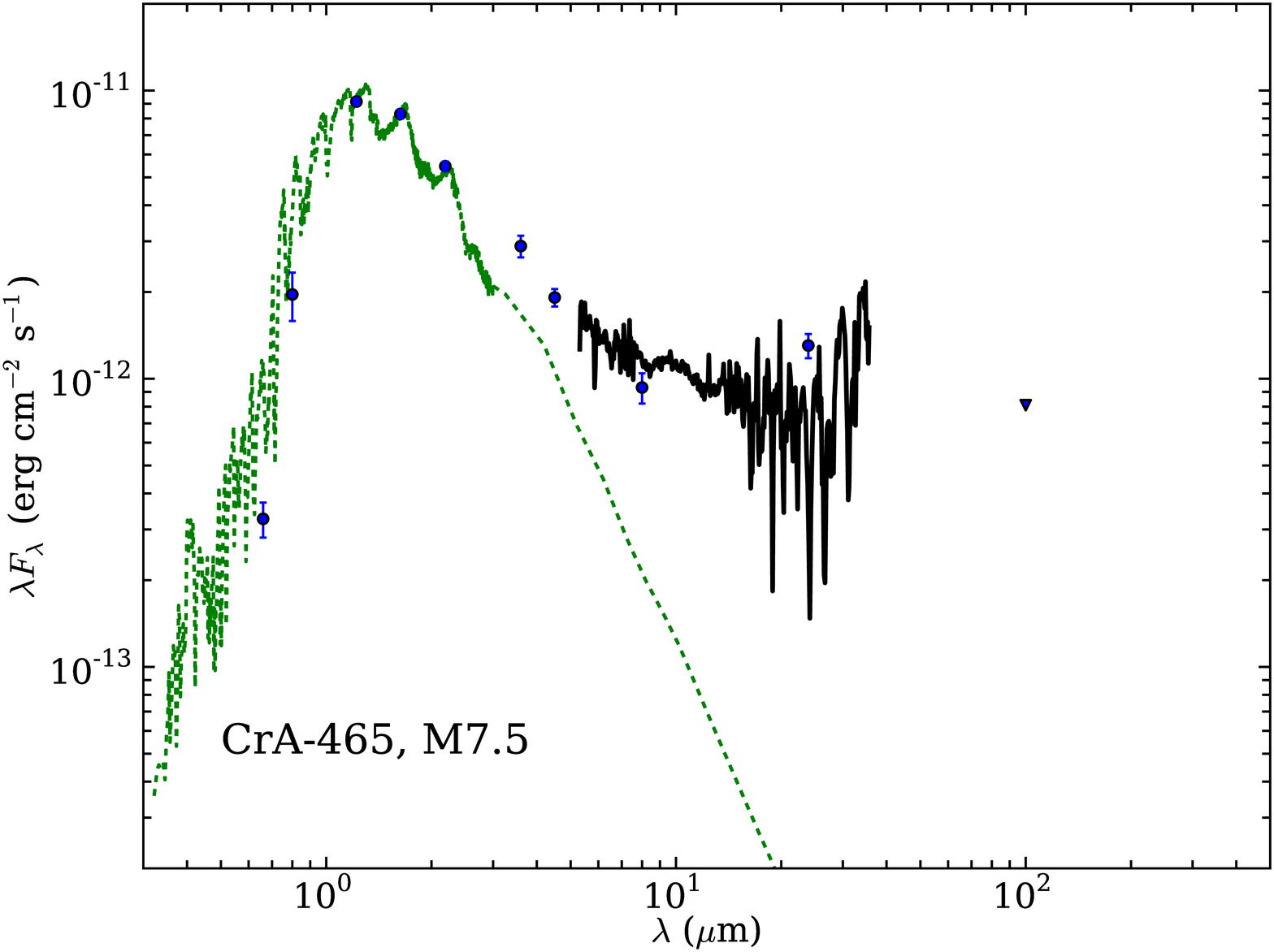,width=0.32\linewidth,clip=} &
\epsfig{file=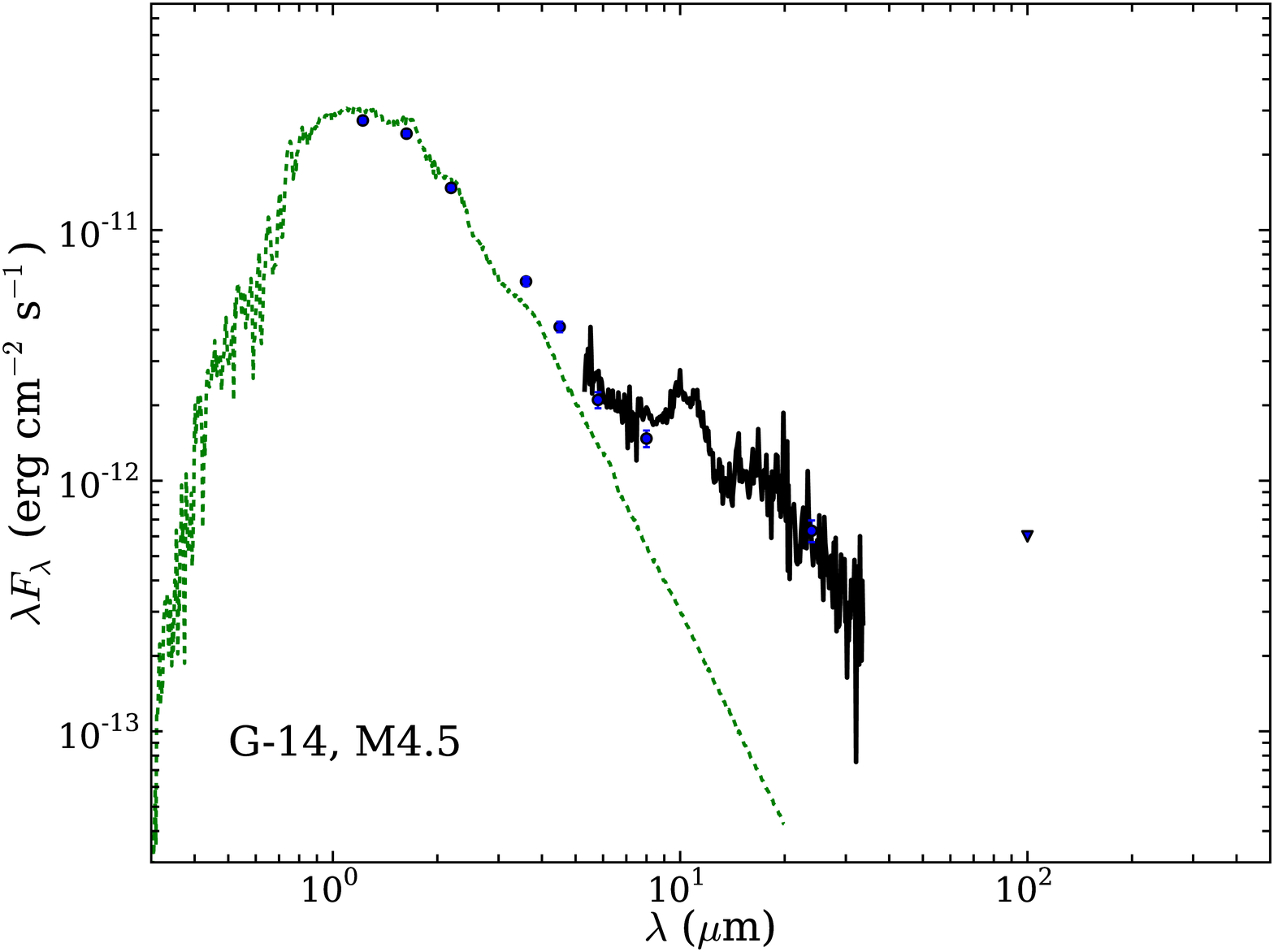,width=0.32\linewidth,clip=}&
\epsfig{file=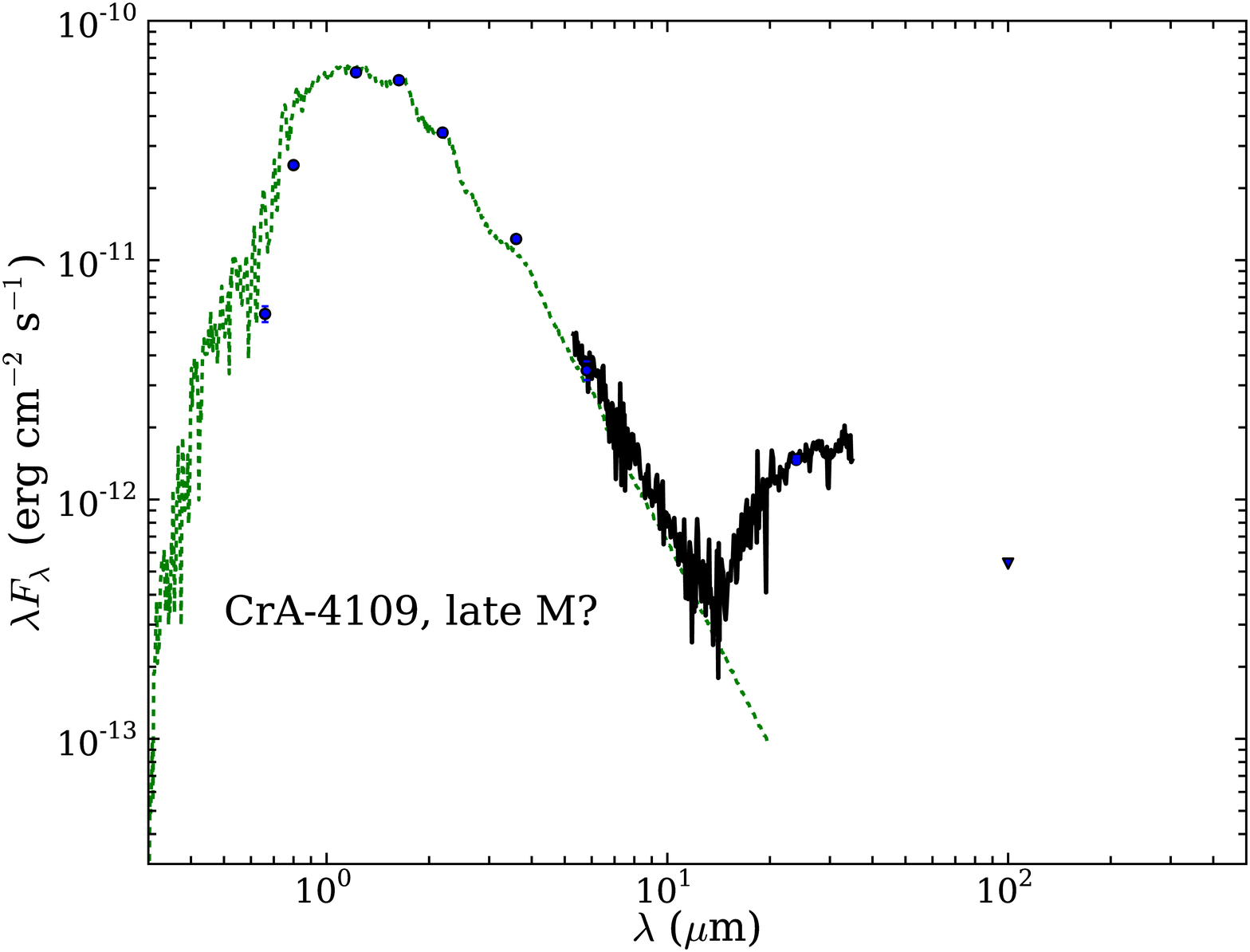,width=0.32\linewidth,clip=} \\
\epsfig{file=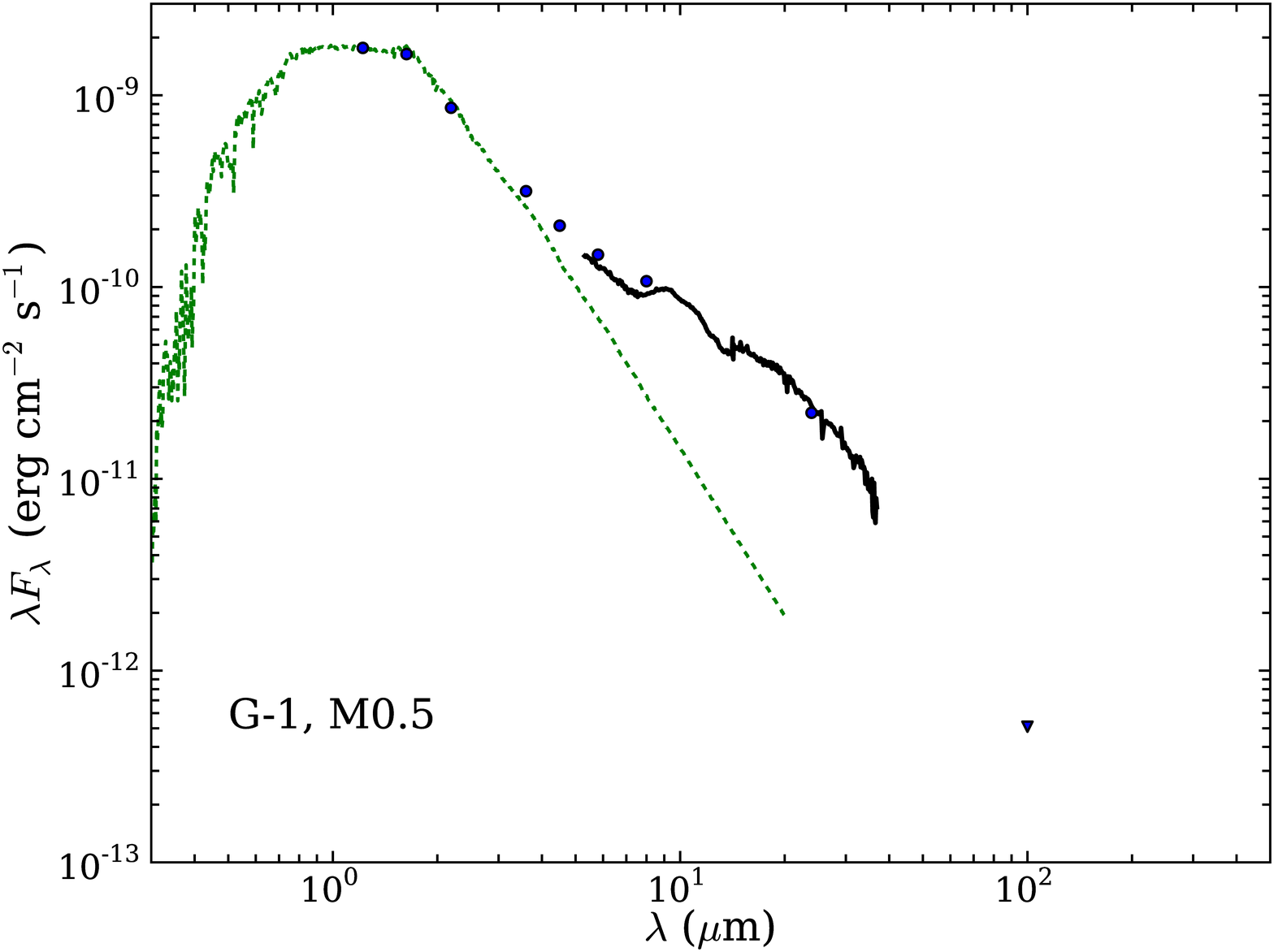,width=0.32\linewidth,clip=} &
\epsfig{file=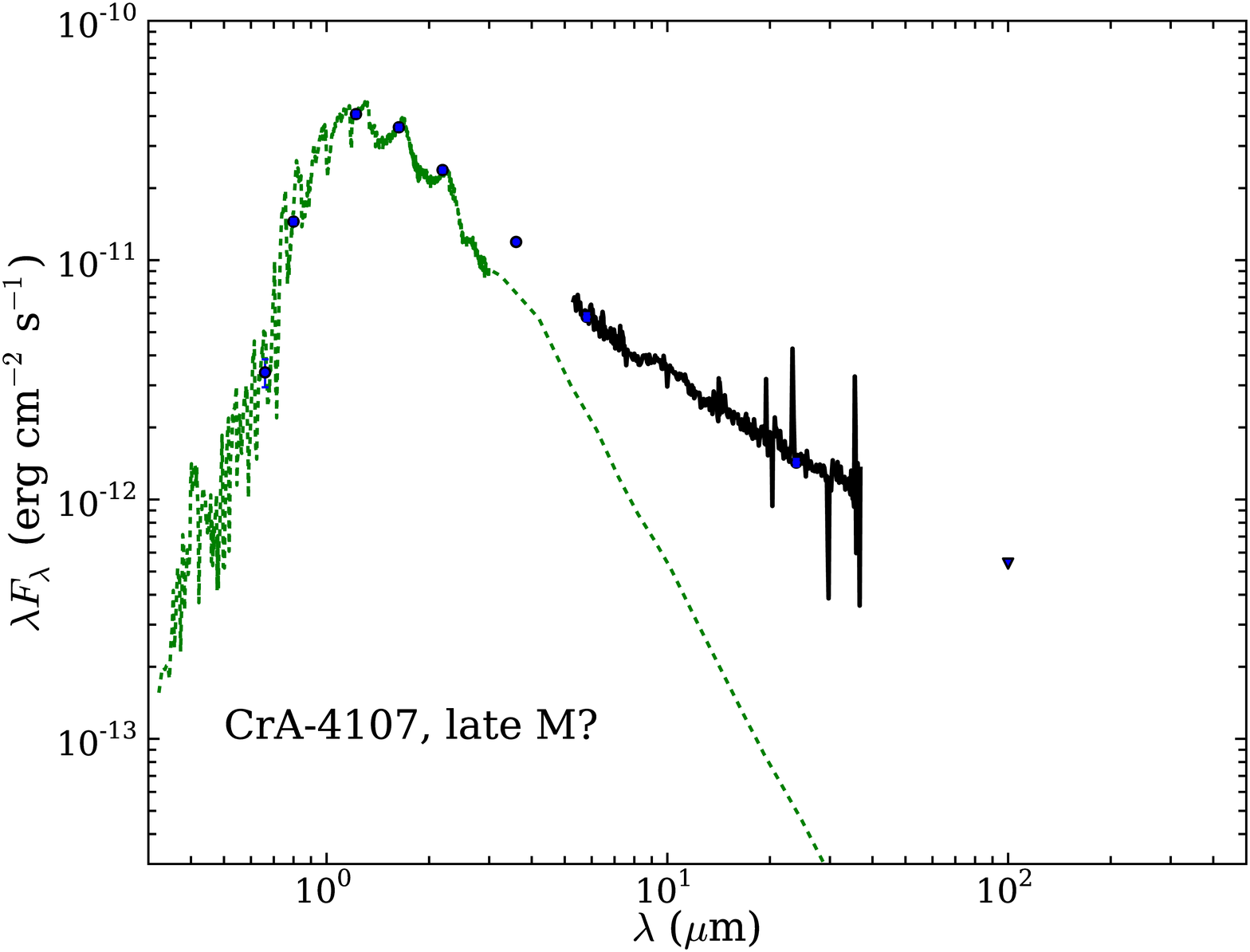,width=0.32\linewidth,clip=} &
\\
\end{tabular}
\caption{SEDs of undetected low-mass TTS with stringent upper limits. Photometry detections are marked as circles, with upper limits
marked as inverted triangles. The spectra correspond to Spitzer/IRS observations,
when available. For comparison, a photospheric MARCS model (Gutafsson et al. 2008)
with a similar spectral type is shown for each object (dotted line). See Table \ref{sed-table} for details regarding 
the photometry data. \label{undet-fig}}
\end{figure*}

\begin{table}
\caption{Upper limits for undetected objects} 
\label{uplim-table}
\begin{tabular}{l c c l}
\hline\hline
Object & $\lambda$ (~$\mu$m) & Upper Limit Flux (Jy) & Comments \\
\hline
CrA-432 & 100 & 0.019 & \\
CrA-4110 & 100 & 0.025 & \\
G-87 & 100 & 0.082 & Transitional disk\\
G-45 & 100 & 0.018 & \\
CrA-465 & 100 & 0.027 & \\
G-32 & 100 & 0.027 & \\
G-14 & 100 & 0.020 & \\
CrA-4109 & 100 & 0.018 & Transitional disk\\
G-1 & 100 & 0.017 & Dust-depleted\\
CrA-4107 & 100 & 0.018 & \\
\\
G-85$^1$    & 160 & 0.69 & Pre-transitional disk\\
CrA-159 & 160 & 0.24 & Truncated disk? \\
\hline
\end{tabular}
\tablefoot{Photometry 3$\sigma$ upper limits for the objects with known IR excess not detected by PACS.
Only objects with significantly low upper limits (compared to their mid-IR emission) are listed here. 
$^1$ G-85 is a marginal (2.5$\sigma$) detection at 160 $\mu$m, so we expect that although uncertain, its
upper limit to the flux at this wavelength is relatively close to the real value.}
\end{table}

Many of the disked TTS listed in the literature and some of the candidate protostars
detected at X-ray (Forbrich \& Preibisch 2007; Sicilia-Aguilar et al. 2008, 2011a;
Currie \& Sicilia-Aguilar 2011; Peterson et al. 2011) are not found in the PACS images. This is
especially true among the faintest very-low-mass objects and the objects with small 
IR excesses.
The presence of emission from substantial cloud material does not allow to place 
reasonable upper limits in many cases (G-32, G-36, G-43, G-65, G-94, G-95, G-112, CrA-205, CrA-4111).
For objects located in clean areas, the PACS non-detections set important
constraints on the properties of several sources (see Table \ref{uplim-table}).

Among the protostellar candidates detected at X-ray wavelengths,
G-45 would be consistent with the object being a protostellar
condensation, undetected in PACS due to the presence of surrounding cloud
emission, although the scarcity of Spitzer data does not allow to fit any model
and further observations are desirable.
Some extended emission is seen near the X-ray objects G-74 and G-128, although not
at the same position. The X-ray emission could be related to potential 
very low-mass objects objects (similar to G-122), or to background extragalactic sources. 

Among undetected TTS, stringent upper limits can be placed for G-1 (also known as HBC 680, M0.5), G-14 (M4.5), G-87 (M1.5), 
CrA-4107, CrA-4109 (both probably late M stars), CrA-4110 (M5), and the BD candidate CrA-465 (M7.5; Figure \ref{undet-fig}).
Some diffuse emission is seen near HBC 679, but since this object did not show any excess emission
at Spitzer wavelengths (Currie \& Sicilia-Aguilar 2011), 
it is most likely of nebular origin. The upper limits of
G-1 confirm a very low level of emission for this object in the far-IR, as it has been
inferred from Spitzer data, being candidate for hosting a globally dust-depleted
disk (Currie \& Sicilia-Aguilar 2011). The upper limits also place strong constraints on
the disk of G-87, another candidate for low dust mass or dust depleted disk. The case of CrA-465 is more
uncertain, since the MIPS images already suggest the presence of some nebular contamination,
and the LABOCA map confirmed the existence of background extended emission, so the 
PACS data probably contain substantial cloud contamination. The detailed properties of
these disks will be discussed using simple disk models in Section \ref{models}.

\section{Analysis and discussion \label{analysis}}

 \begin{figure*}
   \centering
   \resizebox{\hsize}{!}{\includegraphics[width=\textwidth]{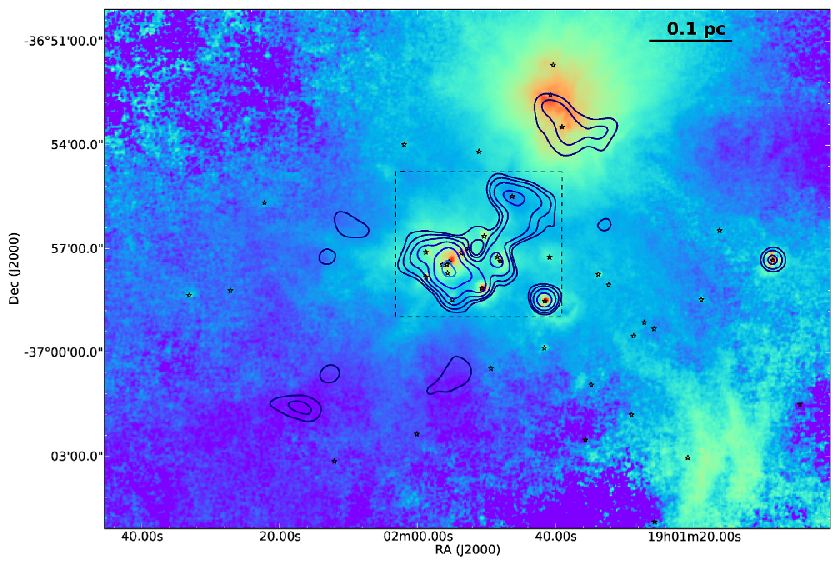}}
\caption{Approximate temperature map, resulting from dividing the 100 $\mu$m by the 160 $\mu$m
images. The known cluster members are marked as stars, and the LABOCA 870 $\mu$m contours at
0.1,0.2,0.4,0.8,1.6,3.2,6.4,12.8 Jy/beam are also displayed. The color gives an approximate idea
of the temperature of the region, with red being the hottest (100 $\mu$m emission stronger
than 160 $\mu$m emission) and violet being the coldest (160 $\mu$m emission stronger than
the 100 $\mu$m emission). Protostars and disks tend to appear hot (red), while the starless
870 $\mu$m regions are clearly colder than the rest of the cloud. The dashed-line box marks the 
region zoomed in in Figure \ref{tempzoom-fig}.  
\label{temp-fig}}
\end{figure*}

\subsection{Protostars, envelopes, and disks \label{proto}}

\begin{figure*}
   \centering
   \resizebox{\hsize}{!}{\includegraphics[width=\textwidth]{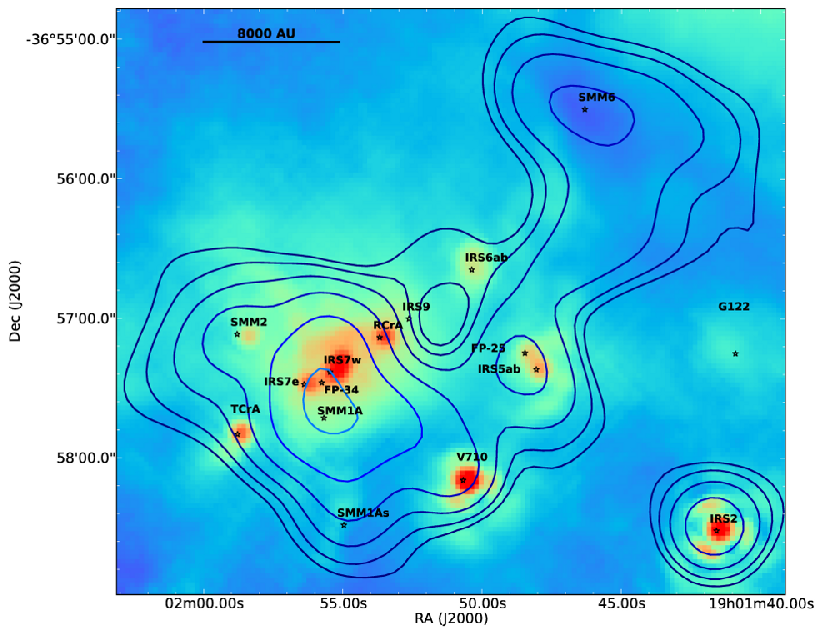}}
\caption{An expanded view of the approximate temperature map from Figure \ref{temp-fig}. The
known cluster members and submillimetre emission regions are marked by stars and labeled. 
Protostars and disks tend to appear hot (red), while the starless
870 $\mu$m regions are clearly colder than the rest of the cloud. FP-25 appears colder than
IRS~5, and IRS~7e is also colder than IRS~7w. There is no significant emission near IRS~9, located
to the east of a submillimetre "hole". The Class 0 candidates, SMM~1~A and SMM~1~As, are clearly
colder than the dominant Class I protostars and the disked objects. The very low-mass protostellar
candidate, G-122, is also colder than more massive Class I sources. The secondary
peak of the LABOCA map, coincident with source SMM~6 from Nutter et al.(2005),
appears as a distinct region even colder than the Class 0 protostars. 
\label{tempzoom-fig}}
\end{figure*}

In order to gain some insight in the mass and temperature of the objects
detected by Herschel/PACS, we have constructed an approximate temperature map by
dividing the 100 $\mu$m image by the 160 $\mu$m one (Figure \ref{temp-fig}).
This map provides important relative information
on a pixel-by-pixel scale, being thus important to characterize the extended structures and
the whole cloud, also in the regions where no Herschel emission is evident. 
Bright sources in the ratioed map correspond to objects with
hotter temperatures, while cool regions that are stronger emitters at 160 $\mu$m than at 100 $\mu$m
appear faint. 
With this exercise, we confirm the existence of a cold, well-defined structure in the position of
the submillimetre source SMM~6 (see Figure \ref{tempzoom-fig}), 
which is also coincident with an extended peak in the
APEX/LABOCA 870 $\mu$m map, being thus a relatively massive
and cold core. Other structures detected at 870$\mu$m but without
evident PACS counterparts also appear as cold regions
(see Figure \ref{temp-fig}), while the emission around HD 176386 and TY~CrA 
appears hot, indicating heating by the nearby intermediate-mass stars
or even the presence of some new embedded protostars.
The differences between IRS~7w vs. IRS~7e and 
IRS~5a/b vs. FP-25 are also evident, with IRS~7w appearing hotter than IRS~7e, and
IRS~5a/b being hotter than FP-25, although without further data it is hard to establish
whether this difference is due to evolution, viewing angle,
or source mass. The very low-mass Class I candidate G-122 appears
also colder than typical, higher-mass Class I sources, and the Class 0 candidates SMM~1~A
and SMM~1~As are also cold and probably extended. 

By observing their SEDs (Figures \ref{proto-fig}-\ref{undet-fig}), the objects can be classified
as protostellar Class I candidates (with SEDs peaking in the mid-IR and silicate features in 
absorption, when an IRS spectrum is available) or Class II sources with disks (with SEDs
peaking in the optical or near-IR and silicate features in emission).
Objects detected only at far-IR/submillimeter wavelengths are Class 0 candidates.
In principle, our Herschel data confirms the previous classification in the literature,
with objects like IRS~7w, IRS~5, V~710, and SMM~2 being Class I candidates, while 
S~CrA, T~CrA, G-85, CrA-159 and similar sources are Class II objects. 
A few objects have uncertain nature (e.g. IRS~2), and in other
cases, the far-IR detection is probably associated with nearby cloud material  (e.g. HD~176386),
remnant envelopes or a embedded nearby companion (e.g. IRAS 18598).

\begin{figure*}
   \centering
   \resizebox{\hsize}{!}{\includegraphics[width=\textwidth]{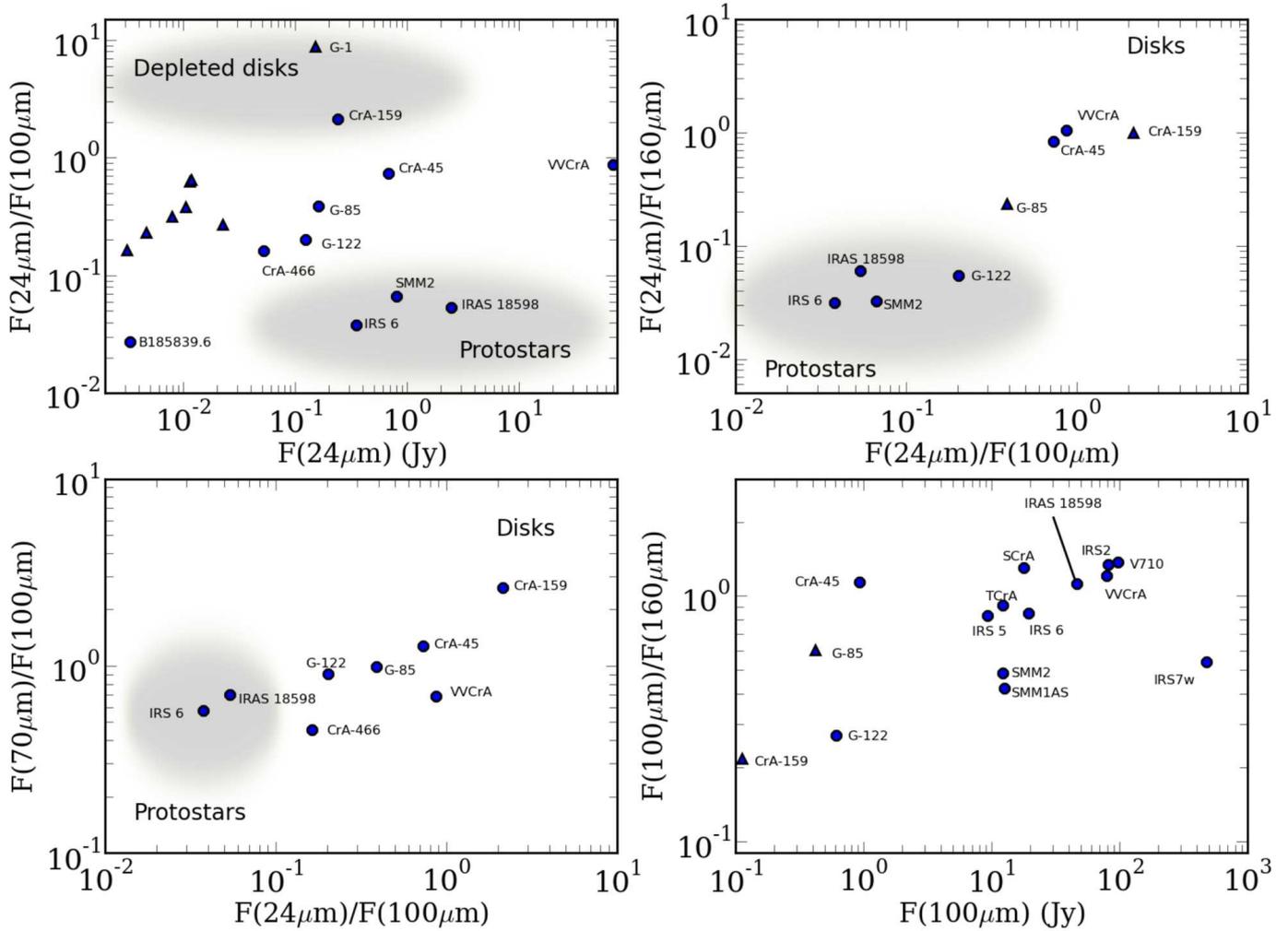}}
\caption{Different fluxes and flux ratios for the CrA members. Lower limits in the ratios
(for the cases with upper limits in the Herschel/PACS fluxes) are displayed as
triangles. Approximate regions for depleted disks, disks, and protostars are marked. 
Note that the emission of IRAS 18598 is dominated by the far-IR counterpart. 
\label{ratios-fig}}
\end{figure*}

To obtain more precise results, we have explored different types of diagnostic
diagrams involving the MIPS 24 and 70~$\mu$m data and the PACS photometry for
the detected objects and those with relevant upper limits (Figure \ref{ratios-fig}),
confirming this classification. The main parameter affecting the flux
ratios is the object temperature. While disks appear relatively flat or with a 
moderate negative slope in the 24-70-100-160~$\mu$m range, protostellar objects 
can show flat, rising, or decreasing slopes in this range, depending on 
their temperatures. The second important parameter is the object mass. 
Massive disks have much flatter slopes than depleted disks,
which becomes clear in the relation
between the 24~$\mu$m and 100~$\mu$m flux (see the loci of the typical disks
like G-85, CrA-45, CrA-466 vs that of the mass-depleted or low-mass disks like 
CrA-159 and G-1). Dust-depleted disks would also have higher 24~$\mu$m/160~$\mu$m ratio, although
none is detected at this wavelength. For protostellar objects, 
their masses, temperatures, and evolutionary state are to some extent degenerate,
 and difficult to uniquely determine without detailed modeling, although the
general trend is that less massive and less evolved objects will tend
to have lower fluxes and lower temperatures (Myers \& Ladd 1993). The distinction between protostars and
disks is thus maximal in the diagrams involving the shorter wavelengths (24~$\mu$m).
The 100~$\mu$m/160~$\mu$m vs 100~$\mu$m diagram is harder to interpret in terms of evolutionary
state, since emission at these wavelengths in both protostars and disks comes mostly from small cold grains,
so objects are rather separated by their luminosity.
The low-mass protostar candidate G-122 appears
in an intermediate location due to its low flux and long wavelength SED peak (low
temperature). Although there are less objects detected at 70 and 160~$\mu$m (in particular,
among the low-mass protostars and low dust-mass disks), these
diagrams also reveal a similar trend. 

\begin{table*}
\caption{Temperature and luminosity for the protostellar candidates} 
\label{proto-table}
\begin{tabular}{l c c c c l}
\hline\hline
Object 		& Class & T (K) & $\beta$ & L$_{bol}$ (L$_\odot$) & Comments \\
\hline
G-122 		& I/0:	& ---   & ---	& 0.005: & No submillimetre data  \\
IRS~2 		& I	& 25	& 1.5	& 0.8  	 &   \\
IRS~5a/b	& I	& 45	& 0	& 0.2  	 &   \\
FP-25		& I	& 20	& 1.5	& 0.05:  & Contamination by IRS~5a/b   \\	
IRS~6a/b	& I	& ---   & ---	& 0.04:  & No submillimetre data  \\	
V~710		& I	& 100 	& 0/1	& 3.1  	 &   \\
SMM~1~A		& 0	& 16	& 2	& 0.5  	 &   \\
IRS~7w		& I	& 25	& 2	& 0.9  	 &   \\
SMM~1~As	& 0	& 16	& 2	& 0.04   &   \\
SMM~2		& I	& 16	& 2	& 0.04 	 &   \\
IRAS~18598	& I	& 20/32	& 1/2	& 0.3  	 &   \\
\hline
\end{tabular}
\tablefoot{Temperature, frequency exponent, and bolometric luminosity for
the sources considered as Class 0 and Class I protostars. Note that for Class I
objects, T and $\beta$ refer to the fit to the longer wavelengths, which can be assimilated 
to the envelope/outer disk, while the luminosity is in all cases the integrated bolometric
luminosity derived from the full SED, without applying any further corrections.
The luminosity integration is performed by assuming that the object can be 
reproduced by a sum of black bodies with different temperatures in order to extrapolate
the emission at longer and shorter wavelengths. Typical errors are in the range 10-20\%. More uncertain
values (usually, due to the uncertainty of the photometry and/or to the lack of observations
at longer wavelengths) are marked by ":". The envelope temperature of G-122 and IRS~6a/b cannot be
determined due to the lack of submillimetre data, while the long wavelength SEDs of V~710 and IRAS~18598
can be fitted with different values of T and $\beta$.}
\end{table*}

The PACS and submillimetre data for two of the sources
associated with the main cluster (SMM~1~A and SMM~1~As; Groppi et al. 2007)
strongly resemble the emission expected for Class 0 objects. SMM~1~A
had been already classified as a potential Class 0 source (Nutter et al. 2005).
Including the Herschel data, we have constructed simple modified black-body models for these
objects, following Ward-Thompson et al. (2002). The emission of a modified
black-body, or grey-body, can be written as

\begin{equation}
	F_\nu = B_{\nu}(T) (1-e^{-\tau_\nu})\Omega,
\end{equation}

where $F_\nu$ is the flux density, $B_{\nu}$(T) is the black-body emission for
a temperature T, $\tau_\nu$ is the frequency-dependent optical depth, and $\Omega$ is
the solid angle subtended by the source.
If we assume that at long wavelengths,
the optical depth follows a power law with frequency, $\tau_\nu \propto \nu^\beta$,
it is possible to derive the temperature of the source. Taking the values
from Ward-Thompson et al.(2002) for $\beta$=2 and $\tau_{200\mu m}$=0.06, we obtain
very good fits to both objects for temperatures around 16 K (see Figure \ref{proto-fig}),
consistent with pre-stellar cores. Nevertheless, further data, including higher resolution 
maps, are required to fully determine the nature of these objects,
since external heating may also play a role, especially in the surroundings
of other brigth protostars (Lindberg \& J{\o}rgensen 2012). 

As a comparison, and although a modified black-body does not provide a good fit to
any of the remaining protostars, we have repeated this exercise for the submillimetre part of
the SEDs of the Class I objects with enough long-wavelength data (see Table \ref{proto-table}). The fits are also displayed
in Figure \ref{proto-fig}, together with the temperature and $\beta$ coefficient information.
Except for SMM~2, which can be well reproduced at long wavelengths
with a similar model to SMM~1~A and SMM~1~As,
all the other objects require higher temperatures (in the range of 20-100 K) and 
different frequency power law exponents ($\beta$=0-2). If we identify this
longer wavelength part of the SED as the remaining envelope material of the source, this
would suggest higher temperatures and different degrees of grain growth, as expected
if these objects are more evolved than the Class 0 candidates.

%717: omega=9e-9
%725: omega=1.5e-7

\subsection{Disks properties in the Coronet cluster \label{models}}

\begin{figure*}
\centering
\begin{tabular}{ccc}
\epsfig{file=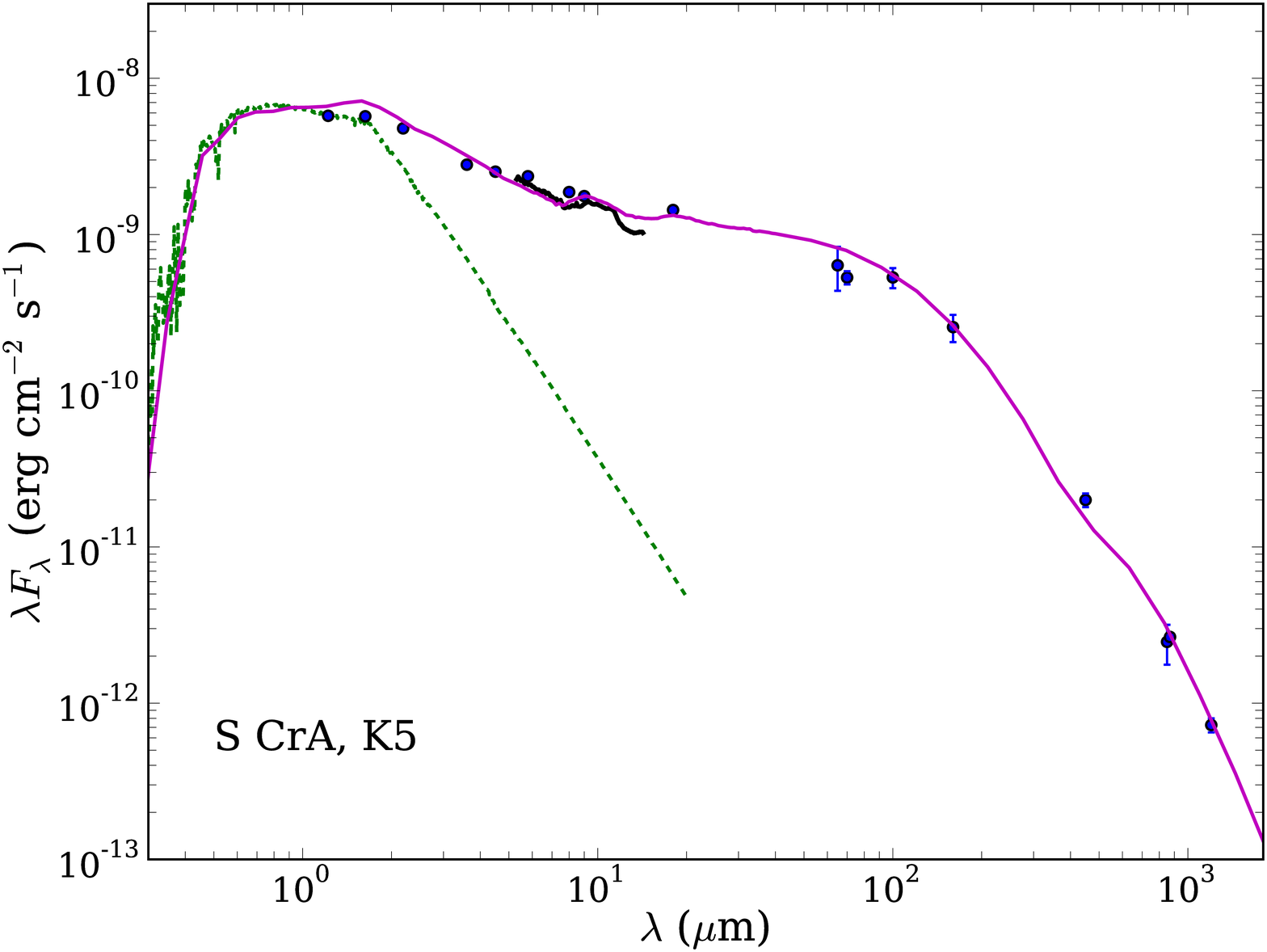,width=0.32\linewidth,clip=} &
\epsfig{file=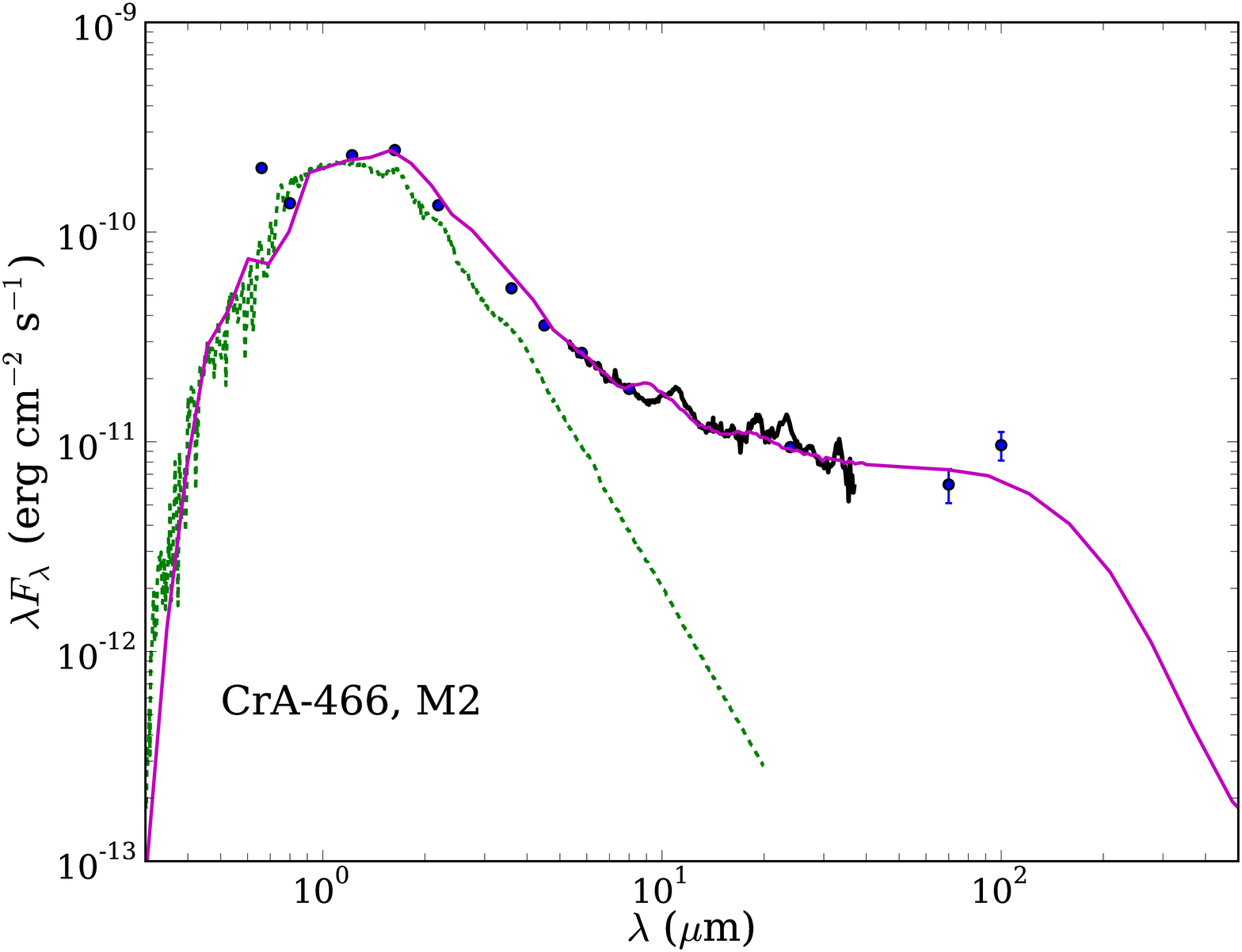,width=0.32\linewidth,clip=} &
\epsfig{file=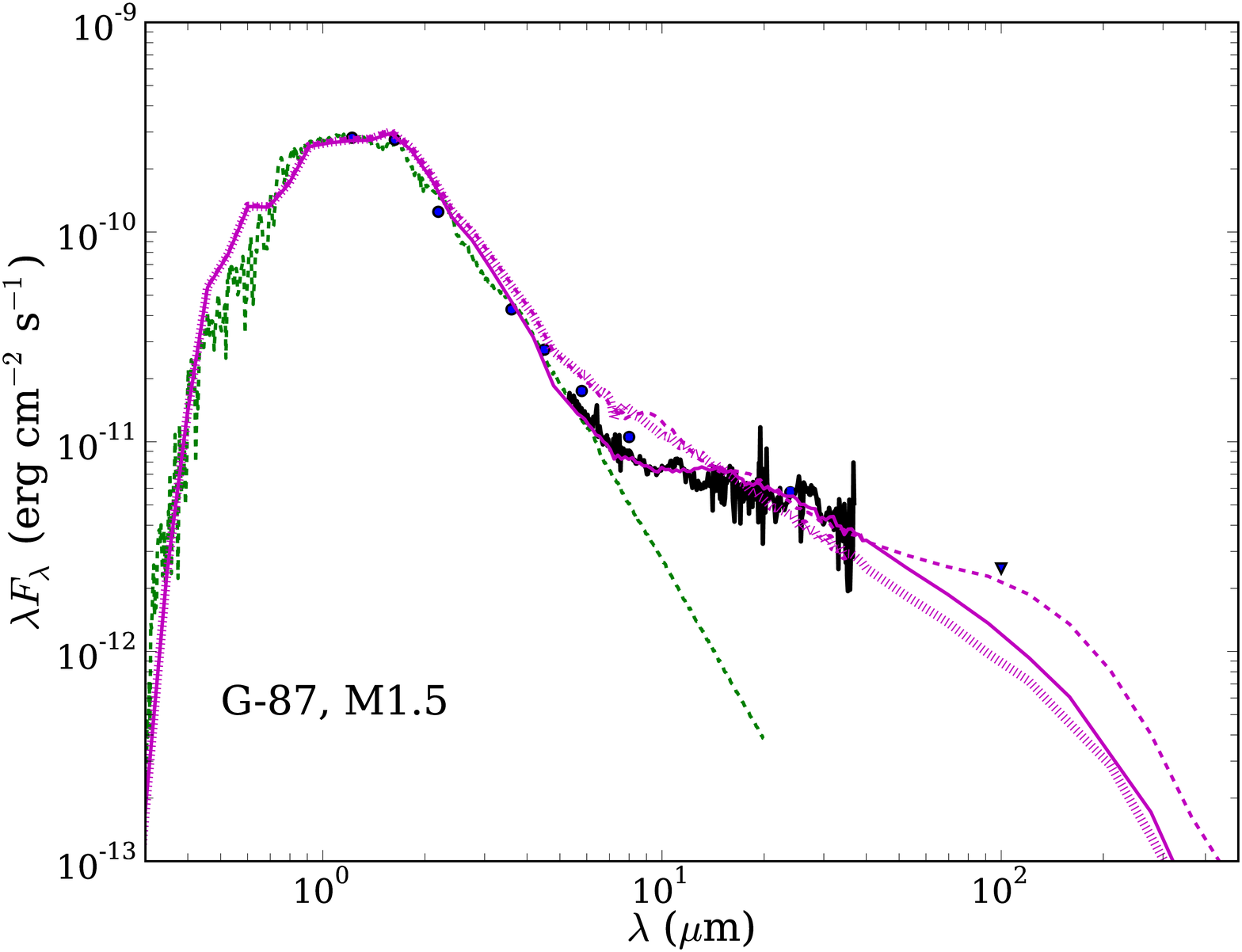,width=0.32\linewidth,clip=} \\
\epsfig{file=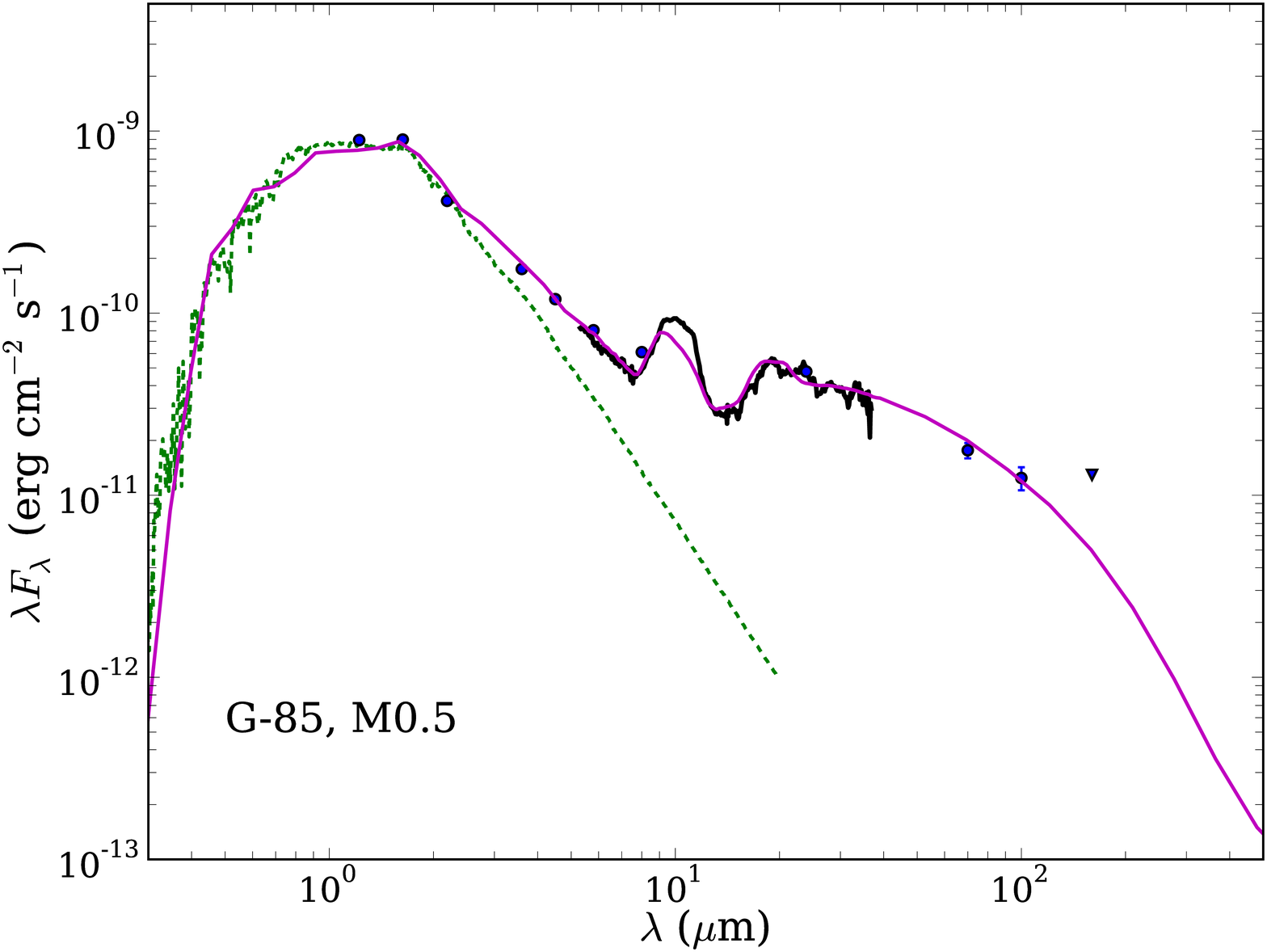,width=0.32\linewidth,clip=} &
\epsfig{file=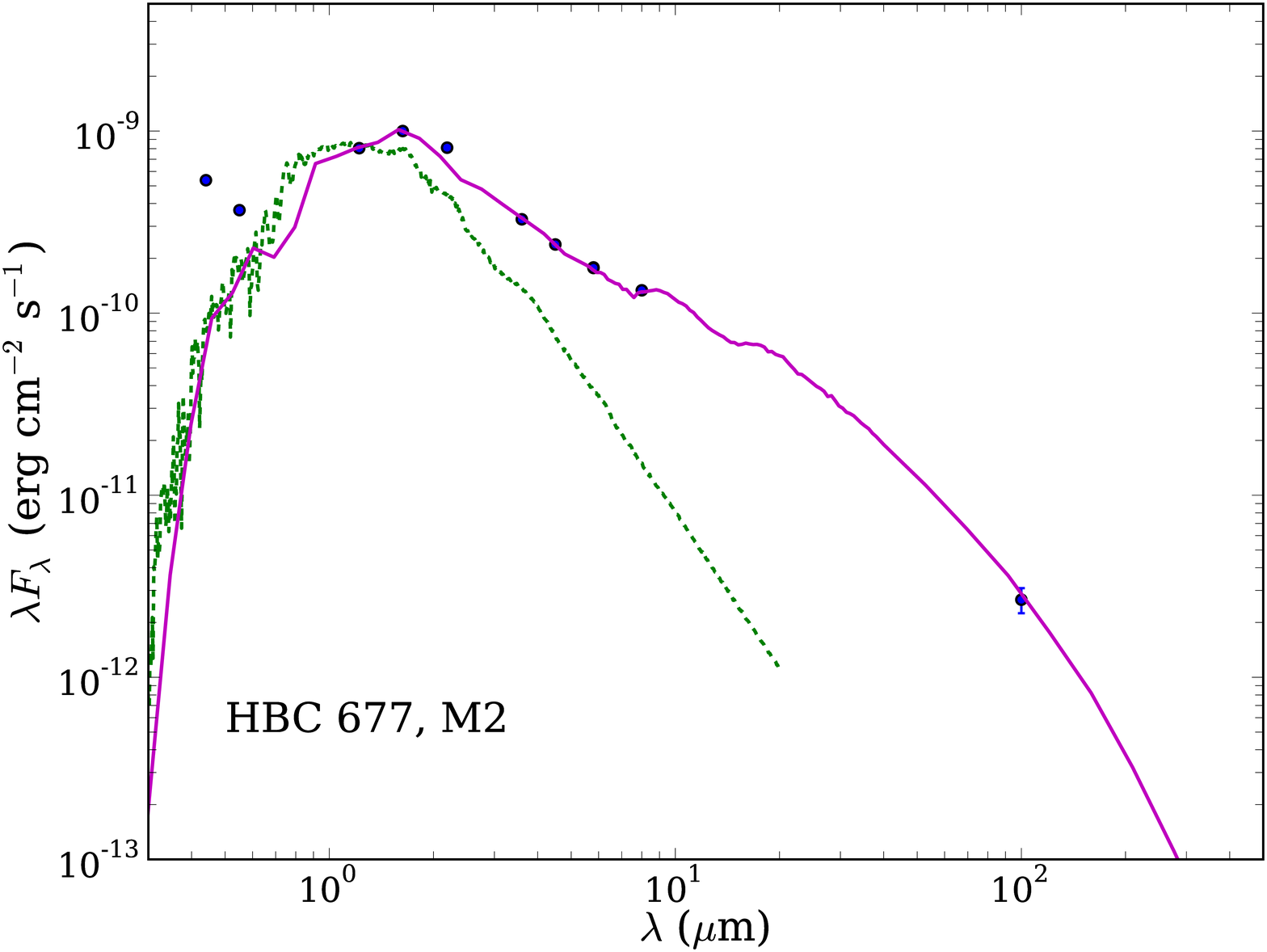,width=0.32\linewidth,clip=} &
\epsfig{file=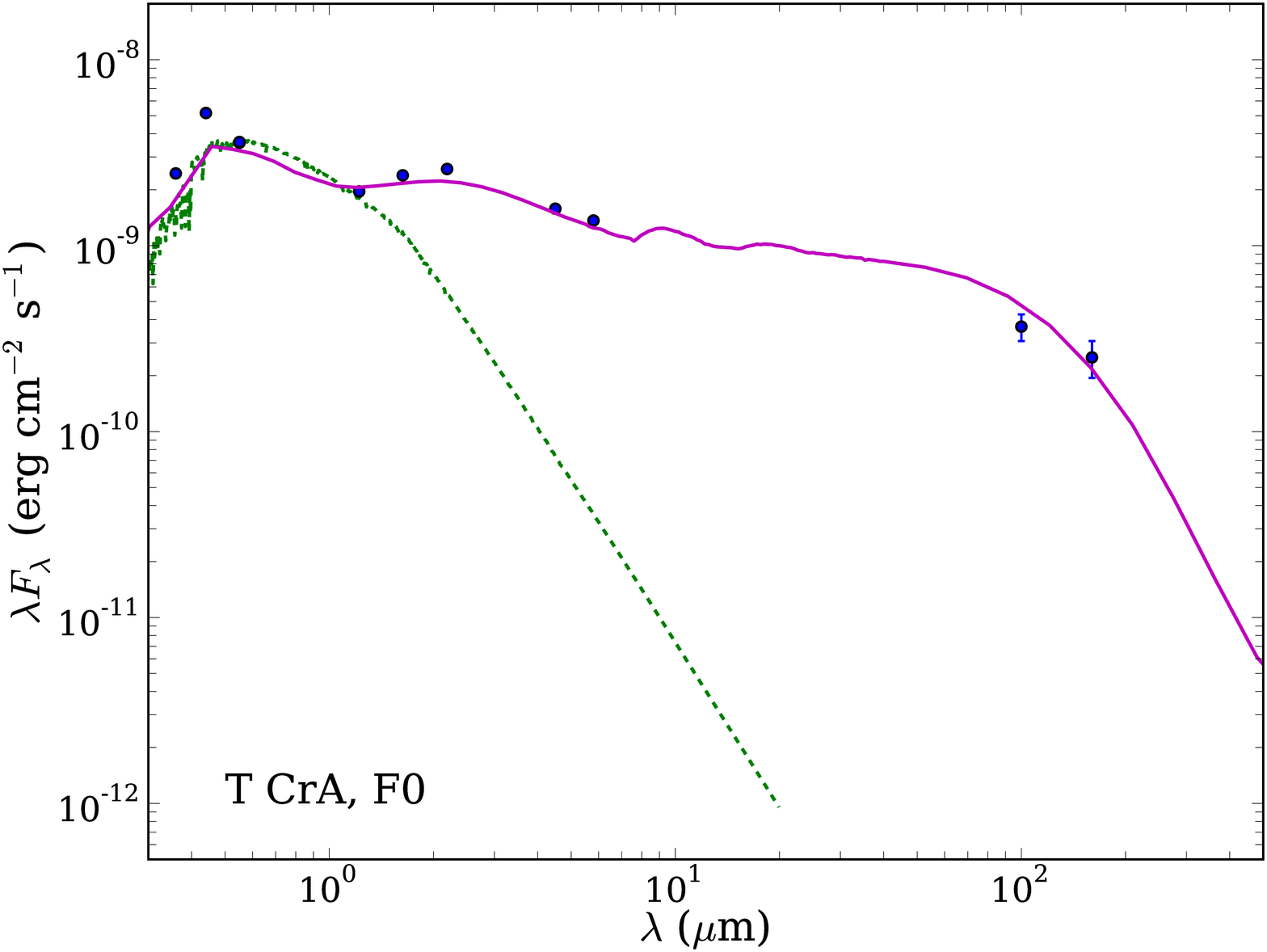,width=0.32\linewidth,clip=} \\
\epsfig{file=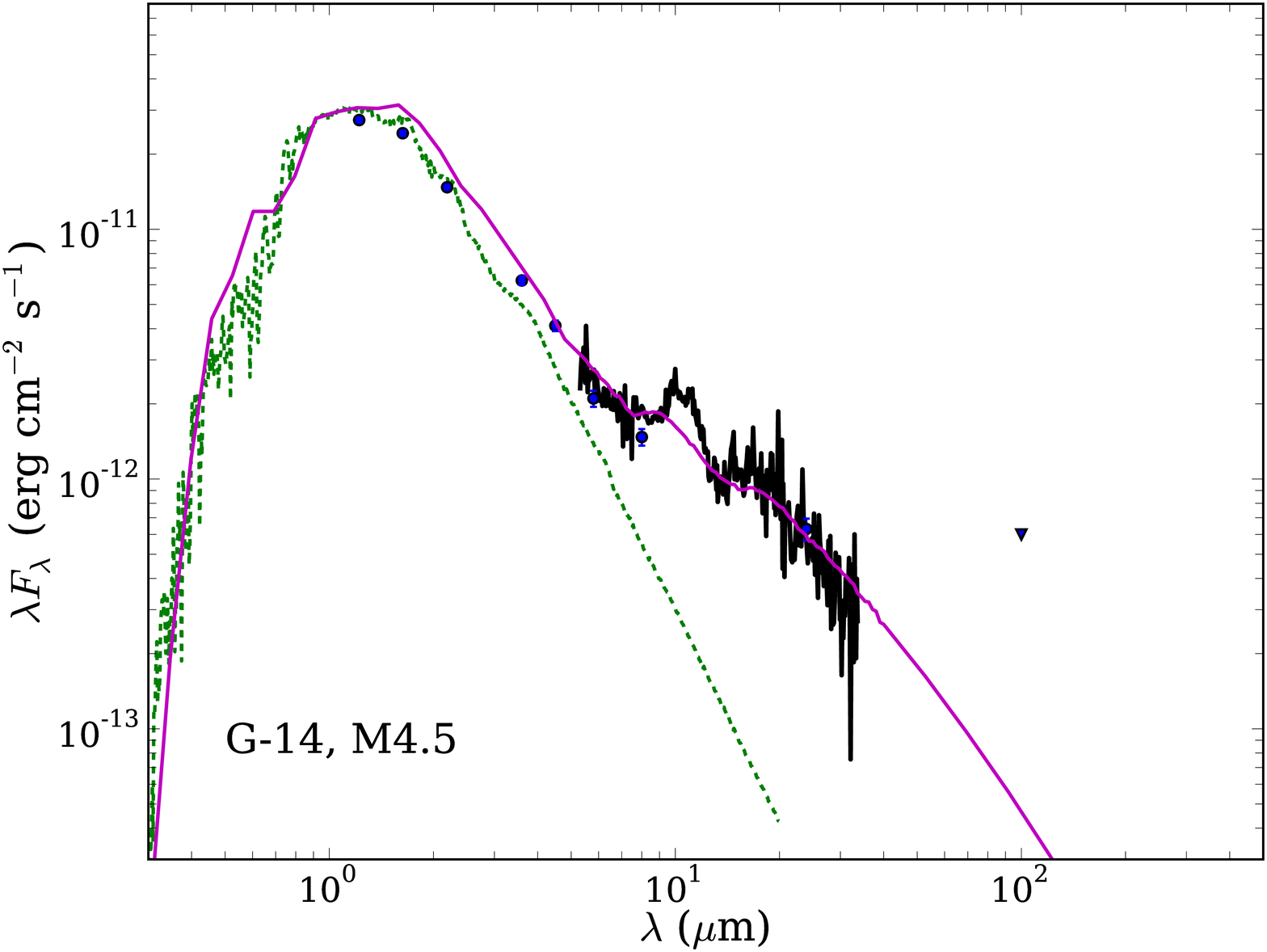,width=0.32\linewidth,clip=} &
\epsfig{file=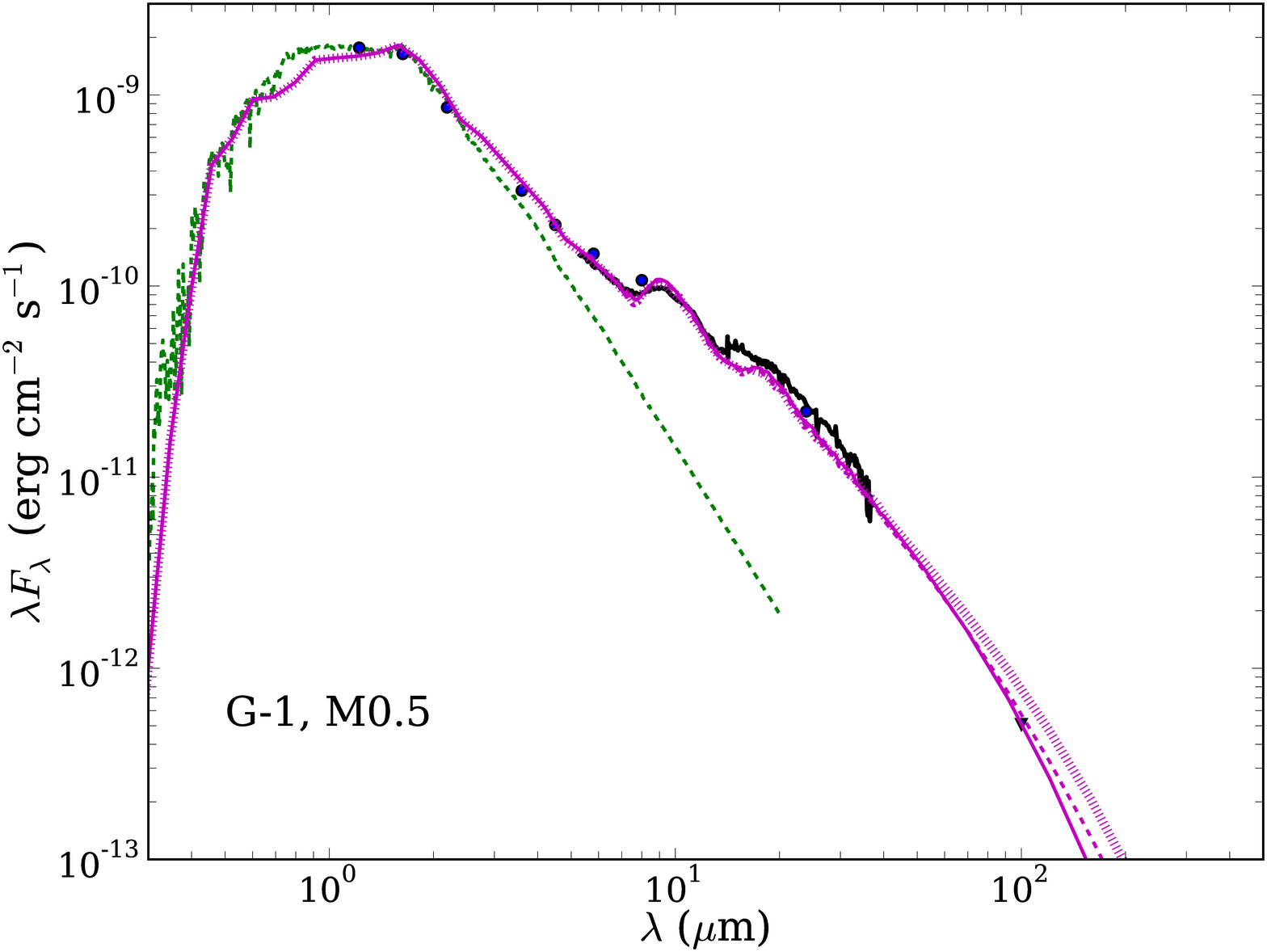,width=0.32\linewidth,clip=} &
\epsfig{file=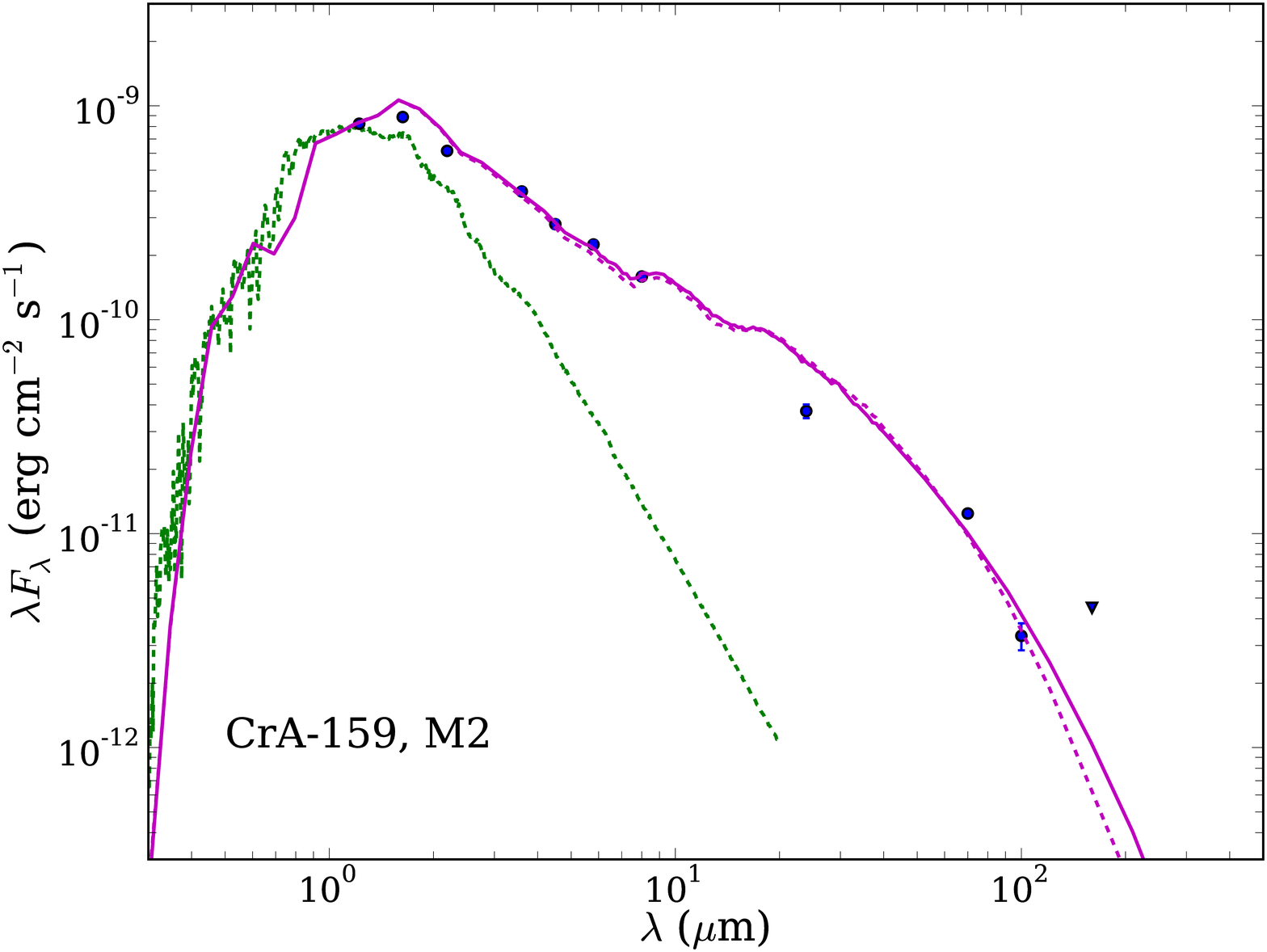,width=0.32\linewidth,clip=} \\
\epsfig{file=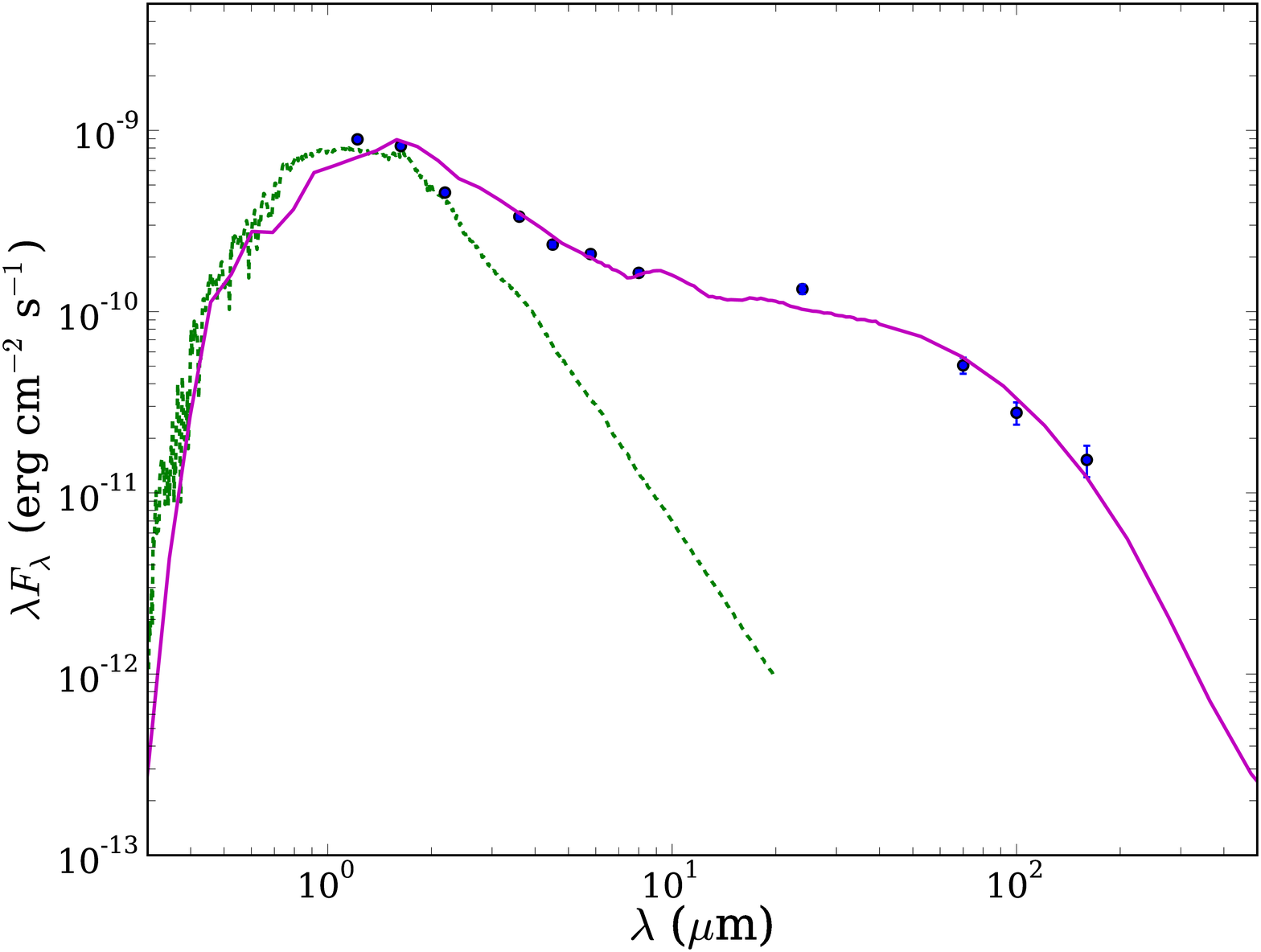,width=0.32\linewidth,clip=} \\
\end{tabular}
\caption{SEDs and disk models.  Photometry detections are marked as circles, with upper limits
marked as inverted triangles. The spectra correspond to Spitzer/IRS observations,
when available. For comparison, a photospheric MARCS model (Gustafsson et al. 2008)
with a similar spectral type is shown for each object (green dotted line). The disk models are shown as magenta
bold lines (or also dotted and dashed lines, in the cases with more than one model). See text and Table \ref{diskmodel-table} 
for the detailed discussion. \label{models-fig}}
\end{figure*}

Although the Coronet cluster is a very young region (1-2 Myr; Sicilia-Aguilar et al. 2011a)
and its compactness would suggest a small age spread between the cluster members around the
cluster core, we observe all types of protoplanetary disks among the solar-type and
low-mass stars (see Figures \ref{interm-fig} and \ref{tts-fig}).
In order to explore the disk structure of the objects, we have modeled their SEDs using
the RADMC 2D code (Dullemond \& Dominik 2004). 
The RADMC code is used to reproduce the SEDs of the different objects, using the observed stellar
parameters to determine the luminosity (effective temperature, stellar radius, T$_{eff}$, R$_*$)
and varying the disk mass, radius, and vertical scale height.
The disks are assumed to be flared with a single flaring law 
at all radii, H$_p$/R$\propto$R$^{1/7}$,
and the vertical scale height at the disk outer radius, H$_{rdisk}$/R$_{disk}$, 
is varied to fit the data. We assume an inner disk
rim at the dust destruction radius (for silicate dust, located at the distance at which
the temperature reaches 1500 K), and a typical grain population with sizes between 0.1 and
100~$\mu$m following a collisional power law distribution with exponent -3.5. We take the outer disk radius
to be 100 AU for the low-mass stars, and 300-400 AU for the intermediate-mass stars\footnote{S~CrA and T~CrA.
In both cases we started with a 100 AU outer disk radius, but the submillimetre and millimetre observations
of S~CrA suggest a slightly larger radius and presence of larger dust grains.}.
The dust component of the disk is assumed to be composed of amorphous grains with similar amounts
of Mg and Fe (J\"{a}ger et al. 1994; 
Dorschner et al. 1995\footnote{See http://www.astro.uni-jena.de/Laboratory/OCDB/newsilicates.html}). 
This simple dust model reproduces the strength of the silicate features very well, although 
we note that the main purpose of this exercise is to understand the global SED shape, and
not the dust composition in the disk atmosphere.
In addition, 25\% of carbon has been included, with a similar size distribution as the silicate
grains. In order to obtain the full disk mass, we consider a gas-to-dust ratio of 100. 
We assume that there is no dust temperature dependence on the grain size, and 
the dust grains are considered to be well mixed (i.e., without size-dependent
differential settling).
The stellar parameters (namely R$_*$ and T$_{eff}$) were estimated by using the temperature-spectral type
relation for Taurus stars (Kenyon \& Hartmann 1995) and varying the radius to reproduce the total
observed luminosity in the optical/near-IR.

These simple models do not account for the many effects expected in protoplanetary disks
(e.g. differential settling and grain growth, inside-out evolution), but our aim is to
understand the global SED shape and properties of the disks. 
Only in cases where no reasonable fit to the observed SED could be achieved with the 
simplified models, we included additional parameters, specifically by considering:
inclusion of large grains/removal of small grains in the dust component, modification of
the inner disk rim to include an inner hole at distances larger than the dust destruction radius,
and variation of the flaring and dust properties between the inner and outer disk.
This is the same procedure that we followed for the disks in Cep OB2 (Sicilia-Aguilar et al.
2011b).
The properties of the best-fitting models are listed in Table \ref{diskmodel-table}.
The models are displayed in Figure \ref{models-fig}. We did not attempt to model
objects that appear problematic due to the presence of companions (VV CrA; Kruger et al. 2011)
or to nebular emission (HD~176386), since 
these simple disks models would not be adequate.

\begin{landscape}
\begin{table}
\caption{Disk models for the Coronet cluster members} 
\label{diskmodel-table}
\centering
\begin{tabular}{l c c c c c c c c  l l}
\hline\hline
Object & T$_{eff}$ (K) & R$_*$ (R$_\odot$) & M$_*$ (M$_\odot$) &  M$_{disk}$  (M$_\odot$) & a$_{min}$-a$_{max}$ (~$\mu$m) & H$_{rdisk}$/R$_{disk}$ & T(R$_{in}$) (K) & R$_{out}$ (AU) &  Comments \\
\hline
S~CrA 	& 4500 & 3.60 &  1.5  &  3.8$\times$10$^{-2}$ & 0.1-10000 &  0.30 & 1500 & 400 & Massive, typical CTTS disk \\
T~CrA	& 7000 & 1.14 &  4.0  &  3.8$\times$10$^{-3}$ & 0.1-100 & 0.33 & 1500 & 300 & Typical CTTS disk\\
CrA-466 & 3500 & 1.06 &  0.4  &  2.8$\times$10$^{-4}$ & 0.1-100 & 0.1 & 1500 & 100 & Typical CTTS disk \\
CrA-45 	& 3700 & 1.62 &  0.6  &  1.5$\times$10$^{-4}$ & 0.1-100 & 0.32 & 1500 & 100 & Typical CTTS disk \\
CrA-159 & 3500 & 1.95 &  0.4  &  3.8$\times$10$^{-6}$ & 0.1-100 & 0.33 & 1500 & 40 & Thick, truncated or strongly depleted. Gap? (continuous line)\\
" 	& 3500 & 1.95 &  0.4  &  5.4$\times$10$^{-6}$ & 0.1-20 & 0.33 & 1500 & 100 & Small grains only. Depleted. Gap? (dashed line) \\
HBC~677 & 3500 & 1.95 &  0.4  &  5.8$\times$10$^{-6}$ & 0.1-100 & 0.32 & 1500 & 100 & Low mass disk, relatively thick\\
G-1  	& 3900 & 2.50 &  0.6  &  $<$1.2$\times$10$^{-6}$ & 0.1-100 &  0.095 & 1500 & 100 & Very dust depleted and settled (dotted line) \\
"  	& 3900 & 2.50 &  0.6  &  $<$3$\times$10$^{-7}$ & 0.1-100 &  0.085 & 1500 & 30 & Settled, truncated (dashed line) \\
"  	& 3900 & 2.50 &  0.6  &  $<$6$\times$10$^{-7}$ & 0.1-20 &  0.095 & 1500 & 100 & Small grains only, depleted (continuous line) \\
G-14	& 3500 & 0.41 &  0.3  &  $<$9$\times$10$^{-7}$ & 0.1-100 &  0.08 & 1500 & 100 & Grain size differences between inner/outer disk\\
G-85$^a$ & 3800 & 1.60 & 0.6  &  1.7$\times$10$^{-4}$ & 0.1-2/0.1-100$^a$ & 0.09/0.095$^a$ & 1500/200$^a$ & 100 & Opt. thin inner disk/gap (pre-transitional) $^a$\\
G-87 	& 3700 & 1.15 &  0.5  &  $<$1.6$\times$10$^{-4}$ & 0.1-100  &  0.03 & 1500 & 100 & Standard disk model (dashed line)\\
" 	& 3700 & 1.15 &  0.5  &   $<$1.6$\times$10$^{-4}$ & 20-1000 &  0.02 & 1500 & 100 & Large grains (dotted line) \\
" 	& 3700 & 1.15 &  0.5  &   $<$1.6$\times$10$^{-4}$ & 20-1000 &  0.02 & 400  & 100 & Best fit: Inner hole plus large grains (continuous line)\\
\hline
\end{tabular}
\tablefoot{RADMC disk models for the Coronet cluster members. The stellar parameters (T$_{eff}$, R$_*$)
are determined from the optical observations and spectroscopy. The disk parameters are
modified to reproduce the observed SED in the whole wavelength range, starting with the simplest 
possible model (see text). The disk masses are estimated assuming a gas-to-dust ratio of 100
and a collisional distribution for the dust with grain sizes a$_{min}$-a$_{max}$.
In case of upper limits at PACS wavelengths, the corresponding
disk masses are upper limits only. $^a$ Marks objects with a distinct inner and outer disks, 
so the given values correspond to the inner and outer disk, respectively.
In case of different models, the comments include a reference to the appropriate figure.}
\end{table}
\end{landscape}

Several of the disks could be well fitted with models that reproduce the typical behavior of
CTTS disks: relatively massive, flared disks with a vertical scale height 
similar to that expected from hydrostatic equilibrium, which suggests
little dust settling. These are the cases of the most massive members (T~CrA and S~CrA) as
well as CrA-45 and CrA-466 among the lower-mass stars, although the high excess at 2-8~$\mu$m
observed in S~CrA and T~CrA suggest the presence of a slightly puffed-up inner
rim and the mid-IR emission of CrA-466 suggest some settling (or a departure from hydrostatic
equilibrium) or inside-out evolution. The fit of CrA-45 is uncertain due to the lack
of information on its spectral type and extinction, and the fact that the mismatches between
the 2MASS, IRAC, and MIPS fluxes suggest some degree of IR variability. Even though all 
these four disks are relatively massive, only the disk around S~CrA exceeded the minimum mass for
the solar nebula (Weidenschilling 1977),
assuming always a standard gas-to-dust ratio of 100. The model fits also proved that, although far-IR
and submillimetre data are needed to determine correctly the disk mass, the 30~$\mu$m IRS data
already puts very strong constraints to the amount of dust. For the object with the most complete
submillimetre data (S~CrA), the SED shape suggests the presence of millimetre-sized
grains. A good fit is attained using a dust distribution
with sizes 0.1-10000 ~$\mu$m and a collisional power law distribution with exponent -3.5.

The Herschel observations reveal the presence of some severely mass depleted 
disks, with disk masses below 10$^{-5}$M$_*$ derived from the Herschel data and the
best fit models, and considering a standard gas-to-dust ratio (100): CrA-159, HBC~677, and G-1. 

The most remarkable case is CrA-159. While being a relatively flared and optically thick
disk with a large vertical scale height and strong near-IR excess, 
the very low PACS flux can only be reproduced assuming 
a very low dust mass. A good fit is achieved assuming that the disk is
truncated to about 40-50 AU. A modification
of the grain distribution (for instance, reducing the amount of $>$20~$\mu$m grains; see
dashed line for CrA-159 in Figure \ref{models-fig}) can explain the low PACS fluxes but 
still results in too high 24~$\mu$m fluxes. The fact that the  24~$\mu$m flux
is lower than expected even for a truncated/low mass model could be a sign of the presence of
a gap, albeit at a larger radius than in typical pre-transitional disks (Espaillat et al. 2010).
A large gap could help to reduce the 
mid-IR flux without changing the grain size distribution and also explain the low mass 
of the disk, but further observations are needed to test this hypothesis.

HBC~677 is very similar to CrA-159, although the lack of MIPS data results in a large
unexplored region in its SED and thus a large unconstrained area in the disk
parameter space. The larger vertical scale height and mass
of this disk (compared to CrA-159) make it consistent with a CTTS disk, albeit
with very low dust mass, compared to typical Taurus disks around 
similar stars (Andrews \& Williams 2005).

The disk around G-1 was not detected by Herschel/PACS, but the object is located
in a region free from extended emission, which allows us to place a very stringent 
upper limit to its 100$\mu$m flux. This sets a strong constraint to the
disk mass, which would be as low as $\sim$1$\times$10$^{-6}$ M$_\odot$.
A simple model tends to overpredict the Herschel fluxes (dotted line
for G-1 in Figure \ref{models-fig}).
Disk truncation could be also invoked, as in CrA-159, to reduce the flux (dashed line), 
which could be a possibility since K\"{o}hler et al. (2008) suggested that this object could be a binary
(0.22" projected separation).
In addition, a small-dust-only distribution in a very low-mass
disk provides a good fit (continuous line for G-1 in Figure \ref{models-fig}).
The low near-IR excess points to a small vertical scale height, although given the
very low-mass of the disk, the object does not need to be dramatically settled 
and could be close to hydrostatic equilibrium.
The Herschel data thus confirms G-1 to be a good example of a globally dust depleted
disk, as it had been previously suggested by the IR observations alone
(Currie \& Sicilia-Aguilar 2011). The IR observations of G-14 are also consistent with a
globally dust depleted disk, although due to the lower luminosity of the object, the
Herschel upper limits do not allow to constrain the full dust content and vertical scale height.

The disk around G-85 is consistent with a pre-transitional disk with an inner gap
(similar to those in Espaillat et al. 2010). 
The strong silicate feature and low near-IR excess requires the presence of small
silicate grains in an optically thin environment, and the large flux at 100~$\mu$m reveals
a relatively large total disk mass. While previous disk models had classified this object
as having a typical CTTS disk (based on broad-band photometry), 
it is not possible to reproduce both the strong silicate
emission and the large mid- and far-IR flux assuming a uniform disk with a single 
dust composition and a single density power law. The best results for a simple model are achieved if the disk is
separated in an inner and an outer part at a distance of about 2 AU, with
the inner part being flared and optically thin and the outer part being relatively
massive and optically thick. The large silicate feature requires the inner disk to be
populated by mostly small grains (here we assume a flat distribution of grains with sizes 0.1-2~$\mu$m),
while the outer disk is consistent with a standard dust grain distribution.
Such filtered grain distributions have been predicted theoretically in cases of
disks with inner gaps and planets (Rice et al. 2006).
The presence of an optically thick inner disk, followed
by a relatively clean gap with small grains and an optically thick outer disk cannot
be excluded either with the present data. 

Although undetected, the Herschel upper limit flux for G-87 puts strong constraints to its disk
structure.
Fitting of the G-87 disk, that has been successively regarded as a transition (Sicilia-Aguilar et al.
2008), primordial (Ercolano et al. 2009), and transition/homologously depleted (Currie \& Sicilia-Aguilar 2011)
disk, reveals that it is consistent with a low mass disk (M$_d<$3.2$\times$10$^{-4}$M$_*$) with an inner hole.
Assuming a standard dust distribution (0.1-100 ~$\mu$m, exponent -3.5) and varying the vertical scale
height results in too large fluxes at $\lambda >$5~$\mu$m for any values that reproduce the
emission at 20-30~$\mu$m (see G-87 in Figure \ref{models-fig}, dashed line),
which would have resulted at least in a marginal detection at 100$\mu$m. The object
has a small silicate feature and a large crystallinity fraction (Sicilia-Aguilar et al. 2008),
which is typical of disks with strong grain growth where most of the amorphous grains are in aggregates
a few microns in size. If we then change the dust distribution to reduce the amount of small
dust (10-1000 or 20-1000 ~$\mu$m, exponent -3.5), the near-IR flux is reduced for thin enough disks,
but the emission in the 5-15~$\mu$m region is still too large (see G-87 in Figure \ref{models-fig}, dotted 
line). The best fit is achieved
including an inner hole of about 0.3-1 AU in size, together with a grain population mostly consistent
with large grains  (10-1000 or 20-1000 ~$\mu$m, exponent -3.5; see G-87 in Figure \ref{models-fig}, continuous 
line). A small fraction of submicron and
partly crystalline grains present in the inner hole could account for the observed silicate feature.
The models thus point towards the transitional nature of this object.

\subsection{The disk-protostar connection? \label{sparseregions}}

\begin{figure}
   \centering
   \resizebox{\hsize}{!}{\includegraphics[width=\textwidth]{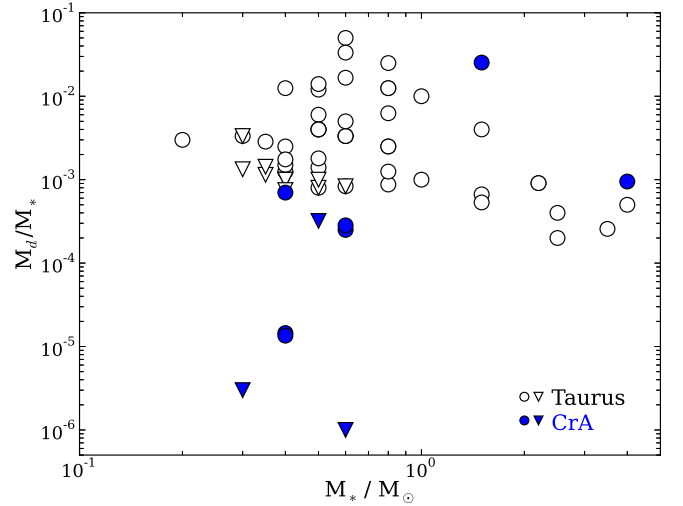}}
\caption{Disk mass versus stellar mass in Taurus (Andrews \& Williams 2005, open symbols)
and in the CrA region (filled symbols). In both cases, inverted triangles mark upper limits. 
In the case of objects with different disk models, we only display the mass derived from
the best-fit model, although the differences in the masses between the models
for each object are minimal, due to the strong constraint provided by the PACS fluxes.
\label{tauruscomp-fig}}
\end{figure}

To summarize the previous sections, despite the youth of the cluster, the Coronet region shows already a rich
and diverse disk population, where the signs of disk evolution (small dust depletion, inside-out
evolution) are common. 
The masses of the disks are in general lower than what is typically observed in clusters
with similar ages. Figure \ref{tauruscomp-fig}
shows a comparison of the disk masses of the objects detected (or strongly constrained) wiht
Herschel versus the stellar mass observed in Taurus (from Andrews \& Williams 2005)
with our results for the CrA region. Despite the small number of objects in the CrA region
(10), we find
that the disks around the M-type stars are significantly less massive than their Taurus
counterparts. In particular, although there are a few Taurus disks with masses as low as those 
measured here, most of the Taurus disks are clearly more massive than the typical Coronet
disks. Given the small size of the region in comparison with the typical sizes of
star-forming Herschel filaments (Arzoumanian et al. 2011), it does not seem likely that
the Coronet/CrA region is significantly older than Taurus. Although not all the disks 
are detected with Herschel, the tendency would be
towards detection of the most massive objects (at least, within a given spectral type
range), which makes this difference even more striking.    
On the other hand, the more massive stars S~CrA and T~CrA have larger masses than
the typical 1-2 M$_\odot$ star in Taurus. 
It is also remarkable that the objects with the largest disk masses or stronger PACS
excesses among the solar-type and low-mass stars 
(S~CrA, VV CrA, CrA-45) are not associated with the main cluster core, but rather on the 
outskirts of the dense cloud. 

There are some differences in the way the Taurus disk masses are determined in
Andrews \& Williams (2005) and the masses derived for the CrA members by fitting
their disks with RADMC. The main differences result from the dust opacity and the dust
distributions considered. While the dust opacity in the millimetre is in both cases
basically the same, changes in the $\beta$ millimetre slope used by Andrews \& Williams (2005) for deriving
their dust masses can introduce a factor of few difference between the derived masses.
We have thus explored the resulting disk mass for the stars T~CrA and CrA-466 by using different
dust distributions with maximum grain sizes between 100 and 10000~$\mu$m, finding variations
up to a factor of $\sim$4 in the derived mass. This would result in a difference of 0.6 dex in 
Figure \ref{tauruscomp-fig}, which would still place the CrA disks around low-mass stars
among the less massive ones, compared to similar Taurus stars. A different mass analysis
(Currie \& Sicilia-Aguilar 2011), including only optical, 2MASS, and Spitzer observations,
also confirms this trend towards lower masses compared to Taurus among the CrA members.
The presence of substantial dust mass locked in larger grains cannot
be excluded without further observations at longer wavelengths, although in this case
the reason why grain growth would have proceeded more efficiently in the Coronet disks than in
Taurus objects is not clear.

One potential explanation of this anomalous behavior of the disk properties could
lay in the line of the arguments by Fang et al.(2012), who found a clear difference in the
disk fraction vs. time in sparse cluster compared to more populous regions. While a exponential
trend with a disk lifetime around 2-3 Myr is a very good approach for the disk fractions
observed in more massive clusters (several hundreds of members) and OB associations
(see also Fedele et al. 2010), a similar diagram for the sparse clusters (including the
Coronet cluster) is harder to interpret. Very young regions have relatively moderate disk
fractions that to a large extent survive over longer periods of time. Some authors have
suggested that such behavior could respond to the lifetimes of disks around binary stars
being much shorter than the lifetimes of disks around single stars (Bouwman et al. 2006). 
In the case of the Coronet cluster, we do not have information about the multiplicity of
most the very low-mass stars, although most of the protostars are in multiple systems
and a high degree of multiplicity has been suggested for many of the known TTS members
(K\"{o}hler et al. 2008).

On the other hand, the Herschel data reveals that, although the region is usually assumed to
be relatively quiet due to the lack of very massive stars, the cluster center corresponds to
a very crowded area, where the region around IRS~7 hosts about 20 objects (counting in multiple systems,
protostars, and the intermediate-mass stars R~CrA and T~CrA) within a region a bit less than 0.1pc in diameter. 
Moreover, the Herschel data suggest strong interaction at least in some of the systems
(IRS~5 complex), which are very likely to affect the formation of their disks. 
Further observations of the region should be used to trace the early interactions of the
systems and to quantify to which extent this may have contributed to shape the disks
of the low-mass cluster members.

\section{Summary and conclusions \label{conclu}}

We have presented Herschel/PACS observations at 100 and 160~$\mu$m of the CrA region.
The observations have covered the main parts of the Coronet cluster and detected many
nearby YSOs,
most of which are related to the central,
denser part of the cluster with the strongest millimetre/submillimetre emission. 
Substantial cloud emission around the
whole region is also revealed by the observations, including some details of
spatially resolved sources. Most of the Herschel compact sources can be traced
back to known submillimetre/millimetre condensations (consistent with protostellar
objects or dense cloud parts) and to mid-IR sources identified as young
T Tauri stars with disks.

 \begin{enumerate}

\item Although Herschel reveals several extended structures following the
path of the extinction maps, most of the star formation in the
region, detectable as compact sources, is taking place in the
central part of the cluster, which shows the higher densities and
extinction (Kainulainen et al. 2009). Given that we are sensitive to
disks around very low-mass stars and very low-mass protostars,
we can safely exclude active star formation in most of the cloud,
which suggest a poor efficiency or cloud densities below the
threshold for star formation. 

\item We are able to trace fine details of the structure of two Class I protostars,
IRS~2 and the IRS~5 complex. The first one appears associated with a bubble-like structure,
suggesting the expansion of a wide-angle wind in the molecular cloud material. 
The components of the IRS~5 multiple protostar, IRS~5a/b and
the X-ray source FP-25, appear surrounded by a filamentary extended structure
that resembles the spiral arms usually associated to multiple systems by theoretical
simulations of binary formation (Bate 2000). The size of the observed filaments is 
about 7 times larger than the typical sizes predicted by the models, although
the models are highly dependent on the initial angular momentum properties and
on the mass of the protostars. Another possible explanation could be a similar
swept-up, heated structure similar to what we find near IRS~2. 
Although IRS~5a/b appears brighter in the near-
and mid-IR and also at 100~$\mu$m, FP-25 is brighter at 160~$\mu$m, suggesting
that it is in a more embedded, colder phase. 

\item The components of the binary protostar IRS~7w/e are also resolved, although both appear
partially blended. IRS~7e, the faintest at wavelengths shorter than 100~$\mu$m, dominates
at 160~$\mu$m. The region between IRS~7w and
IRS~7e is well resolved at 100~$\mu$m, showing a decrease in emission. 
The Class 0 candidate source SMM~1~A is located to the South of both. Interestingly,
the dip in far-IR emission between the three sources is coincident with the X-ray source FP-34
(Forbrich \& Preibisch 2007), which could be thus associated with a third embedded
object or maybe to a jet/shock from any of the two protostars (there is further
evidence of X-ray emission from optical jets in the region, see object
G-80 in Sicilia-Aguilar et al. 2008).

\item The region presents objects in very early evolutionary stages.
We confirm the sources SMM~1~A and SMM~1~As as Class 0 candidates with
similar grey-body temperature (16 K). On the other hand,
the strong submillimetre emission to the North of the IRS~7
complex, named SMM~6 in Nutter et al. (2005) and also detected in our
LABOCA maps (Sicilia-Aguilar et al. 2011a) appears as a dark
lane in the 100$\mu$m images and brigther at 160~$\mu$m, 
being consistent with a starless core
or maybe a pre-stellar condensation.

\item The Herschel/PACS data also reveals several
very low-mass protostellar objects. The submillimetre source SMM~2
(Nutter et al. 2005) is probably a M-type Class I object. 
The mid-IR and X-ray source G-122 (Sicilia-Aguilar et al. 2008)
is consistent with an even fainter very low-mass protostar. We also find a second
condensation to the East of G-122, which is near the Herschel/PACS detection limit,
but resembles to G-122 in color and size.

\item The structure between the stars HD~176386 and TY~CrA also reveal interesting
characteristics. While the two objects display a double cavity in
near- and mid-IR, the Herschel/PACS emission reveals several peaks in the area
in between the stars, which suggest that it could be related to heated, dense,
compressed material (maybe a first step in triggered star formation). 
It is noticeable that the Herschel peak in this area does
not coincide with the peak of 870~$\mu$m LABOCA emission, which is mostly
centered (but extended) around HD~176386 and should mark the location
of the densest cloud material. The whole region shows also strong radial
striations, which could be a sign of large-scale collapse motions.

\item We have also detected dust emission from the disks around several of the
cluster members. In order to interpret the SEDs, we have constructed
radiative transfer models for the disks using the RADMC code
(Dullemond \& Dominik 2004). While the disk emission around the solar-type and intermediate-mass
stars is consistent with flared, massive disks (S~CrA, T~CrA), the disks
around the lower-mass stars (spectral types M0 or later) have remarkably
low masses (M$_d$/M$_*$$<$10$^{-3}$M$_\odot$/yr). 
In some cases, the shape of the SED suggest either disk truncation
(as in the case of CrA-159) or strong dust depletion (as in the case of HBC~677
and G-1). Some degree of inside-out evolution, like the presence of inner
holes or gaps, is also required to explain the full SEDs of some of the disks. In
particular, G-85, which has low near-IR flux, strong silicate feature, and
relatively high far-IR flux , is best fitted assuming the presence of a gap at few AU,
being thus consistent with a pre-transitional disk.
G-87  is best fitted by assuming a disk with strong grain growth
and an inner hole, as we had previously inferred from its IRAC/IRS data
(Sicilia-Aguilar et al. 2008). 

\item While the masses of the disks around solar- and intermediate-mass stars appear
similar or higher than their Taurus counterparts, all M-type stars have
disk masses that are systematically lower than the disk masses of other stars 
with similar ages (e.g. Taurus; Andrews \& Williams 2005). This confirms 
our previous results inferred from the Spitzer
observations (Sicilia-Aguilar et al. 2008; Currie \& Sicilia-Aguilar 2011)
that were initially interpreted as signs of evolution.
The reduced mass of the disks could be either related to very
marked early evolution, or be a result of the initial conditions in the
star-forming cloud affecting the structure of the disks. Given the youth of the
cluster, and that the present star-formation seems to be rich in multiple
systems and shows signs of interactions between the stars and the parental cloud,
the effect of initial conditions may have played an important role
in shaping the disks. Observations in other clusters, revealing differences
in the disk fraction vs. time in sparse regions compared to more massive
clusters may also point in this direction (Fang et al. 2012).

\end{enumerate}

Acknowledgments: This work is based on Herschel/PACS data.
PACS has been developed by a consortium of institutes led by MPE (Germany) and including UVIE (Austria); 
KU Leuven, CSL, IMEC (Belgium); CEA, LAM (France); MPIA (Germany); INAF-IFSI/OAA/OAP/OAT, LENS, SISSA (Italy); 
IAC (Spain). This development has been supported by the funding agencies BMVIT (Austria), ESA-PRODEX (Belgium), 
CEA/CNES (France), DLR (Germany), ASI/INAF (Italy), and CICYT/MCYT (Spain). This work has made use of the SIMBAD database.
It also makes use of data products from the Two Micron All Sky Survey, 
which is a joint project of the University of Massachusetts and the Infrared Processing and 
Analysis Center/California Institute of Technology, funded by the National Aeronautics and 
Space Administration and the National Science Foundation.
We thank C. P. Dullemond for his help providing the RADMC 2D code, and the anonymous referee
for providing useful comments that helped to clarify this paper.
A.S.A acknowledges support by the Spanish ``Ram\'{o}n y Cajal" program from the Spanish MICINN,
grant number RYC2010-06164.

Facilities: Spitzer, Herschel.

\clearpage

\onecolumn


\begin{thebibliography}{}

%\bibitem[]{}

\bibitem[Acke \& van den Ancker(2004)]{2004A&A...426..151A} Acke, B., \& van den Ancker, M.~E.\ 2004, \aap, 426, 151 

\bibitem[Andr{\'e} et al.(2010)]{2010A&A...518L.102A} Andr{\'e}, P., Men'shchikov, A., Bontemps, S., et al.\ 2010, \aap, 518, L102 

\bibitem[Andrews \& Williams(2005)]{andrews05} Andrews, S.,\& Williams, J., 2005, ApJ, 631, 1134

\bibitem[Arzoumanian et al.(2011)]{2011A&A...529L...6A} Arzoumanian, D., Andr{\'e}, P., Didelon, P., et al.\ 2011, \aap, 529, L6 

\bibitem[Bate(2000)]{2000MNRAS.314...33B} Bate, M.~R.\ 2000, \mnras, 314, 33 

\bibitem[Bast et al.(2011)]{2011A&A...527A.119B} Bast, J.~E., Brown, J.~M., Herczeg, G.~J., van Dishoeck, E.~F., \& Pontoppidan, K.~M.\ 2011, \aap, 527, A119 

\bibitem[Beckwith et al.(2000)]{2000prpl.conf..533B} Beckwith, S.~V.~W., Henning, T., \& Nakagawa, Y.\ 2000, Protostars and Planets IV, 533 

\bibitem[Bibo et al.(1992)]{1992A&A...260..293B} Bibo, E.~A., The, P.~S., \& Dawanas, D.~N.\ 1992, \aap, 260, 293 

\bibitem[Chen et al.(1997)]{chen97} Chen, H.; Grenfell, T. G.; Myers, P. C.; Hughes, J. D., 1997, ApJ, 478, 295

\bibitem[Chini et al.(2003)]{chini03}Chini, R. et al. 2003, A\&A, 409, 235

\bibitem[Currie \& Sicilia-Aguilar(2011)]{curriesa11} Currie, T., Sicilia-Aguilar, A., 2011, ApJ 732, 24

\bibitem[De Muizon et al.(1980)]{demuizon80} De Muizon, M., Rouan, D., Lena, P., Nicollier, C., Wijnbergen, J., 1980, AA, 83, 140

\bibitem[Dorschner et al.(1995)]{dorschner95} Dorschner, J, Begemann, B., Henning, T., J\"{a}ger, C., \& Mutschke, H., 1995, A\&A, 300, 503 

\bibitem[Dullemond \& Dominik(2004)]{dullemond04radmc} Dullemond, C., \& Dominik, C., 2004, A\&A, 417, 159

\bibitem[Eiroa et al.(1997)]{eiroa97} Eiroa, C., Palacios, J., Casali, M., 1997, ASPC 119, 107

\bibitem[Espaillat et al.(2010)]{2010ApJ...717..441E} Espaillat, C., et al.\ 2010, \apj, 717, 441 

\bibitem[Fang et al.(2012)]{2012arXiv1209.5832F} Fang, M., van Boekel, R., Bouwman, J., et al.\ 2012, A\&A in press, arXiv:1209.5832 

\bibitem[Favata et al.(2002)]{favata02} Favata, F.; Fridlund, C. V. M.; Micela, G.; Sciortino, S.; Kaas, A. A., 2002, ASPC 277, 467

\bibitem[Forbrich et al.(2006)]{2006A&A...446..155F} Forbrich, J., Preibisch, T., \& Menten, K.~M.\ 2006, \aap, 446, 155 

\bibitem[Forbrich \& Preibisch(2007)]{forbrich07} Forbrich, J.; Preibisch, T., 2007, A\&A 475, 959

\bibitem[Garmire \& Garmire(2003)]{garmire03} Garmire, G., \& Garmire, A., 2003, Astron. Nachr. 324, 153 

\bibitem[Garmire et al.(2003)]{garmireetal03} Garmire, G. P., Bautz, M. W., Ford, P. G., Nousek, J. A., \& Ricker, G. R., 2003 Proc. SPIE 4851,28.

\bibitem[Groppi et al.(2004)]{groppi04} Groppi et al., 2004, ApJ, 612, 946;

\bibitem[Groppi et al.(2007)]{2007ApJ...670..489G} Groppi, C.~E., Hunter, T.~R., Blundell, R., \& Sandell, G.\ 2007, \apj, 670, 489 

\bibitem[Gustafsson et al.(2008)]{gustafsson08marcs} Gustafsson, B., et al. 2008, A\&A 486, 951

\bibitem[Gutermuth et al.(2009)]{2009ApJS..184...18G} Gutermuth, R.~A., Megeath, S.~T., Myers, P.~C., et al.\ 2009, \apjs, 184, 18  

\bibitem[Hamaguchi et al.(2005)]{hamaguchi05a} Hamaguchi, K., et al. 2005, ApJ 618, 360

\bibitem[Harju et al.(1993)]{harju93} Harju, J., Haikala, L.K., Mattila, K., Mauersberger, R., Booth, R.S., Nordth, H.L., 1993, \aap 278, 569

\bibitem[Hennemann et al.(2012)]{2012A&A...543L...3H} Hennemann, M., Motte, F., Schneider, N., et al.\ 2012, \aap, 543, L3 

\bibitem[Henning et al.(1994)]{henning94} Henning, Th.; Launhardt, R.; Steinacker, J.; Thamm, E., 1994, A\&A, 338, 223

\bibitem[J\"{a}ger et al.(1994)]{jaeger94} J\"{a}ger, C., Mutschke, H., Begeman, B., Dorschner, J., Henning, Th., 1994, AA, 292, 641 

\bibitem[Kainulainen et al.(2009)]{kai09} Kainulainen, J., Beuther,  H., Henning, T., \& Plume, R.\ 2009, \aap, 508, L35

\bibitem[Kataza et al.(2010)]{kataza10}Kataza, H., Alfageme, C., Cassatella, A., Cox, N., Fujiwara, H., Ishihara, D., Oyabu, S., Salama, A., Takita, S., and Yamamura, I., 2010, AKARI-IRC Point Source Catalogue Release note Version 1.0

\bibitem[Kley \& Burkert(2000)]{2000ASPC..219..189K} Kley, W., \& Burkert, A.\ 2000, Disks, Planetesimals, and Planets, 219, 189 

\bibitem[K{\"o}hler et al.(2008)]{2008A&A...488..997K} K{\"o}hler, R., Neuh{\"a}user, R., Kr{\"a}mer, S., et al.\ 2008, \aap, 488, 997 

\bibitem[Kruger et al.(2011)]{2011ApJ...729..145K} Kruger, A.~J., Richter, M.~J., Carr, J.~S., et al.\ 2011, \apj, 729, 145 

\bibitem[Lindberg \& J{\o}rgensen(2012)]{2012arXiv1209.4817L} Lindberg, J.~E., \& J{\o}rgensen, J.~K.\ 2012, arXiv:1209.4817 

\bibitem[L\'{o}pez-Mart\'{\i} et al.(2005)]{lopezmarti05} L\'{o}pez-Mart\'{\i}, B.,  Eisl\"{o}ffel, J., Mundt, R., 2005, AA, 444, 175 

\bibitem[L{\'o}pez Mart{\'{\i}} et al.(2010)]{2010A&A...515A..31L} L{\'o}pez Mart{\'{\i}}, B.,  et al. \ 2010, \aap, 515, A31 

\bibitem[Loren(1979)]{loren79} Loren, R. B., 1979, ApJ, 227, 832

\bibitem[Marraco \& Rydgren(1981)]{marraco81}Marraco \& Rydgren 1981, AJ, 86, 62

\bibitem[Meyer \& Wilking(2009)]{2009PASP..121..350M} Meyer, M.~R., \& Wilking, B.~A.\ 2009, \pasp, 121, 350 

\bibitem[Myers \& Ladd(1993)]{1993ApJ...413L..47M} Myers, P.~C., \& Ladd, E.~F.\ 1993, \apjl, 413, L47 

\bibitem[Neuh\"{a}user et al.(2000)]{neuhauser00} Neuh\"{a}user, R., et al., 2000, A\&A, 146, 323

\bibitem[Nisini et al.(2005)]{nisini05} Nisini, B.; Antoniucci, S.; Giannini, T.; Lorenzetti, D., 2005, A\&A, 429, 543

\bibitem[Nutter et al.(2005)]{2005MNRAS.357..975N} Nutter, D.~J., Ward-Thompson, D., \& Andr{\'e}, P.\ 2005, \mnras, 357, 975 

\bibitem[Ott(2010)]{2010ASPC..434..139O} Ott, S.\ 2010, Astronomical Data Analysis Software and Systems XIX, 434, 139 

\bibitem[Peterson et al.(2011)]{peterson11} Peterson, D.E., et al. 2011, ApJSS 194, 43

\bibitem[Pilbratt et al.(2010)]{2010A&A...518L...1P} Pilbratt, G.~L., Riedinger, J.~R., Passvogel, T., et al.\ 2010, \aap, 518, L1 

\bibitem[Poglitsch et al.(2010)]{2010A&A...518L...2P} Poglitsch, A., Waelkens, C., Geis, N., et al.\ 2010, \aap, 518, L2 

\bibitem[Ratzka et al.(2008)]{2008poii.conf..519R} Ratzka, T., Leinert, C., Przygodda, F., \& Wolf, S.\ 2008, The Power of Optical/IR Interferometry: Recent Scientific Results and 2nd Generation, 519 

\bibitem[Rice et al.(2006)]{rice06} Rice, W., Armitage, P., Wood, K., Lodato, G., 2006, MNRAS 373, 1619 

\bibitem[Roussel(2012)]{roussel12scanamorphos} Roussel, H., 2012, arXiv:1205.2576

\bibitem[Sicilia-Aguilar et al.(2008)]{sic08cra}  Sicilia-Aguilar, A.; Henning, Th.; Juh\'{a}sz, A.; Bouwman, J.; Garmire, G.; Garmire, A., 2008, \apj, 687, 1145

\bibitem[Sicilia-Aguilar et al.(2010)]{sic10}  Sicilia-Aguilar, A., Henning, Th., Hartmann, L., 2010, ApJ, 710, 597

\bibitem[Sicilia-Aguilar et al.(2011a)]{2011ApJ...736..137S} Sicilia-Aguilar, A., Henning, T., Kainulainen, J., \& Roccatagliata, V.\ 2011a, \apj, 736, 137 

\bibitem[Sicilia-Aguilar et al.(2011b)]{2011ApJ...742...39S} Sicilia-Aguilar, A., Henning, T., Dullemond, C.~P., et al.\ 2011b, \apj, 742, 39 

\bibitem[Stutz et al.(2009)]{2009ApJ...707..137S} Stutz, A.~M., Rieke, G.~H., Bieging, J.~H., et al.\ 2009, \apj, 707, 137 

\bibitem[Taylor \& Storey(1984)]{taylor04} Taylor \& Storey, 1984, MNRAS, 209, 5

\bibitem[Tobin et al.(2011)]{2011ApJ...740...45T} Tobin, J.~J., Hartmann, L., Chiang, H.-F., et al.\ 2011, \apj, 740, 45 

\bibitem[Torres et al.(2006)]{2006A&A...460..695T} Torres, C.~A.~O., Quast, G.~R., da Silva, L., et al.\ 2006, \aap, 460, 695 

\bibitem[Walter(1986)]{walter86} Walter, F., 1986, ApJ, 306, 573

\bibitem[Walter et al.(1997)]{walter97} Walter, F., et al., 1997, AJ, 114, 1544.

\bibitem[Wang et al. (2004)]{wang04} Wang, H., Mundt, R., Henning, Th., Apai, D., 2004, ApJ 617, 1191

\bibitem[Weidenschilling(1977)]{1977Ap&SS..51..153W} Weidenschilling, S.~J.\ 1977, \apss, 51, 153 

\bibitem[Wilking et al.(1985)]{wilking85} Wilking et al., 1985, ApJ, 293, 165

\bibitem[Williams et al.(1994)]{1994ApJ...428..693W} Williams, J.~P., de Geus, E.~J., \& Blitz, L.\ 1994, \apj, 428, 693 


%\bibitem[]{}

\end{thebibliography}
\end{document}